\documentclass[useAMS,usenatbib]{mn2e}
\usepackage{amsmath}%\
\usepackage{amsfonts}%
\usepackage{amssymb}
\usepackage{graphicx, natbib, wasysym, subfiles, aas_macros}

\renewcommand{\vec}[1]{\ensuremath{\boldsymbol{#1}}} % vectors are bold italic
\DeclareMathAlphabet{\textbfsf}{\encodingdefault}{\sfdefault}{bx}{sl}
\usepackage{ulem}
\usepackage{color}

\title[Unstable accretion in  TW Hya]{Unstable accretion in TW Hya: 3D simulations and comparisons with observations}

\author[Romanova et al.]{\parbox{\textwidth}{M. M. Romanova$^{1,2}$\thanks{E-mail of
corresponding author: \texttt{romanova@astro.cornell.edu}}, C. C.  Espaillat$^{3,4}$,  J. Wendeborn$^{3,4}$, J.-F. Donati$^{5}$,
P. P. Petrov$^{6}$,
  R. V. E. Lovelace$^{1,2}$}
\vspace{0.4cm}\\
\parbox{\textwidth}{ % affil list
$^{1}$Department of Astronomy, Cornell University, Ithaca, NY 14853-6801, USA\\
$^{2}$Carl Sagan Institute, Cornell University, Ithaca, NY 14853-6801, USA\\
$^{3}$ Department of Astronomy, Boston University, 725 Commonwealth Avenue, Boston, MA 02215, USA \\
$^{4}$ Institute for Astrophysical Research, Boston University, 725 Commonwealth Avenue, Boston, MA 02215, USA \\
$^{5}$ CNRS/Universite de Toulouse
IRAP - Observatoire Midi-Pyrenees
14 Av E Belin
31400 Toulouse Cedex
France \\
$^{6}$ Crimean Astrophysical Observatory, p/o Nauchny, 298409, Republic of Crimea\\}}
\date{\today}

\pagerange{\pageref{firstpage}--\pageref{lastpage}} \pubyear{2002}

\begin{document}
\label{firstpage}

\maketitle

\begin{abstract}

\noindent  We investigate the origin of photometric variability in the classical T Tauri star TW Hya by comparing light curves obtained by \textit{TESS} and ground-based telescopes with light curves created using three-dimensional (3D)  magnetohydrodynamic (MHD) simulations.  TW Hya is modeled as a rotating star with a dipole magnetic moment, which is slightly tilted about the rotational axis. We observed that for various model parameters, matter accretes in the unstable regime and produces multiple hot spots on the star's surface, which leads to stochastic-looking light curves similar to the observed ones. 
Wavelet and Fourier spectra of observed and modeled light curves show  multiple  quasiperiodic oscillations  (QPOs)
with quasiperiods from less than 0.1 to 9 days.
 Models show that variation in the strength and tilt of the dipole magnetosphere  leads to different periodograms,
 where the period of the star may dominate or be hidden. 
The amplitude of QPOs associated with the stellar period can be  smaller than that of other QPOs if the tilt of the dipole magnetosphere is small and when the unstable regime is stronger.
In models with small magnetospheres, the short-period QPOs associated with rotation of the inner disc dominate and can be mistaken for a stellar period.
We show that longer-period (5-9 days) QPOs  can be caused by waves forming beyond the corotation radius.

\end{abstract}

\begin{keywords}
accretion discs, hydrodynamics, planet-disc interactions,
protoplanetary discs
\end{keywords}

\section{Introduction}
\label{sec:Introduction}

Classical T Tauri stars (CTTSs) show photometric variability on different time scales, from seconds to decades  (e.g.,  \citealt{HerbstEtAl1994,HerbstEtAl2002,HartmannEtAl2016,FischerEtAl2023}).  
Observations of multiple CTTSs in clusters revealed that 
only a few CTTSs show periodic light curves. Most show quasi-periodic or stochastic-looking light curves (e.g., \citealt{AlencarEtAl2010,CodyEtAl2014,CodyEtAl2022,CodyHillenbrand2018,RobinsonEtAl2022}). Many light curves show bursts, which may indicate stochastic accretion (e.g., \citealt{StaufferEtAl2014}).

To better understand the mass accretion process in CTTSs,  a multi-epoch, multi-wavelength spectroscopic and photometric monitoring
campaign of four CTTSs (TW Hya, RU Lup, BP Tau, and GM Aur) was carried out in 2020/2021 (Epoch 1) and
2022/2023 (Epoch 2) as part of the \textit{UV Legacy Library of Young Stars as Essential Standards} (ULLYSES) $HST$ Director's Discretionary Time Program \citep{RomanDuvalEtAl2020} and the \textit{Outflows and Disks Around Young Stars: Synergies for the Exploration of
ULLYSES Spectra} (ODYSSEUS) program (\citealt{EspaillatEtAl2022}). Light curves were obtained using several ground-based telescopes at multiple wavelengths (see Tab. 1 in \citealt{WendebornEtAl2024b}).
All stars show significant variability in their light curves at different time scales; some time scales are associated with the period of the star (like in GM Aur), while in RU Lupi, the stochastic component dominates \citep{WendebornEtAl2024a,WendebornEtAl2024b}. Those authors conclude that there is a strong connection between mass accretion rate and photometric variability. Therefore, it is important to understand which type of accretion processes can produce such light curves.

CTTSs typically have a strong, $\sim kG-$scale, magnetic field (e.g., \citealt{Johns-Krull2007,DonatiLandstreet2009}), which stops
 the accretion disc and the matter is expected to accrete onto the star in two funnel streams forming two ordered spots near magnetic poles (e.g., \citealt{Konigl1991,BouvierEtAl2007a,HartmannEtAl2016}).
Early global three-dimensional (3D) magnetohydrodynamic (MHD) simulations of accretion onto a star with a tilted dipole magnetosphere confirmed the theory and have shown
the magnetospheric accretion in two funnel streams \citep{RomanovaEtAl2003,RomanovaEtAl2004} which form two banana-shaped spots at the star's surface, and the light curve is almost sinusoidal.
Some observations suggest that accretion occurs through multiple streams (e.g., \citealt{InglebyEtAl2013,
JohnstoneEtAl2014,RobinsonEspaillat2019}).

Global 3D MHD simulations performed at a broader range of parameters and finer grid resolution have shown that in models with a small tilt of the dipole magnetosphere,  the matter often accretes in the unstable regime, where it penetrates through the  magnetosphere  due to the magnetic Rayleigh-Taylor (interchange) instability  \citep{RomanovaLovelace2006}. 
Earlier, this instability was proposed by \citet{AronsLea1976} for mixing matter and the field in the external parts of the magnetosphere.  However, the penetration of filaments (or tongues) into the deep layers of the magnetosphere was not expected. This unstable regime has been studied in detail by \citet{KulkarniRomanova2008,KulkarniRomanova2009}, and \citet{RomanovaEtAl2008} in 3D MHD simulations. Simulations show that matter accretes in several unstable tongues, which form at the inner edge of the disc, creating frequent and randomly located hot spots on the stellar surface.
 Fourier and wavelet analysis show that in a strongly unstable regime, QPO associated with rotating tongues may dominate over the period of the star
producing false ``periods".  

 In models with smaller magnetospheres and slower rotating stars, an ``ordered unstable regime"  has been observed where matter accretes in one or two unstable tongues that rotate with the angular velocity of the inner disc.  The period of their rotation 
 may be much smaller than the period of the star (e.g., \citealt{RomanovaKulkarni2009}).  
\citet{BlinovaEtAl2016} studied the unstable regime at multiple parameters and derived a boundary between stable and unstable regimes.
An unstable accretion regime has also been observed in more recent 3D MHD simulations of other groups (e.g., \citealt{TakasaoEtAl2022,ParfreyTchekhovskoy2023,ZhuEtAl2024}).

Several groups used the paradigm of unstable regime to explain the light curves 
of CTTSs. 
For example, \citet{SiwakEtAl2011} analyzed the light curves of TW Hya obtained with the \textit{MOST} satellite and concluded that irregular
 light variations could be caused by the stochastic variability of hot spots induced by unstable accretion. 
\citet{TakamiEtAl2016} analyzed spectra of the active star RW Aur A and suggested that it may switch between stable and unstable regimes during its evolution when comparing the complex variability of different spectral lines to that simulated
by \citet{KurosawaRomanova2013}.

Overall, the unstable regime is frequently used to explain the quasi-periodic or stochastic light curves of CTTS. However,  the observed light curves have never been compared with those obtained by dedicated numerical models of stars accreting in the unstable regime. Here, we choose one of the CTTS stars (TW Hya) targeted by the ODYSSEUS campaign and develop a 3D MHD model of a star with the parameters of TW Hya.
We analyze and compare the observed and modeled light curves and QPOs. 

In Sec. \ref{sec:Observations} , we present observational data of TW Hya and describe our numerical model in Sec. \ref{sec:num-model}. In Sec. \ref{sec:results}, we show the results of simulations and compare them with observations. In Sec. \ref{sec:Discussion} , we discuss different points and conclude in Sec. \ref{sec:Conclusions}.

\begin{figure}
     \centering
     \includegraphics[width=0.49\textwidth]{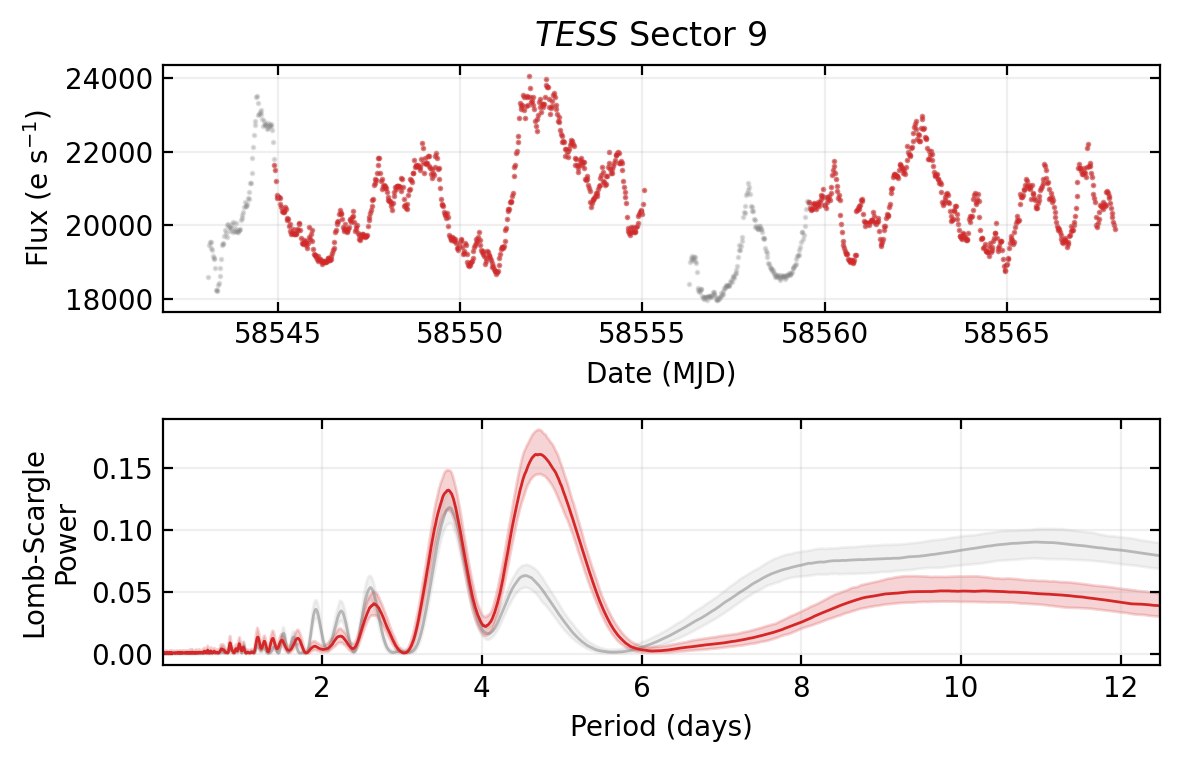}
   \includegraphics[width=0.49\textwidth]{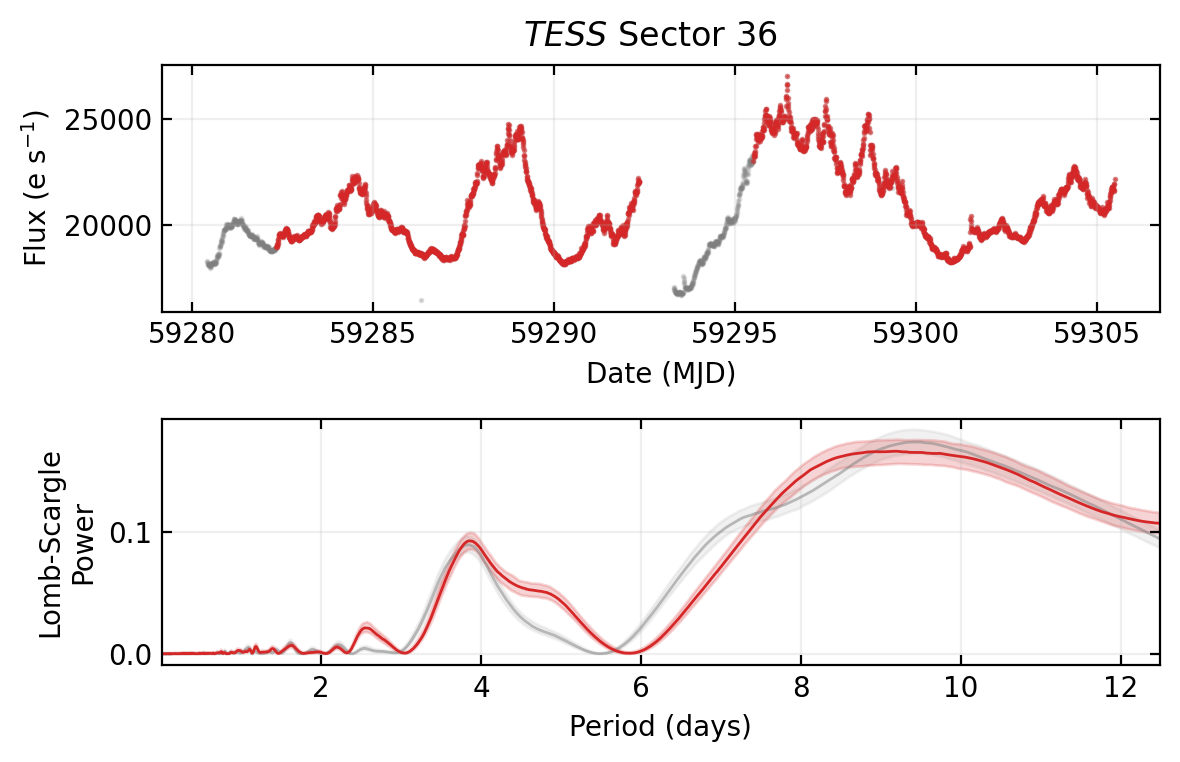}
   \includegraphics[width=0.49\textwidth]{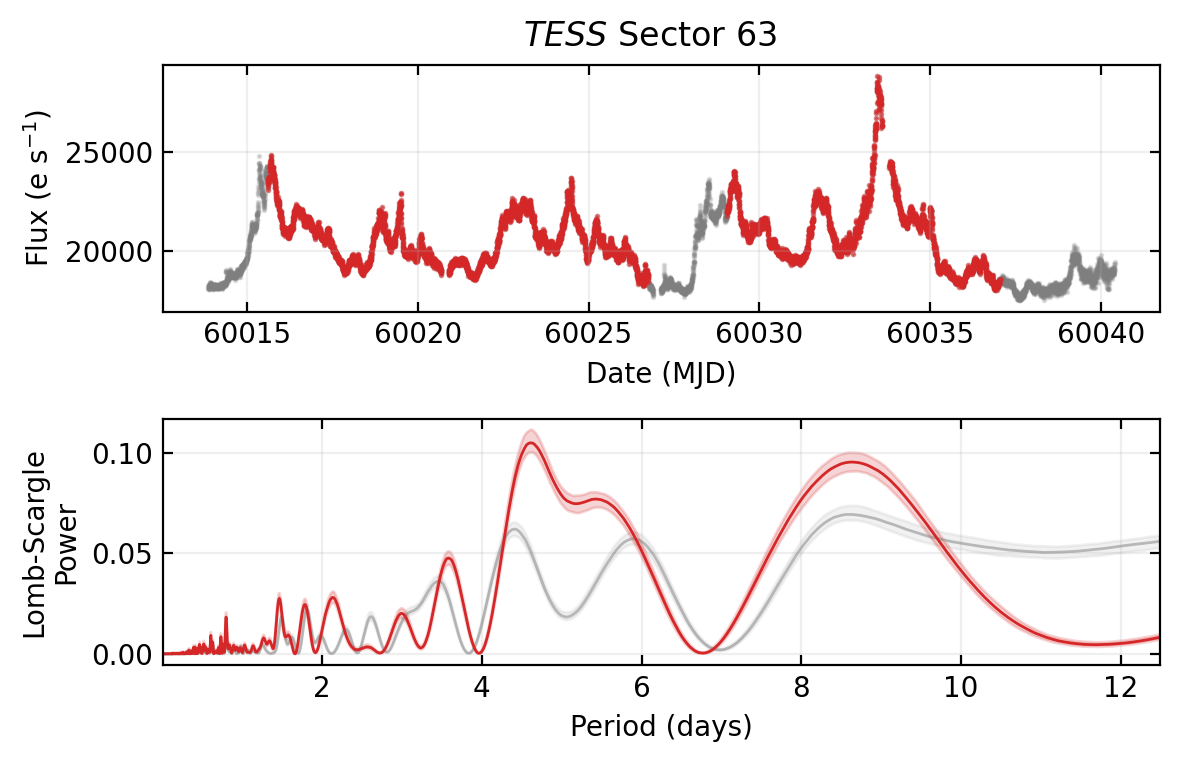}
    \caption{\textit{Top two panels:} Top: \textit{TESS} light curve in Sector 9 for TW Hya. Grey points have been flagged by either the default \textit{TESS} pipeline or by tglc (typically due to high background flux), while red points are data that have not been flagged. Bottom: Lomb-Scargle periodograms for the associated light curves. The red curve is calculated using the non-flagged data, while the grey curve is calculated using all data, including flagged points. Shaded regions are 1-sigma uncertainties calculated using a bootstrap approach. \textit{Middle two panels:} The same but for Sector 36.  \textit{Bottom two panels:} The same but for Sector 63. 
     \label{tess-3}}
\end{figure}

\begin{figure}
     \centering
     \includegraphics[width=0.49\textwidth]{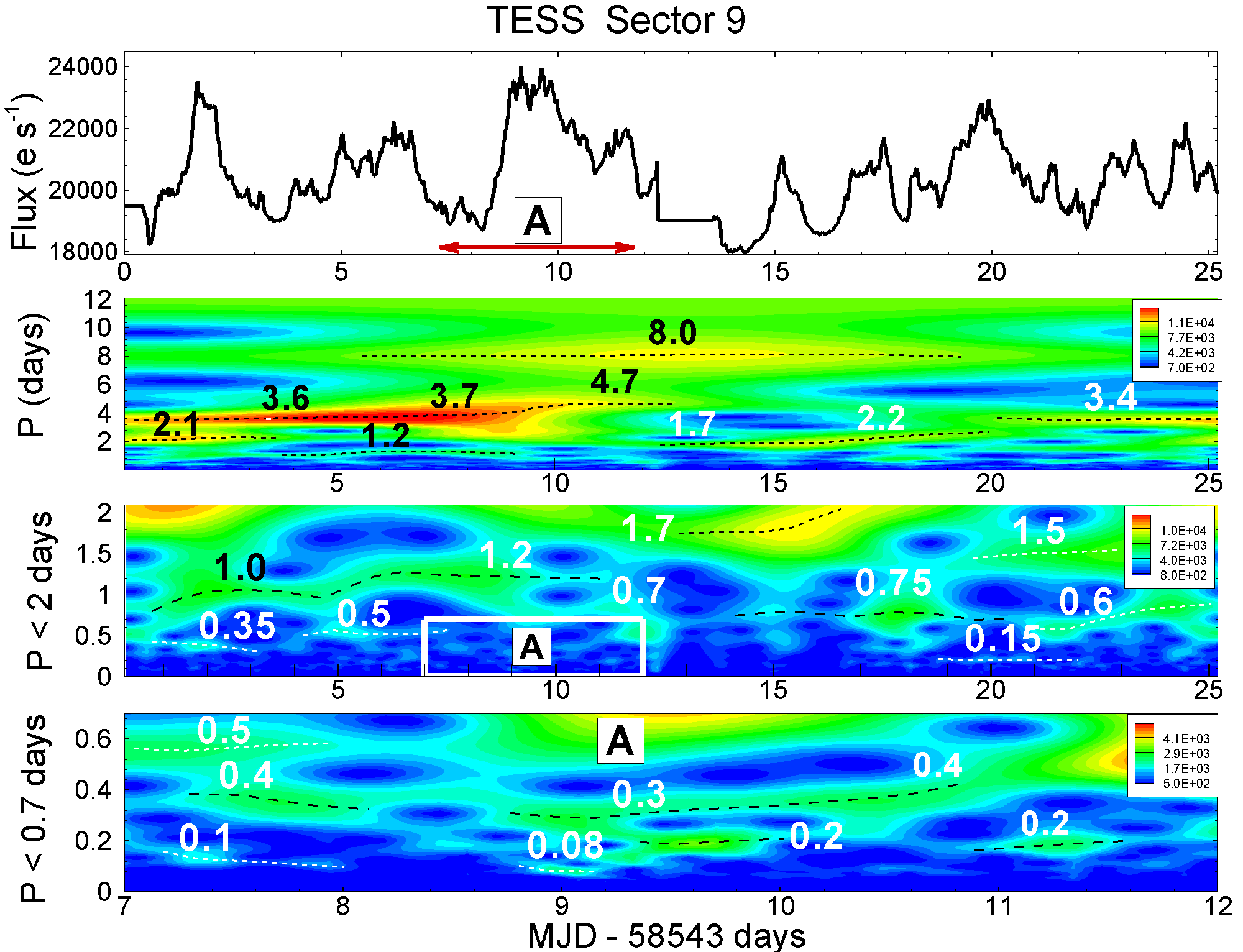}
\medskip
     \includegraphics[width=0.49\textwidth]{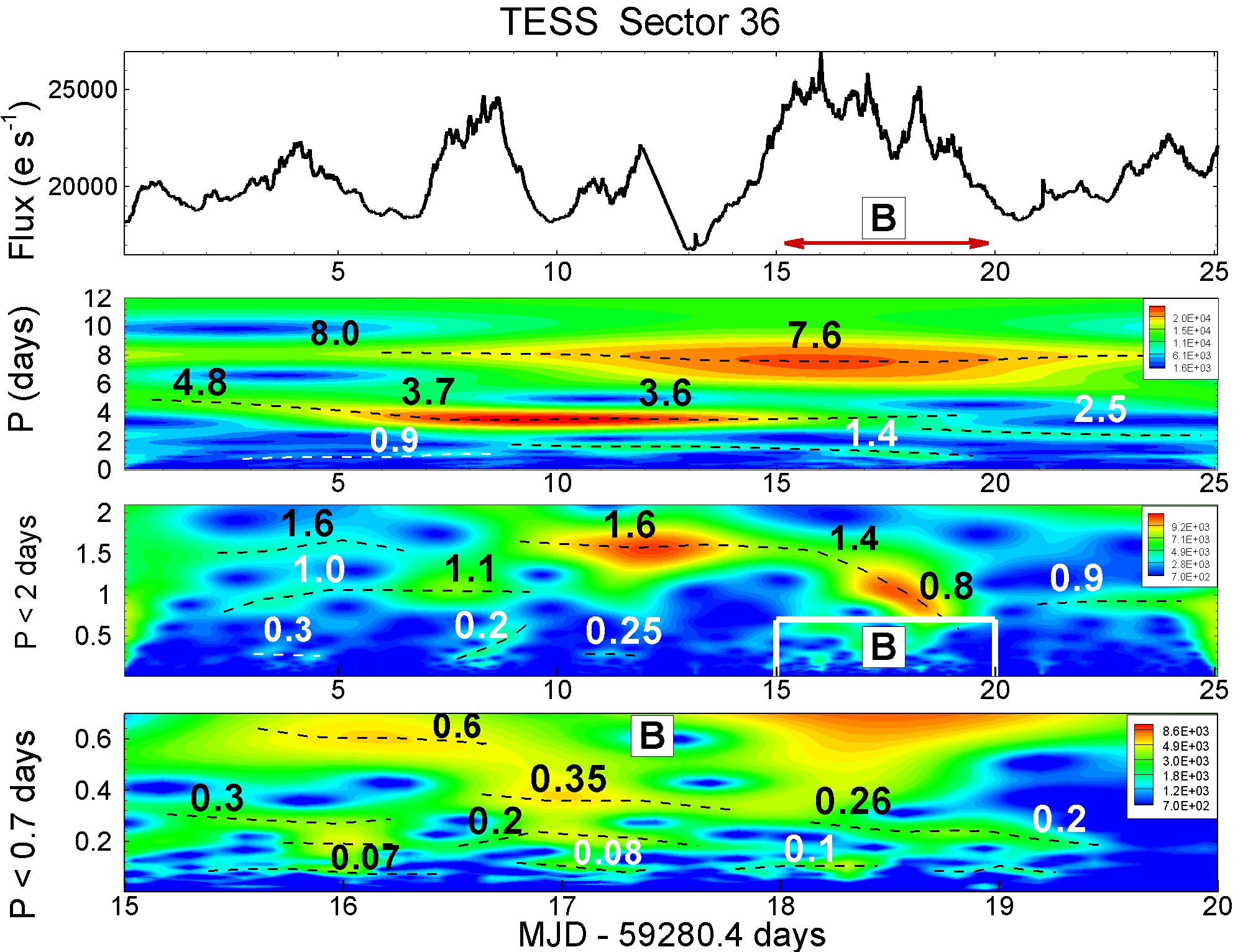}
\medskip
   \includegraphics[width=0.49\textwidth]{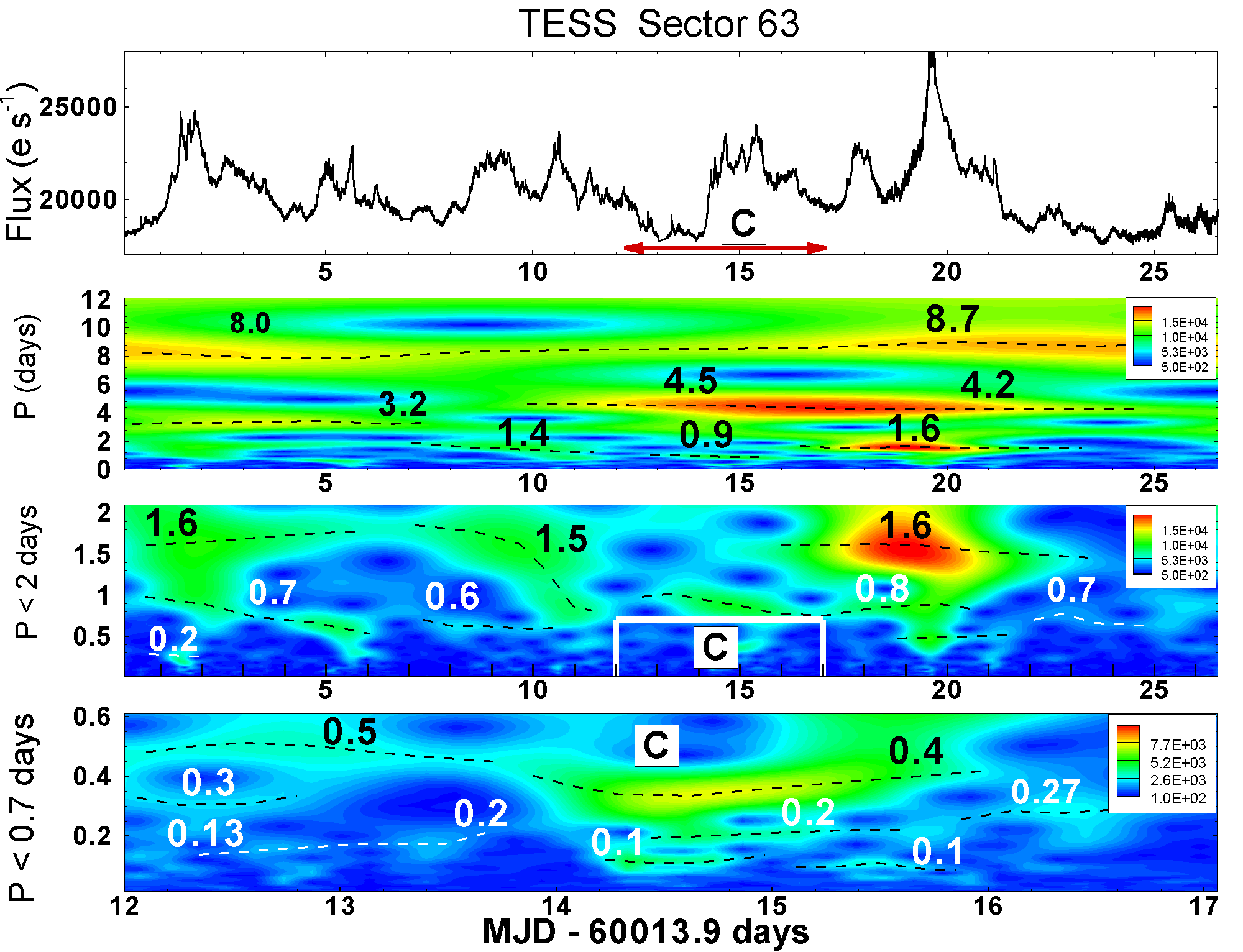}
    \caption{Morlet wavelet analysis of the light curves obtained by \textit{TESS} during Sectors 9, 36, and 63 observations. 
 \textit{Top panels:} the light curves, obtained with homogeneous grid which include flagged points and interpolation. 
 \textit{2nd panels from top:} Morlet wavelet obtained for light curves. \textit{3rd panels from top:} A part of wavelets for periods of $P<2$ days. \textit{Bottom panels:} A part of wavelets for periods of $P<0.7$ days taken during the 5-days interval (marked in red in top panels).
     \label{TESS-lc-w-all}}
\end{figure}

\begin{figure*}
     \centering
     \includegraphics[width=0.9\textwidth]{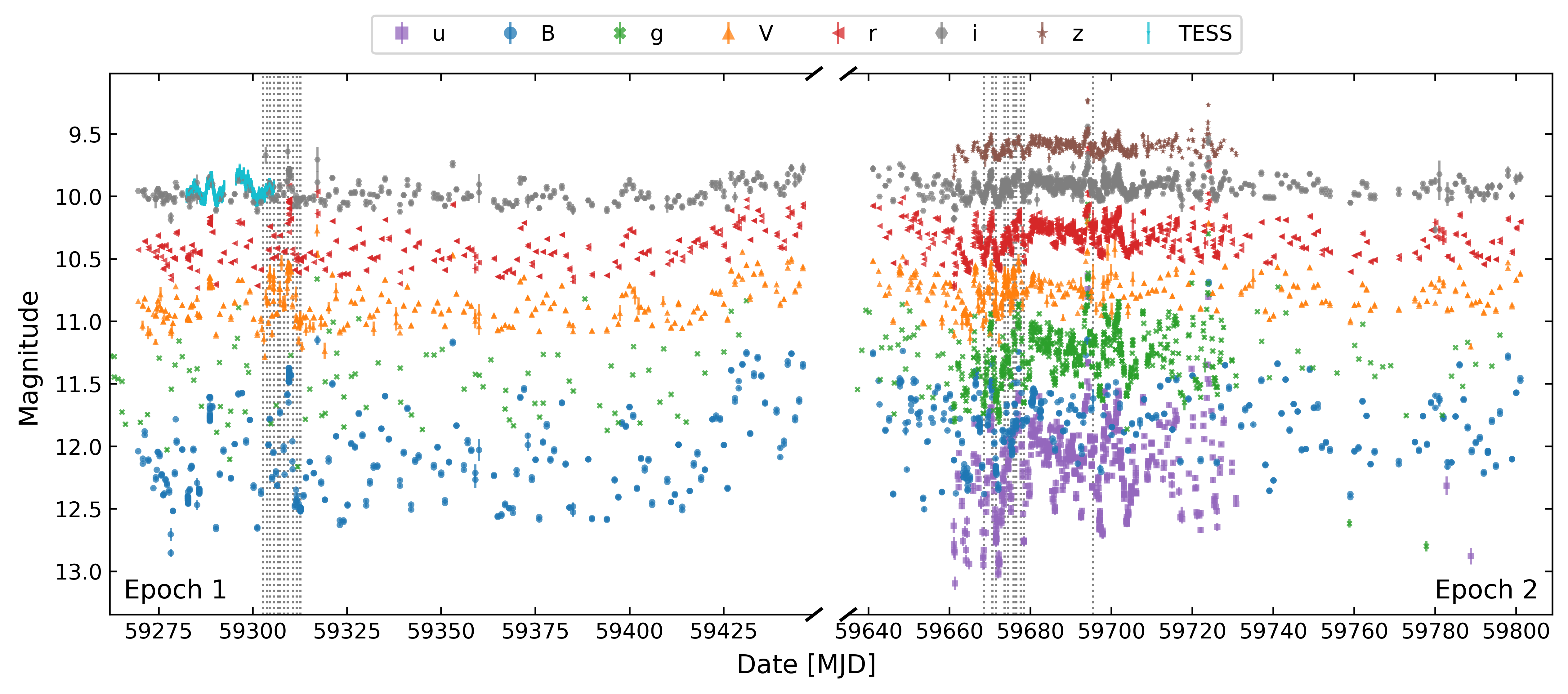}
     \includegraphics[width=0.9\textwidth]{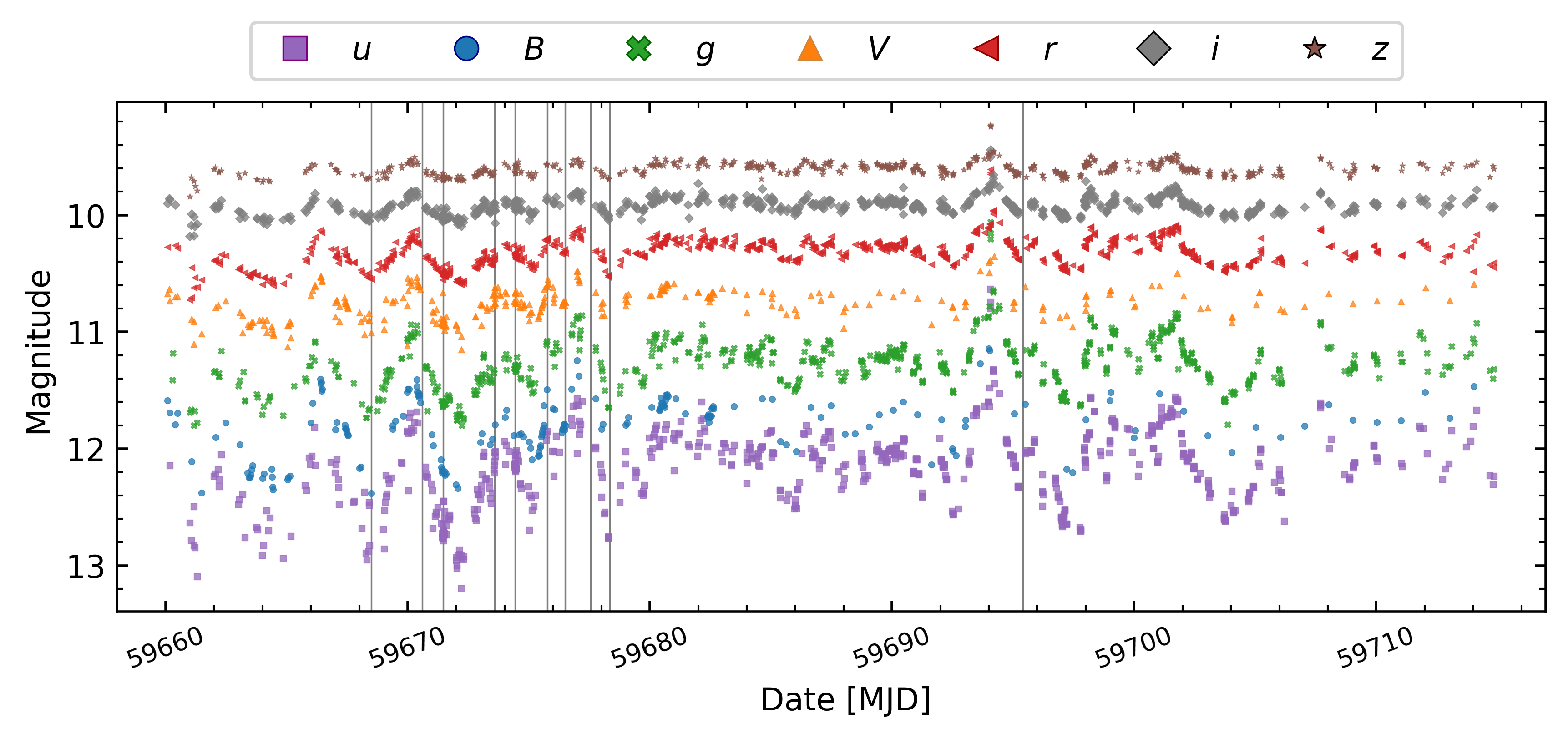}
    \caption{\textit{Top panel:} Optical light curves for TW Hya obtained during Epochs 1 (2021, left) and  Epoch 2 (2022, right)  of observations. Symbols corresponding to uBgVriz data are labeled in the key.
Data obtained by \textit{TESS} are shown in Epoch 1. 
More details on the sources of the photometry are given in Table 1 of \citet{WendebornEtAl2024b}.
\textit{Bottom panel:} A part of Epoch 2 light curves (MJD 59660 -59715) where the data were recorded more frequently. The vertical lines correspond to the times of the \textit{HST} ULLYSES observations. 
 \label{fig:multi-all}}
\end{figure*}

\begin{figure}
     \centering
     \includegraphics[width=0.49\textwidth]{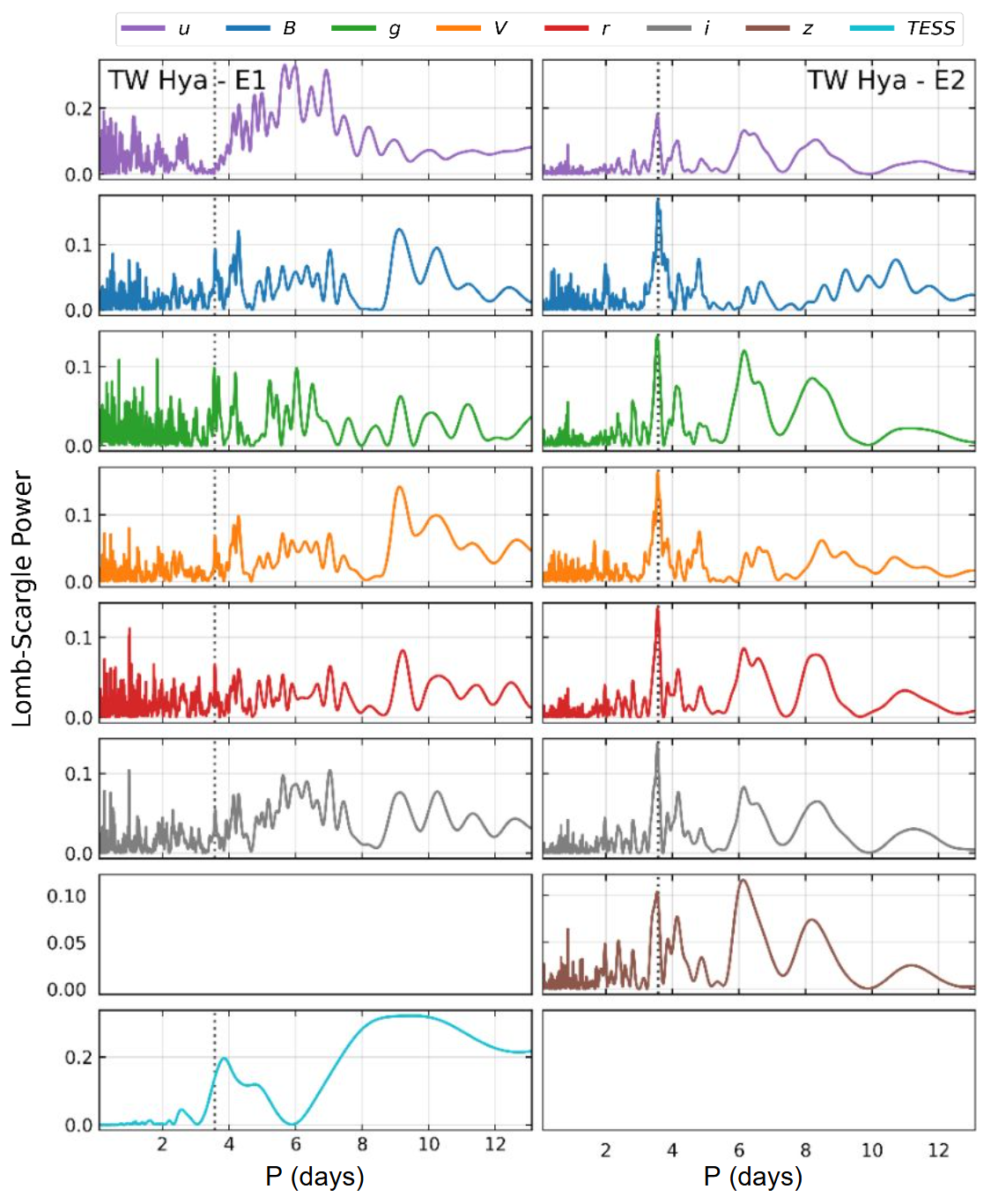}
    \caption{\textit{Left panels:} Lomb-Scargle periodograms for TW Hya  during Epoch 1. 
 \textit{Right panels:} The same but for Epoch 2. From \citet{WendebornEtAl2024b}.
     \label{fig:Scargle-E1-E2}}
\end{figure}

\begin{figure*}
     \centering
     \includegraphics[width=0.49\textwidth]{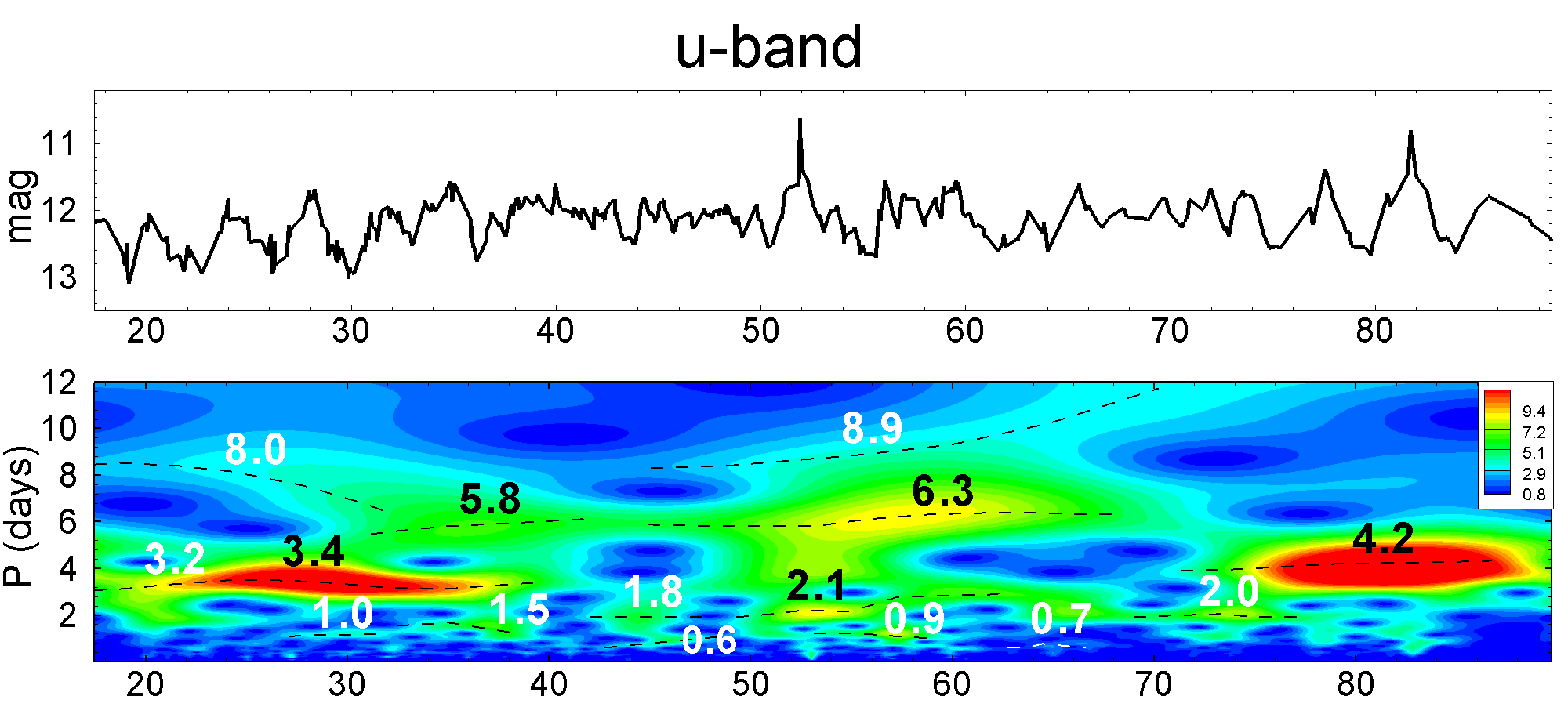}
     \includegraphics[width=0.49\textwidth]{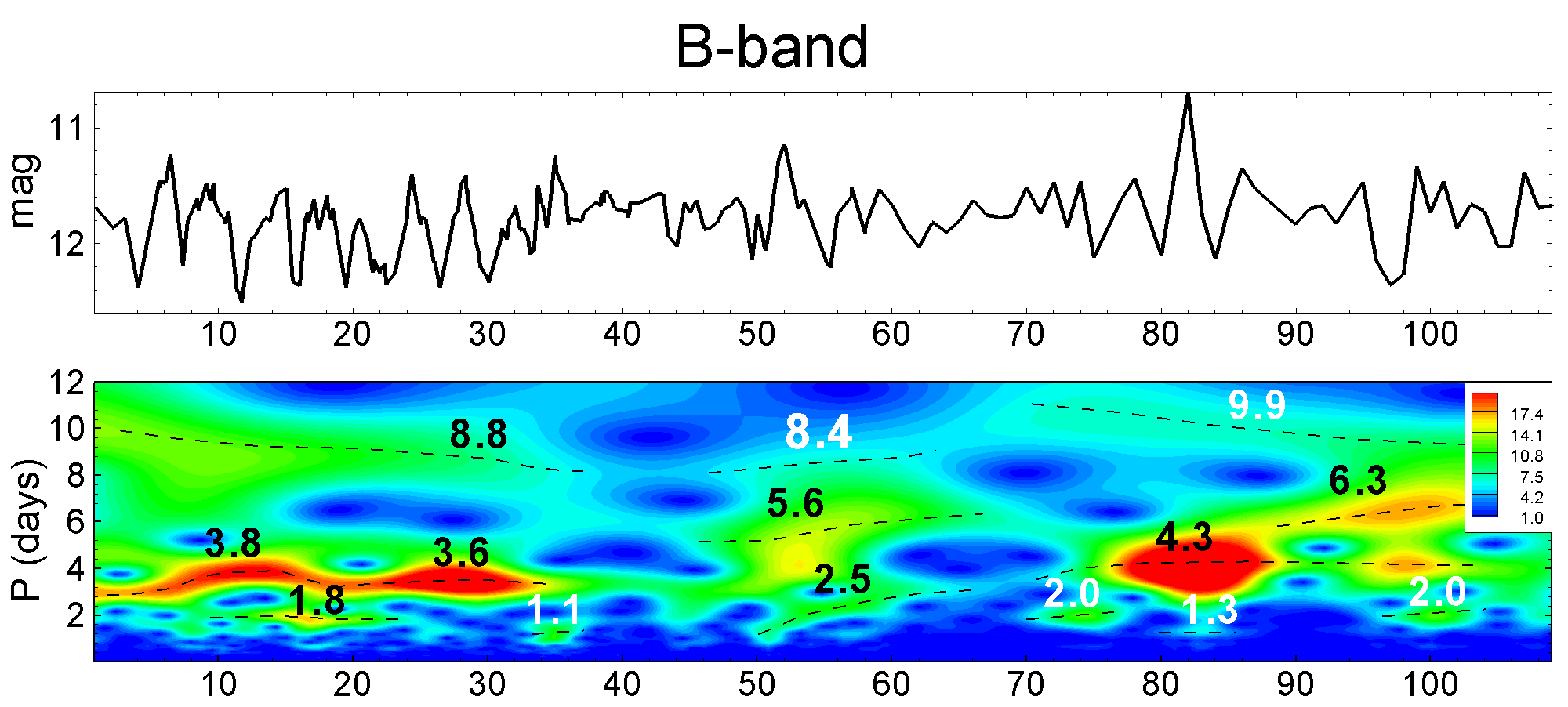}
\includegraphics[width=0.49\textwidth]{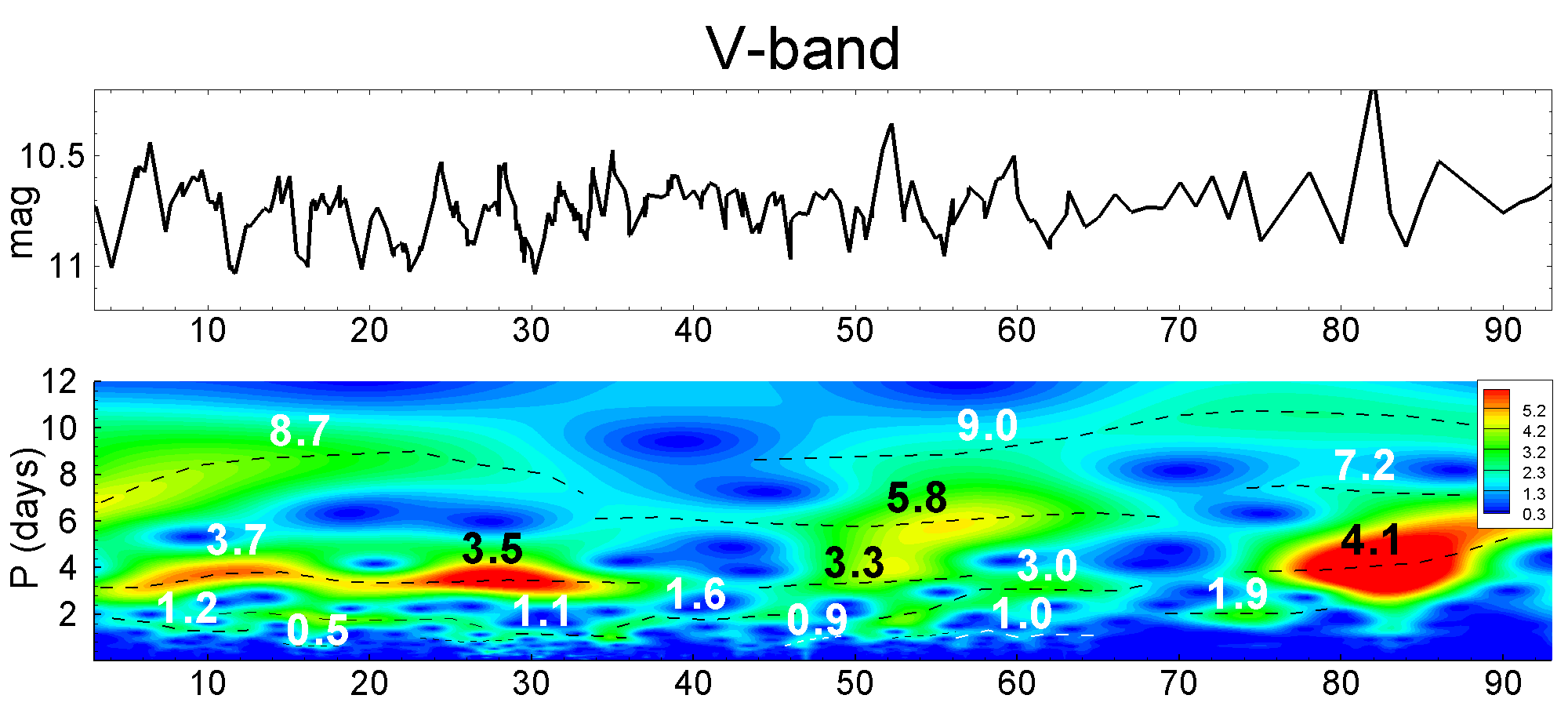}
\includegraphics[width=0.49\textwidth]{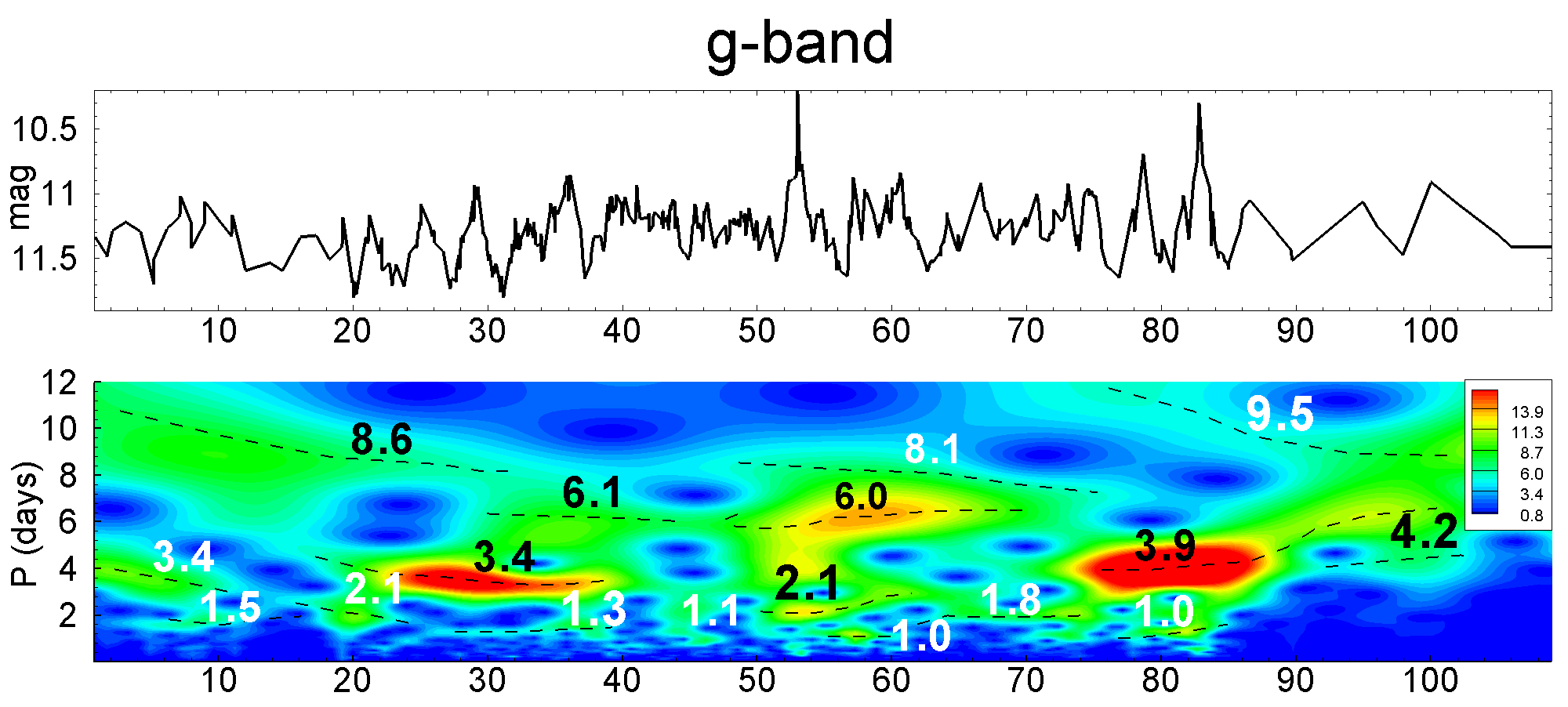}
     \includegraphics[width=0.49\textwidth]{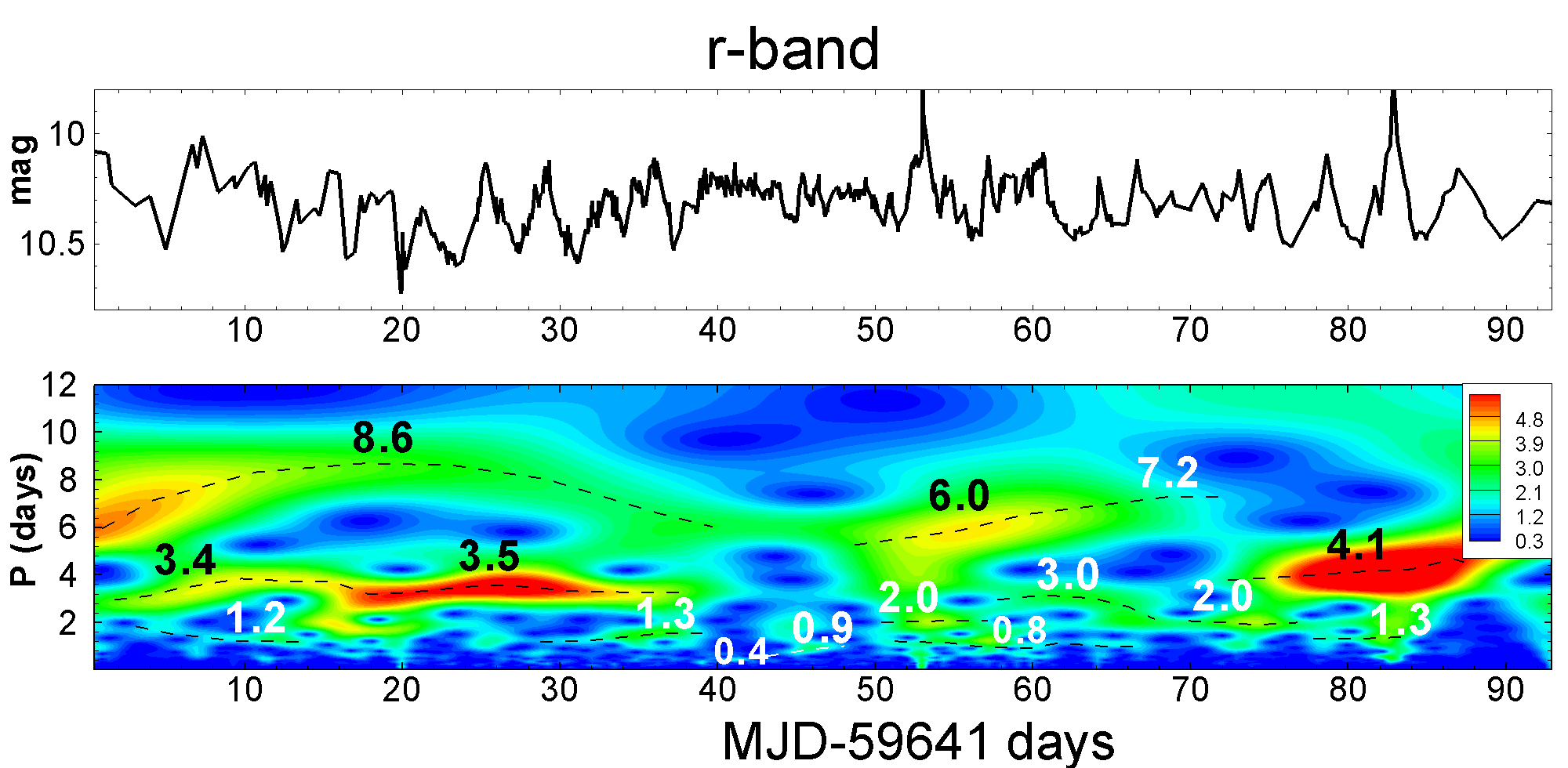}
     \includegraphics[width=0.49\textwidth]{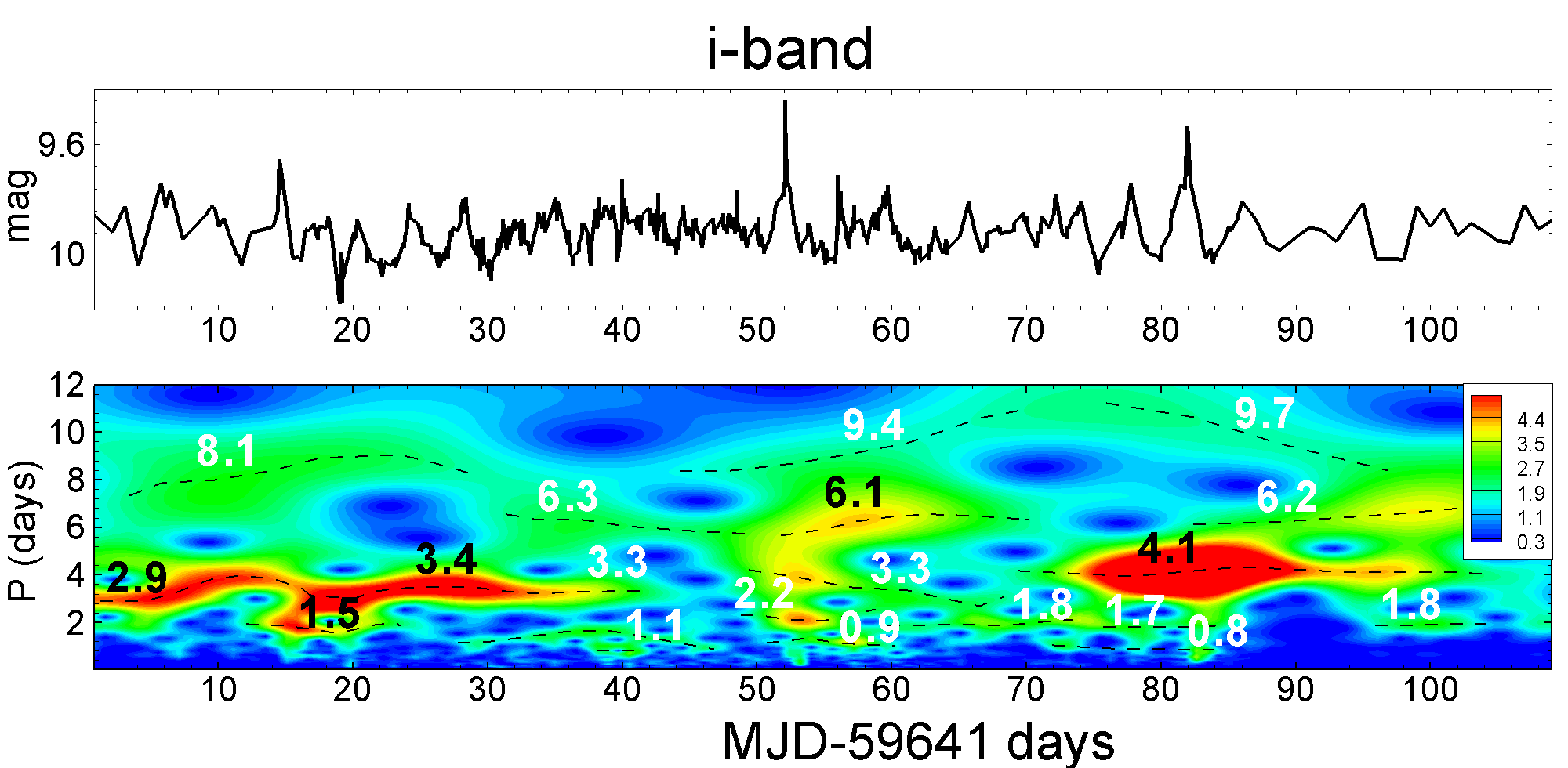}
 \caption{Light curves and wavelets obtained from multiwavelength observations by multiple ground-based telescopes during a part of Epoch 2.
%The time interval is 5942 - 59715 days (except for the u-band, which is 59660 - 59715 days).
 \label{fig:multi-wavelength}}
\end{figure*}

\section{Observations of TW Hya}
\label{sec:Observations}

TW Hya is a bright, nearby CTTS of the spectral type K6/K7 (when measured at blue/optical wavelengths; \citealt{ManaraEtAl2014}).
It retains a large massive disc at the age of 8--10 Myr \citep{SokalEtAl2018} and also contains a number of gaps and rings (e.g. \citealt{CalvetEtAl2002}, \citealt{AndrewsEtAl2016}). 
The disc and the magnetosphere are both seen nearly pole-on (e.g., \citealt{QiEtAl2004}; \citealt{DonatiEtAl2011}). 
Table \ref{tab:TWHya-param} summarizes the observational parameters of TW Hya.  

\subsection{Searching for a period of TW Hya}

Many attempts have been made to determine the rotation period of TW Hya using photometric data obtained with
ground-based telescopes.
\citet{Rucinski1988} did
not find a significant period from observations in 1986, but reported a possible 2-day period from earlier data.
 \citet{HerbstKoret1988} found a 1.28-day period. \citet{Mekkaden1998} obtained a 2.2-day period and pointed out the presence of sudden brightenings, which could be due to short-lived
hot spots that can occur at any photometric phase and
could mask the periodicities. \citet{KastnerEtAl1999}
examined the optical V-band photometry from \textit{Hipparcos}, and showed that neither
of the above periods produced a regular folded light curve.
They suggested that quasi-random flaring, instead of a hot
spot, could be causing the appearance and disappearance of
periods in different data sets.
\citet{AlencarBatalha2002} analyzed spectra of TW Hya and performed
periodogram analysis of the veiling and
veiling-corrected line intensity variations using the Scargle
(1982) periodogram estimator.
They  obtained three prominent periodicities at 1.4±0.1, 2.85±0.25
and 3.75 ± 0.45 days. A similar study by \citet{BatalhaEtAl2002} pointed
to 4.4 ± 0.4 days as the stellar rotation period.  \citet{LawsonCrause2005} found a period of 2.80 days.
So, ground-based observations suggested a range of periods for TW Hya.

\citet{SetiawanEtAl2008} detected 
periodic (3.56 days) radial velocity variability in their
spectroscopic observations,  interpreted as the rotation of a 10 Jupiter mass planet. 
However, \citet{HuelamoEtAl2008} has shown that this period can be explained by  
the rotation of a cool stellar spot, and they measured a stellar rotation period of 3.57 days. 
%However, variability associated with accretion and hot spots continued providing different quasiperiods.

The search for QPOs associated with magnetospheric accretion and hot spots became even more intense and interesting with data from space telescopes, which provide frequent and continuous observations.
\citet{RucinskiEtAl2008} analyzed photometric variability in TW Hya using data from  the \textit{MOST} 
(Microvariability \& Oscillations of STars)  space telescope
on timescales from a fraction of a day to 7.5 weeks.
 A 3.7-day period was observed in the continuous 11-day observations performed with 0.07 days
time resolution in 2007. 
 However, this
periodicity was absent in the second, 4 times longer MOST run in 2008. Instead, a spectrum of 
quasiperiods
within the 2–9 days range was observed.

\citet{SiwakEtAl2011} studied the variability of TW Hya using the light curves obtained by the
\textit{MOST} satellite and the All Sky Automated Survey
(ASAS) project over 40 days in 2009 with a temporal resolution of 0.2 days. A wavelet
analysis of the combined MOST–ASAS data provided a rich picture of QPOs with periods of 1.3–10 days,  similar to those discovered in the 2008 data. The authors concluded that the observed shortest oscillation period may indicate a stellar rotation period of
1.3 or 2.6 days, synchronized with the disc at 4.5 or $7.1R_*$, respectively.

\citet{SiwakEtAl2014} presented an analysis of the 2011 photometric observations of TW Hya by the \textit{MOST} satellite. 
The light variations were dominated
by a strong, quasi-periodic 4.18-day oscillation with superimposed chaotic-looking flares.
They conclude that the QPOs were probably produced by stellar rotation, with one large hot spot created
by a stable accretion funnel, while the flaring component could be produced by small hot spots created at
moderate latitudes by unstable accretion tongues.

\citet{SiwakEtAl2018} report on photometric variability observed by MOST during 2014, 2015, and 2017. 
In 2014 and 2017, the light curves showed stable 3.75- and 3.69-day QPOs,
respectively. Both values appear closely related to the stellar rotation period,
as they might be created by changing the visibility of a hot spot formed near the magnetic pole.
 These major light variations were superimposed on a chaotic,
flaring-type activity caused by hot spots resulting from unstable accretion – a situation reminiscent
of that in 2011 when TW Hya showed signs of a moderately stable accretion state.
In 2015, only drifting QPOs were observed, similar to those present in 2008–2009
data and typical for magnetized stars accreting in a strongly unstable regime.

Observations of TW Hya with the \textit{Transiting Exoplanet Survey Satellite} (\textit{TESS}) were performed in Sectors 9, 36 and 63. 
%Observations in Sector 36 overlapped with Epoch 1 of the ODYSSEUS program. 
Scargle-Lomb periodograms of \textit{TESS} light curves show QPOs with a period close to the star's period. However, QPOs with other periods are also observed and often have higher amplitudes \citep{WendebornEtAl2024b}. The ODYSSEUS program's ground-based telescopes observed TW Hya in 2021 (Epoch 1) and 2022 (Epoch 2). Scargle-Lomb periodograms show a QPO with a stellar period in Epoch 2. However, no QPO with a stellar period was observed in Epoch 1 \citep{WendebornEtAl2024b}.
Therefore, observations of TW Hya show the star's period during some observing runs but not during others.

\subsection{Unstable regime. Fastness parameter}

The unstable regime has been studied in multiple 3D MHD simulations by 
\citet{BlinovaEtAl2016}. They conclude that the ``strength" of instability and the boundary
between stable and unstable regimes depends 
on the fastness parameter $\omega_s=\Omega_s/\Omega_K=(r_m/r_{\rm cor})^{3/2}$, where $r_m$ is the magnetospheric (or truncation radius), where matter is stopped by the magnetosphere, and $r_{\rm cor}$ is the corotation radius, where the angular velocity of the star matches the Keplerian angular velocity of the disc; $\Omega_K$ is the Keplerian angular velocity of the disc at $r=r_m$. At a small tilt of the dipole $\theta=5^\circ$, they found a boundary at $\omega_s\approx 0.6$ which corresponds to $r_m/r_{\rm cor}\approx 0.71$, while at the large tilt $\theta=20^\circ$, the boundary is at   $\omega_s\approx 0.54$ which corresponds to  $r_m/r_{\rm cor}\approx 0.66$.
They also observed that in smaller magnetospheres $r_m\lesssim 4.2 R_*$, matter accretes in one or two ordered tongues if $\omega_s\approx 0.45$ ($r_m/r_{\rm cor}\lesssim 0.59$). They rotate with the period of the inner disc and may provide a false period of the star, which can be much shorter than the actual period.

The corotation radius is determined by the period of the star and its mass. The period of TW Hya is known as $P_*\approx 3.56$ days \citep{SetiawanEtAl2008,HuelamoEtAl2008}. We present the corotation radius in the following form:  
\begin{equation}
 r_{\rm cor}=\bigg[\frac{GM_*P_*^2}{(2\pi)^2}\bigg]^{1/3} \approx 9.10 R_\odot\bigg(\frac{M_*}{0.8M_\odot}\bigg)^{1/3}\bigg(\frac{P_*}{3.56 {\rm days}}\bigg)^{2/3}~,
 \end{equation}
 where $M_*$ is the mass of the star.

There are different estimates of TW Hya radius: 
 $R_*=0.85\pm0.25 R_\odot$ \citep{HughesEtAl2007}; $R_*=0.93 R_\odot$ \citep{RobinsonEspaillat2019}; $R_*=1.1 R_\odot$ \citep{RheeEtAl2007}; $R_*=1.16\pm0.13 R_\odot$ \citep{DonatiEtAl2024}; $R_* = 1.29\pm0.19 R_\odot$ \citep{GRAVITY2020}
We adopt a radius of $R_*=0.93 R_\odot$, mass $M_*=0.79M_\odot$, and stellar period $P=3.5$ days and obtain a corotation radius $r_{\rm cor}\approx 9.6 R_*$, which we use in most of our simulations. 
  
There are different theoretical approaches for finding the magnetospheric radius. 
In one of them, it is suggested that the matter of the disc stops when magnetic pressure balances the total 
matter pressure:    
${B_d^2}/{8\pi}=p+\rho v^2$, where $B_d$ is the magnetic field at the disc-magnetosphere boundary, $p-$ and $\rho v^2-$ are the thermal and ram pressure of matter in the disc.
In accretion disk $p\ll\rho v^2$, $v\approx v_{\phi}$, so that
${B^2}/{8\pi}\approx\rho v_{\phi}^2$, where $v_{\phi}=(GM_*/r)^2$ is the Keplerian
velocity of the disc. Substituting the magnetic moment
$\mu_*{\equiv}B_*R_*^3$, we obtain a widely used formula  (e.g., \citealt{LambEtAl1973,ElsnerLamb1977}):
\begin{eqnarray}
\nonumber \noindent r_m=k r_A=k\frac{\mu_*^{{4}/{7}}}{(2GM_*)^{1/7}\dot{M}^{2/7}}= 
k\times 4.09 R_\odot  \bigg(\frac{M_*}{0.8 M_\odot}\bigg)^{-1/7}  \\  
\times \bigg(\frac{\dot M}{2.51\times 10^{-9} M_\odot/yr}\bigg)^{-2/7}
\bigg(\frac{B_{* \rm eq}}{500 G}\bigg)^{4/7}\bigg(\frac{R_*}{R_\odot}\bigg)^{12/7}~,
\label{eq:r_m}
\end{eqnarray}
where coefficient $k\sim 1$.   \citet{BessolazEtAl2008} derived 
 $k\approx 0.77-1$ by comparing the theoretical value with the results of 2D (axisymmetric) simulations. \citet{KulkarniRomanova2013} derived $k=0.55-0.72$ from similar comparisons but in 3D simulations.
\citet{BlinovaEtAl2016} derived $k\approx 0.5-0.9$ in stable regime and  $k\approx 0.8-0.93$ in the \textit{unstable} regime from multiple 3D simulations  (see details in Sec. A of their paper)\footnote{The formulae for $r_m$ is also slightly different from Eq. \ref{eq:r_m}. The coefficient $k$ helps to adjust patameters to standard dependencies shown in Eq. \ref{eq:r_m}. }.

%\footnote{The formulae for $r_m$ is slightly different from Eq. \ref{eq:r_m}. The coefficient $k\approx 0.55-0.72$ helps to adjust patameters to %standard dependencies shown in Eq. \ref{eq:r_m}. 
%Coefficient $k=0.8$ is relevant to models with larger magnetospheres.}. 

In our research, we derive the magnetospheric radius from numerical models but use theoretical formulae for comparisons.
The magnetospheric radius mainly depends on the accretion rate and magnetic moment (or, magnetic field and radius) of the star. Below, we 
briefly discuss the results of observational measurements of these values of TW Hya. 

\subsection{The magnetic field of TW Hya}

\citet{DonatiEtAl2011} used the spectropolarimetry method to measure the magnetic field of TW Hya
from optical spectra secured with ESPaDOnS at the Canada-France-Hawaii Telescope (CFHT, \citealt{Donati2003}). 
They concluded that the field is predominantly octupolar of 2.5-2.8 kG with a smaller dipole component of a few hundred Gauss.
More recently, \citet{DonatiEtAl2024} repeated observations of TW Hya in 2019, 2020, 2021, and 2022 
using the Zeeman-Doppler Imaging\footnote{They observed TW Hya in the 
near-infrared 
with the SPIRou \citep{DonatiEtAl2020} high-resolution spectropolarimeter and velocimeter at the 3.6-m CFHT.}  and obtain that the large-scale field mainly consists of a 1 kG dipole tilted at about $20^\circ$  to the rotation axis,
whereas the small-scale field reaches strengths of up to 3-4 kG. 
They show  that the strength of the dipole component varies from 990 G to 1190 G,  the tilt of the dipole moment varies from $17^\circ$ up to $23^\circ$, and the phase varies too (see their Table 3). 

\subsection{Accretion rate}

The accretion rate of TW Hya obtained from
continuum and line analyses was determined as
$\dot M\approx 4\times 10^{-10} - 6\times 10^{-9} M_\odot {\rm yr}^{-1}$
 (e.g., \citealt{MuzerolleEtAl2000,AlencarBatalha2002,HerczegEtAl2002,DonatiEtAl2011,
RobinsonEspaillat2019,WendebornEtAl2024a, WendebornEtAl2024c}). 
An accretion rate   $\dot M\approx 2.51\times 10^{-9} M_\odot {\rm yr}^{-1}$ has been derived from optical spectra taken mostly with the ESPaDOnS instrument \citep{HerczegEtAl2023}.
Those authors noted that this accretion rate might be underestimated by a factor of up to 1.5 because of uncertainty in the bolometric correction and another factor of 1.7 because of excluding the fraction of accretion energy that escapes in lines, especially Ly$\alpha$
\footnote{
According to \citet{ArulananthamEtAl2023},
the major part of the accretion energy escapes in $Ly_\alpha$ radiation, and this factor may be significant (see also \citealt{FranceEtAl2014}).}.  If these factors are taken into account, then 
$\dot M\approx 6.40\times 10^{-9} M_\odot {\rm yr}^{-1}$. 

\citet{KastnerEtAl2002} analyzed the high-resolution X-ray spectrum of TW Hya and concluded that 
if accretion powers the X-ray emission, then the accretion rate should be    $\sim 10^{-8} M_\odot {\rm yr}^{-1}$.
\citet{NayakEtAl2024} observed TW Hya with the  Ultra-Violet Imaging
Telescope (UVIT). 
 Based on C iv line luminosity, they estimated an accretion luminosity (0.1 $L_\odot$) and a
the mass accretion rate of 
$2.2\times 10^{-8} M_\odot {\rm yr}^{-1}$. 
Due to the wide range of accretion rates measured for TW Hya, we consider different possible accretion rates in this paper.

\begin{figure}
\centering
\includegraphics[width=0.45\textwidth]{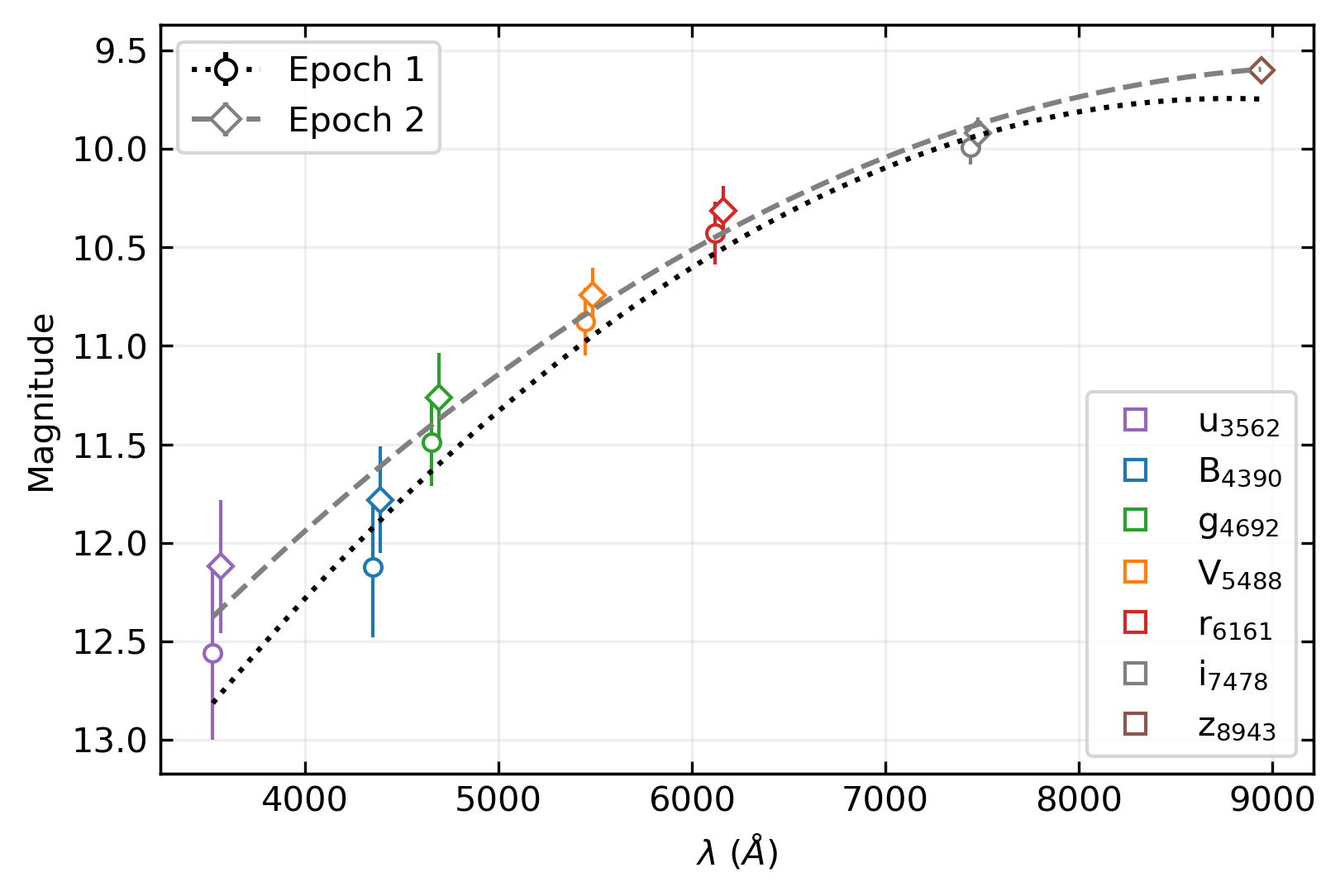}
\caption{The time-averaged stellar magnitude versus wavelength of the band for different bands observed in Epoch 1 (circles and the lower, short-dashed curve) and 2 (diamonds and upper, long-dashed curve). Short- and long-dashed lines show the best fit for Epochs 1 and 2, respectively.  
\label{fig:wavelength-dep}}
\end{figure}

\begin{table*}
\begin{tabular}[]{ c  | c}
 \hline
Parameter of TW Hya                                       &  Value   \\
\hline          
                Mass                                 &      $0.6 M_\odot$$^{a1}$      ~  $\textbf{0.79} M_\odot$ $^{a2}$         ~  $0.8 M_\odot$ $^{a3}$            \\
 Radius   &  $R_*=0.85\pm0.25$ $^{b1}$ ~ ${\bf 0.93} R_\odot$ $^{b2}$, ~ $R_*=1.1 R_\odot$ $^{b3}$,  ~  $1.16\pm0.13  R_\odot$ $^{b4}$,~  $1.22 R_\odot$ $^{b5}$, ~ $1.29 R_\odot$ $^{b6}$       \\
                Period                                 &       ${\bf 3.56}$ d   $^{c1, c2}$   ~  $3.606\pm 0.015$ d   $^{c3}$                 \\ 
%             $R_{\rm co}$                            &      $ 9.75 R_*$           \\
\hline
                Mag field (dipole)                           & a few hundred Gauss  $^{d1}$    \\
                 Mag. field (octupole, almost aligned)                 &      2.5-2.8 kG                                 \\ 
\hline
                Mag. field (dipole)                     &     990-1190 G $^{d2}$                           \\
                Mag. field (small-scale fields)                 &      3-4 kG                                 \\ 
%              $R_{\rm m}$                             &      $ (3-4) R_*$                         \\
              Tilt of the dipole magnetosphere $\theta$     &      $(17-23)^\circ\pm(6-10)^\circ$                \\
\hline              
  Inclination angle $i$                  &      $5-15^\circ$   $^{e1}$, ~$7^\circ$ $^{e2}$~$5.6^\circ$  (for the disc) $^{e3}$\\   
%         Matter flux                         &      $  (10^{-8} - 10^{-9}) M_\odot yr^{-1}$   \\     
%                log(Matter flux)                         & $-8.9\pm0.4$ (Donati et al. 2011)~~ $-9\pm0.6$ (Cieza et al. 2010)                 \\
%                Matter flux                         &      $ (2.51 \times 10^{-9} - 2.2 \times 10^{-8}) M_\odot yr^{-1}$   \\
%    Luminosity (from HST)                     &     from $ 3.8\times 10^{32} erg s^{-1}$ to $7.7\times 10^{32} erg s^{-1}$       \\
\hline
\hline 
\end{tabular}
\caption{Observational parameters of TW Hya. a1: \citet{VenutiEtAl2019},   a2: \citet{ManaraEtAl2014},  a3: \citet{BaraffeEtAl2015}; b1:  \citep{HughesEtAl2007}, b2: \citet{RobinsonEspaillat2019}, b3:  \citet{RheeEtAl2007},~ b4: \citet{BaraffeEtAl2015},~\citet{DonatiEtAl2024},~ b5: \citet{VenutiEtAl2019}, 
b6: \citet{GRAVITY2020};   c1: \citet{SetiawanEtAl2008}, c2:\citet{HuelamoEtAl2008};  c3: \citet{DonatiEtAl2024}; d1: \citet{DonatiEtAl2011}; d2: \citet{DonatiEtAl2024}, e1: \citet{SiwakEtAl2018}, e2: \citet{QiEtAl2004}, e3: \citet{TeagueEtAl2019}.
  \label{tab:TWHya-param}}
\end{table*}

\subsection{TESS observations of TW Hya}

TW Hya was observed by \textit{TESS} (\citealt{RickerEtAl2015}) in 2019, 2021, and 2023 (Sectors 9, 36, 63, respectively).
In Sector 9, observations were performed during $T=25.20$ days with a temporal resolution of $\Delta T=0.0195$ days (0.468 hours=28.08 min) in most cases. In Sector 36,
$T=25.08$ days and $\Delta T\approx 0.0078$ days (0.1872 hours=11.23 min). In Sector 63 $T=26.53$ and $\Delta T =0.0039$ (0.0936 hours=5.61 min).

The light curves show variability on different time scales (see Fig. \ref{tess-3}). 
The Lomb-Scargle periodograms are shown below the light curves for each sector. Red lines show periodograms for non-flagged data. In Sector 9, there are two prominent peaks with periods of  3.5-3.6 days (associated with the star's rotation) and a higher peak at 4.7 days. There is also a smaller amplitude peak at 2.6 days.  In Sector 36, one peak is at 3.8 days, and another peak is at 9 days.  In Sector 63, there is one broad peak at 4.5-5.5 days and a peak of similar amplitude at 8.7 days. There are also several smaller amplitude peaks at shorter periods. 
Periodograms for the light curves, including flagged data (gray color curves), look somewhat different and often show different peaks.  

We also used the Morlet-6 wavelet to analyze \textit{TESS} light curves. First, we carefully checked light curves for stellar flares using earlier developed approaches (e.g., \citealt{SiwakEtAl2010,SiwakEtAl2018,GuntherEtAl2020}). We did not find any notable flares\footnote{In earlier studies of TW Hya,  stellar flares were found in some sets of \textit{MOST} observations, and not in others \citep{SiwakEtAl2018}.}. Next, we interpolated the light curves to a strictly uniform grid (the data in \textit{TESS} light curves are recorded at slightly non-equal time intervals). The new grid is much denser than the original one: it has temporal resolutions of 1.87 min,  0.625 min, and 0.216 min in Sectors 9, 36, and 63, respectively. These grids are   15-26 times denser than the original \textit{TESS} grid, which helped to resolve large and small bursts seen in the original light curve. 
 Next, we calculated the Morlet-6 wavelet using this uniform grid.  
The top panels of Figure \ref{TESS-lc-w-all} show the \textit{TESS} light curves, which include flagged data. Linear interpolation was used in parts of the curve with no data points.
The second panel from the top (for each sector) shows the wavelet of the light curve. Wavelets typically show quasi-periods of
 $P\approx 3.5-4.8$ days which last $T\approx 5-15$ days.  
This quasi-period is close to the stellar period but varies in time. 
There are also oscillations with periods of $P\approx 1-2$ days. They usually last for $T\approx 3-5$ days.
We stretched the wavelet in the vertical direction ($P<2$ days) and  resolved oscillations with even lower periods,
 $P\approx 0.2-0.9$ days which last $T\approx 1-3$ days  (see 3rd panels from the top in Fig. \ref{TESS-lc-w-all}).
  We observe that oscillations with shorter periods last a shorter time. 
We also took a part of the wavelet with $P<0.7$ days with a duration of 5 days (see bottom panels for each sector). Wavelets show QPOs with 0.2-0.3 days, which last approximately one day. There are also QPOs with periods 0.07-0.1 days, which last for a part of the day.

There are also QPOs with long quasiperiods of $P\approx 7.5 - 8.8$ days\footnote{The long-period QPOs may be connected with a short time-interval of observations. However, we observe this type of QPOs in wavelets obtained from much longer sets of ground-based observations (see Sec. 2.6 ). }.   They have a higher amplitude during $T\approx 10-15$ days but  can be seen at lower amplitudes throughout the observations. We discuss the possible origins of these quasiperiods later in the paper.

\subsection{Multiwavelength observations of TW Hya}
\label{sec:multi}

TW Hya was observed during a several weeks observational campaign by a set of ground-based telescopes
\footnote{Light curves were obtained with Las
Cumbres Observatory Global Telescope (LCOGT), American Association of Variable Star Observers (AAVSO), and All-Sky
Automated Survey for Supernovae (ASAS-SN).  See Table 1 of \citet{WendebornEtAl2024b} for details of observations.} 
 in different wavebands (u, B, V, g, r, i, z) during two Epochs: Epoch 1 (2020--2021) and Epoch 2 (2022--2023, see details in \citealt{WendebornEtAl2024b}). Fig. \ref{fig:multi-all} shows the light curves in the different photometric bands.  

The Lomb-Scargle periodogram (see Fig. \ref{fig:Scargle-E1-E2}, taken from \citealt{WendebornEtAl2024b}) shows that in Epoch 1 (left panels) 
multiple periods were observed. However,  the peak of QPO associated with the stellar period is smaller than the other peaks. In Epoch 2, the period of the star typically dominates. However,  the peaks at approximately 6.5 and 8 days often have comparable amplitudes. A number of peaks with periods shorter than the period of the star are also observed.

Quasiperiods are expected to vary with time, and we performed a Morlet-6 wavelet analysis of the light curves. 
For the wavelet analysis, we took an interval of time from Epoch 2, where data were recorded more frequently and regularly (mJD 59641-59767).
 We subtracted $MJD=59641$ days in each spectral band and obtained a total of 109 days in bands B, g, i, and fewer days in other bands due to restricted intervals of observations.
 We processed the light curves in the following way. First, we removed outliers (the points where the flux suddenly dropped, typically during one moment of time). 
Next, we interpolated each light curve into a high-resolution uniform time grid.
For example, the initial number of data points in B-band was 311. In the new grid, the number of points is 
 4,960 with a grid resolution of 46 min. This grid reproduced high- and small-amplitude bursts observed in the original light curve with high precision.    
 Next, we calculated the Morlet wavelets in each spectral band. 

Fig. \ref{fig:multi-wavelength} shows the wavelets and light curves in different wavebands.
The wavelets look similar to those obtained from the \textit{TESS} light curves.
One can see a quasiperiod of 3.2-4.0 days associated with the star's rotation, which lasts 10-20 days, and also QPOs with smaller periods, which last shorter intervals of time. Oscillations with smaller periods last for shorter intervals of time. In addition, the wavelets show the presence of QPOs with a more extended period of 5.3-8.5 days, with typical periods of $P\approx 6.4$ days and $P\approx 8.5$ days. These quasiperiods change with time and last  10-20 days. Similar long-period QPOs were observed in \textit{TESS} wavelets.  

Light curves in  Fig. \ref{fig:multi-wavelength} show that the stellar magnitude decreases systematically with wavelength and the radiated energy increases systematically with wavelength. 
 Fig. \ref{fig:wavelength-dep}  shows the dependence of the time-averaged stellar magnitude on the wavelength of the wavebands. Vertical bars show the scatter in each waveband. One can see that the time-averaged stellar magnitude increases from 12.6 mag to 10.0 mag in Epoch 1 and from 12.1 mag to 9.9 mag in Epoch 2.  
The scatter gradually decreases from the u-band to bands with longer wavelengths.  

\section{Numerical model of TW Hya}
\label{sec:num-model}
 
To model TW Hya, we took 
our earlier developed 3D MHD ``cubed sphere" code  \citep{KoldobaEtAl2002}. The code has been used for modeling accretion onto magnetized stars with a tilted dipole magnetosphere (e.g., \citealt{RomanovaEtAl2003,RomanovaOwocki2015}), tilted dipole and rotational axes \cite{RomanovaEtAl2021} and for modeling unstable regime of accretion  (e.g., \citealt{KulkarniRomanova2008,BlinovaEtAl2016}).  The numerical model is almost identical to that used in \citet{BlinovaEtAl2016} (see Sec. 3 of their paper). Here, we briefly describe our numerical model and provide more details in Appendix \ref{sec:model}.

\subsection{Setup of the model and dimensionalization}
\label{sec:setup}

We place a star of mass $M_*$ and radius $R_*$ to the center of the simulation region.  We place the cold, dense disc in the equatorial plane and a hot, rarefied corona above and below the disc and in the rest of the simulation region. The disc has an initial aspect ratio  $h/r=0.1$ determined at the inner edge of the disc, which is placed at a distance $R_d$ from the star. The density of the disc at $R_d$ is 100 times larger than that in the low-density region around the star. 
Initially, the disc and
corona are in the rotational hydrodynamic equilibrium, where the corona above the disc rotates with the Keplerian velocity of matter
in the disc (see 
\citealt{RomanovaEtAl2002} for details). This helps to decrease initial magnetic braking  at the disc-corona boundary (which is present otherwise). 
A star has a dipole magnetic field with equatorial strength of $B_*$. The magnetic moment of the dipole is tilted about the rotational axis of the star  at an angle $\theta$. The rotational axis  of the star coincides with that of the disc. The simulation region spreads from the inner boundary $R_{\rm in}$ up to the outer boundary $R_{\rm out}\approx 30 R_{\rm in}$.
We used a grid with $N_r=112$ grids in radial direction and $N\times N=51\times51$ grids in each of 6 sides of the inflated cube. A set of 3D MHD equations has been solved using the Godunov-type method described in earlier works \citep{KoldobaEtAl2002,RomanovaEtAl2003}.  
We describe initial and boundary conditions and other details of the numerical setup in the Appendix \ref{sec:model}.

Equations are solved in dimensionless form using dimensionless parameters $\tilde A$. We introduce reference parameters $A_0$ and, after simulations, convert dimensionless values to dimensional ones.
We take the reference scale,
$R_0=R_*/0.35$, where $R_*$ is the radius of the star; 
the reference mass,
$M_0=M_*$.  Reference velocity is Keplerian velocity at $R_0$,
$v_0=(GM_0/R_0)^{1/2}$;
 period of rotation at $r=R_0$: $P_0=2\pi R_0/v_0$.
To obtain the physical dimensional values $A$, the dimensionless
values $\widetilde{A}$ should be multiplied by the corresponding
reference values $A_0$ as $A=\widetilde{A}A_0$. 
For TW Hya, we take the following parameters:  mass $M_0=M_*=0.79 M_*$. Radius $R_*=0.93 R_\odot$ and $R_0=R_*/0.35=2.66 R_\odot$. Then reference velocity $v_0=238$ km/s, and reference period $P_0=0.565$ days. See other reference values in Appendix \ref{sec:model}.

\smallskip

In the model, we used the period of the star  $P_*=3.5$ days (which is close to period $3.56$ days obtained by \citealt{SetiawanEtAl2008} and \citealt{HuelamoEtAl2008}), and obtain the corotation radius $r_{\rm cor}\approx 9.6 R_*$. We use it as a base for all simulations. In our model, we rotate a star with such a period that the corotation radius is located at 9.6 radii of the inner boundary. 
%The dimensionless value of the corotation radius $r_{\rm cor}=R_{\rm cor}/R_*$ is an important parameter of the model.
 Another important parameter is the dimensionless magnetospheric parameter $\mu$ which we vary in the code and which determines the final magnetospheric radius  (see more in Sec. \ref{sec:model}).
 %We vary this parameter between values $\mu=0.5$  and $\mu=2$, which provide the smallest and largest sizes of the magnetosphere.

We  calculated the kinetic energy flux at the stellar surface, $F=v_n \rho v^2/2$ (where $v-$is the total velocity and $v_n-$is its normal component), and assumed that all energy of the falling matter is converted into isotropic radiation. Then, we calculated the radiation flux towards the observer at two inclination angles: $i=5^\circ$ and $i=15^\circ$.  

\begin{figure*}
\centering
 \includegraphics[width=0.7\textwidth]{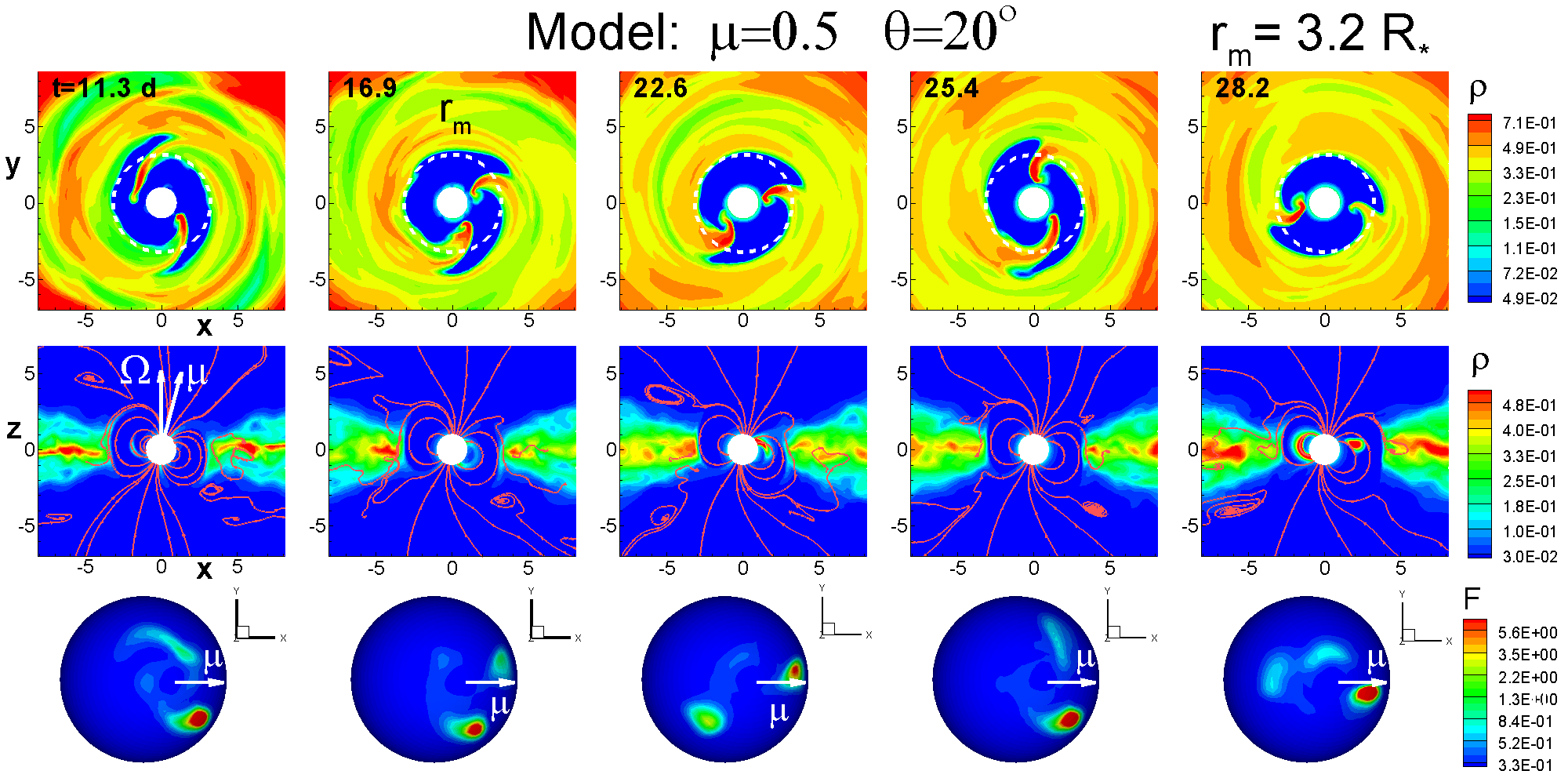}
 \includegraphics[width=0.9\textwidth]{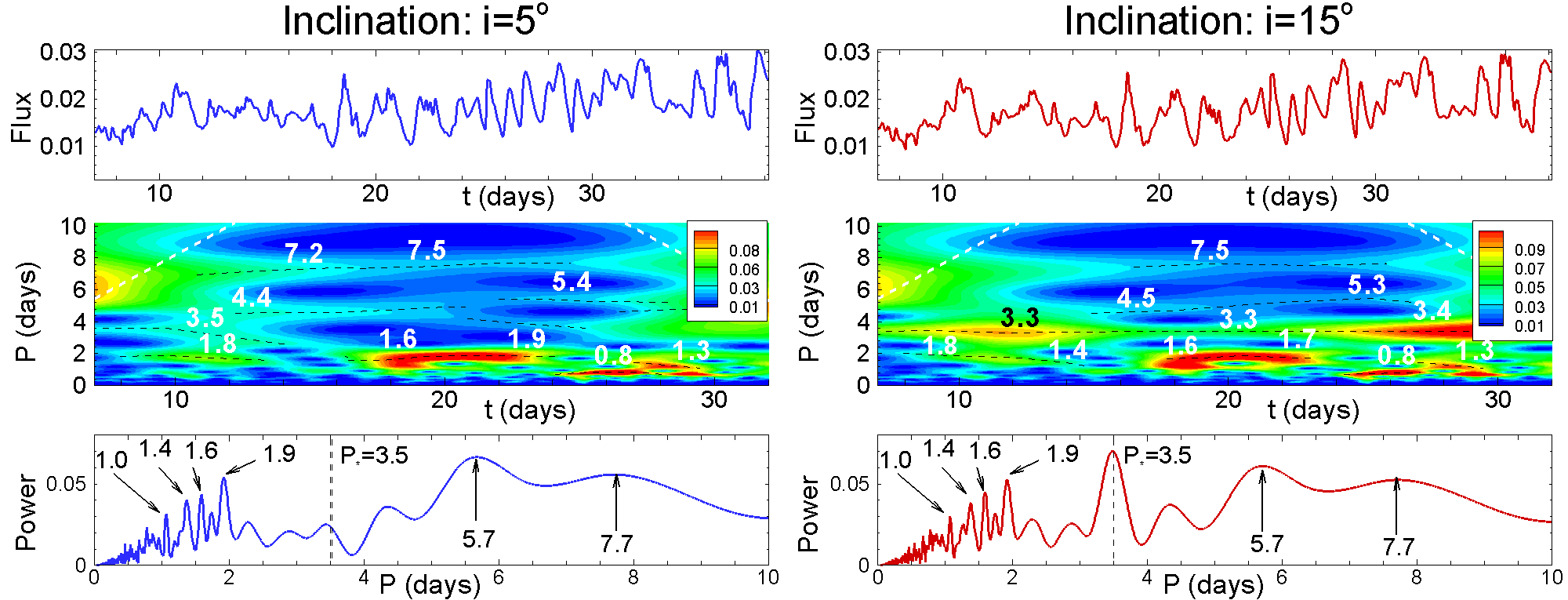}
 \caption{Results of simulations in the model $\mu0.5\theta 20$ where the magnetospheric radius (measured in radii of the inner boundary) is the smallest. \textit{Top row:} slices of density distribution in the equatorial plane
at sample moments of time $t$. \textit{2nd row from the top:} The same, but in the $xz-$ plane, the plane where the vectors of the angular momentum
$\Omega$ and magnetic moment $\mu$ are located. Red lines are sample magnetic field lines.  \textit{3rd row:}
The flux of energy distribution on the star's surface as seen from the pole. \textit{4th row:} fluxes towards the observer at inclination angles $i=5^\circ$ (left) and  $i=15^\circ$ (right).  \textit{5th row:} Morlet wavelet of the observed fluxes. \textit{Bottom row:} 
Fourier spectrum of the observed fluxes. 
\label{fig:2d-d0.5-t20}}
\end{figure*}

\begin{figure*}
\centering
 \includegraphics[width=0.8\textwidth]{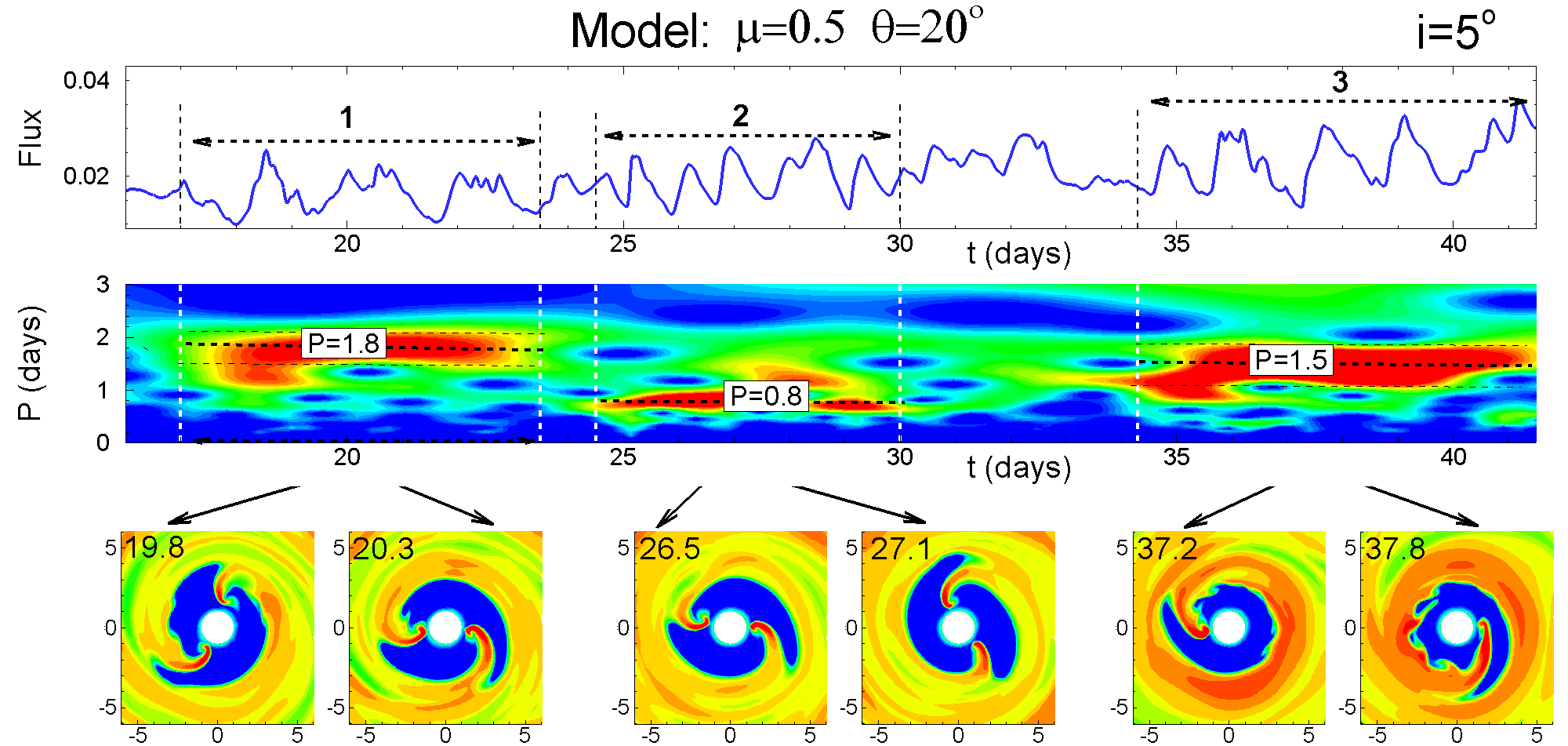}
 \caption{Figure explains the origin of the short-period QPOs in the model  $\mu0.5\theta 20$. \textit{Top row:} A part of the light curve is shown where episodes of short-period QPOs are selected and marked with numbers 1, 2, and 3. The middle panel shows part of the wavelet up to a period of 3 days, where episodes of short-period QPOs are shown and marked. Numbers show an approximate value of quasiperiod. The bottom row shows that two or one tongues dominate during these episodes of short-period QPOs. 
\label{fig:2d-d0.5-t20-part}}
\end{figure*}

\section{Results}
\label{sec:results}

Simulations were performed at several values of the magnetospheric parameters $\mu$ , which determined the final size of the magnetosphere. We calculated models at $\mu=0.5, 1, 1.5, 2$. The size of the magnetosphere increases systematically with $\mu$. We obtain the smallest magnetospheres at $\mu=0.5$ and the largest at $\mu=2$. We obtain the size of the magnetosphere $r_m$ from simulations and show it in stellar radii. In all models, we obtain an unstable regime of accretion.  In models with smaller $\mu$, we obtain a strongly unstable regime, where unstable tongues penetrate closer to the stellar surface,  while in models with $\mu=1.5$ and $2$, we obtain a mildly unstable regime, where unstable tongues are stopped by the inner magnetosphere at larger distances from the star, and matter accretes in funnel streams, going around the inner magnetosphere. 
We typically  perform simulations at two tilts of the dipole moment, $\theta=5^\circ$ and $20^\circ$, but we also calculate models with intermediate angles $\theta=10^\circ$ and $15^\circ$ for comparisons.  We show distances in stellar radii and time and periods in days.
Other variables are shown in dimensionless form but can be converted to dimensional values using reference values from Tab. \ref{tab:refval}.

\subsection{Small-sized magnetosphere, strongly unstable regime}

First, we consider a model with magnetospheric parameter $\mu=0.5$ and the tilt of the dipole magnetosphere $\theta=20^\circ$. Fig. \ref{fig:2d-d0.5-t20} shows the simulation results. The top panels show equatorial slices of density distribution in the inner part of the simulation region. Matter accretes in an unstable regime where equatorial ``tongues" carry matter from the inner disc towards the star.   Two tongues dominate, which corresponds to an ordered unstable regime of accretion. The second from the top row of panels shows the density distribution and sample magnetic field lines in the ${\vec\mu}$-${\vec\Omega}$ plane,  where $\vec\mu$ and   $\vec\Omega$    are unit vectors showing the directions of the magnetic moment and the star's angular momentum, respectively. The 3rd row of panels shows the flux distribution in hot spots viewed pole-on. Two spots are observed, and typically, one spot dominates because one tongue carries more matter than the other. Spots are shown in the pole-on projection. 

In the unstable regime, the inner boundary varies its shape, so we determine the approximate value of the magnetospheric radius by placing a circle which corresponds to the sharp density drop at the disc-magnetosphere boundary (see also Fig. A1 from \citealt{BlinovaEtAl2016}). The dashed circles in the top panels show the location of the magnetospheric boundary, which is $r_m\approx 3.2 R_*$. At our corotation radius of $r_{\rm cor}=9.6 R_*$, the ratio $r_m/r_{\rm cor}\approx 0.33$ corresponds to a strongly unstable regime.

 We calculated the radiation flux from spots towards the observer at angles of $i=5^\circ$ and  $i=15^\circ$ (see left and right panels in the 4th row). 
The 5th row shows the Morlet wavelet spectrum of the light curves.  At the inclination angle of $i=5^\circ$ (left panel), variability with quasiperiods of 1.5-1.9 days dominates. In addition, quasi-periods of 0.8-1.3 days, 3.5, 4.4, 5.4, 7.2-7.5 days are present in the wavelets. At the observer's angle  $i=15^\circ$, QPOs with 1.5-1.9 days and other QPOs are also present. However, the QPOs with 3.3-3.4 days (close to the stellar rotational period) have also a high amplitude. The bottom panels show corresponding Fourier analysis of the light curves. The left panel shows that the period of the star is not present, but several shorter periods with maxima at 1.9, 1.6, 1.4, and 1.0 days dominate. In addition, there is a broader peak at 5.7 days and an even wider one at 7.7 days. The right panel shows that at $i=15^\circ$ the main peak associated with the period of the star dominates, though peaks at the shorter and longer periods are also present.       

This model shows that at a relatively small magnetosphere, an ordered unstable regime dominates (e.g., \citealt{RomanovaKulkarni2009,BlinovaEtAl2016}) where matter accretes predominantly in one or two tongues that rotate with the angular velocity of the inner disc and produces short-period QPOs in light curves, wavelets, and Fourier spectrum. 

\smallskip

We took a part of the light curve from the model $\mu0.5\theta20$ and considered three episodes where quasiperiods are observed (see top row of Fig.  \ref{fig:2d-d0.5-t20-part}). The second row shows the corresponding wavelet where the QPO periods are marked. We see a good correlation between episodes of short-period QPOs observed in the light curves and wavelet.  The bottom panels show that  two tongues dominate  during episodes 1 and 2, and one tongue dominates during episode 3. 
They originate due to the rotation of one or two unstable tongues in an ordered unstable regime (e.g., \citealt{RomanovaKulkarni2009,BlinovaEtAl2016}). 

We calculated the radii in the disc corresponding to Keplerian rotation with QPO periods of $P=1.8$ days (Sector 1 in the plot) and $P=1.5$ days (Sector 3). These radii are $4.23 R_*$ and $3.74 R_*$ and are shown as dashed black circles in the bottom panel of Fig.  \ref{fig:2d-d0.5-t20-part}.  The dashed line shows the region of the disc where tongues form. These QPOs correspond to the rotation period of the strongest filament. Note that in Sector 3, the accretion rate is a bit higher, the inner disc comes closer to the star, and the period of QPOs is lower compared with Sector 1. The QPO in Sector 2 has twice as a lower period, $P=0.8$ days, and can be explained by the rotation of two equal tongues forming at the inner disc at the radius  $3.91 R_*$ (corresponding to $P=1.6$ days). The duration of QPOs is 5-7 days which corresponds to 3-4 Keplerian rotations of the inner disc.  This model demonstrates the direct connection between QPO frequencies and Keplerian velocity 
and the location of the inner disc.

In TW Hya, this regime does not dominate in the currently presented light curves. However, in some earlier observations, the strong peaks at short-period QPOs and possible ``stellar period" of 1.3-2.8 days were suggested by other groups (e.g., \citealt{AlencarBatalha2002,LawsonCrause2005,SiwakEtAl2011}). Their observations may correspond to times when TW Hya switched to an ordered unstable regime.

 In another example,  \citet{ArmeniEtAl2024} studied the photometric variability of another star - RU Lupi, which also shows the stochastic-looking light curves, and concluded that this star may accrete in the ordered unstable (in another terminology - the magnetic boundary layer regime - MBL, \citealt{RomanovaKulkarni2009}). In this star, the magnetospheric radius is expected to be small, $r_m\approx 2 R_*$ \citep{ArmeniEtAl2024}, and an ordered unstable/MBL regime is expected.

\begin{figure*}
     \centering
     \includegraphics[width=0.7\textwidth]{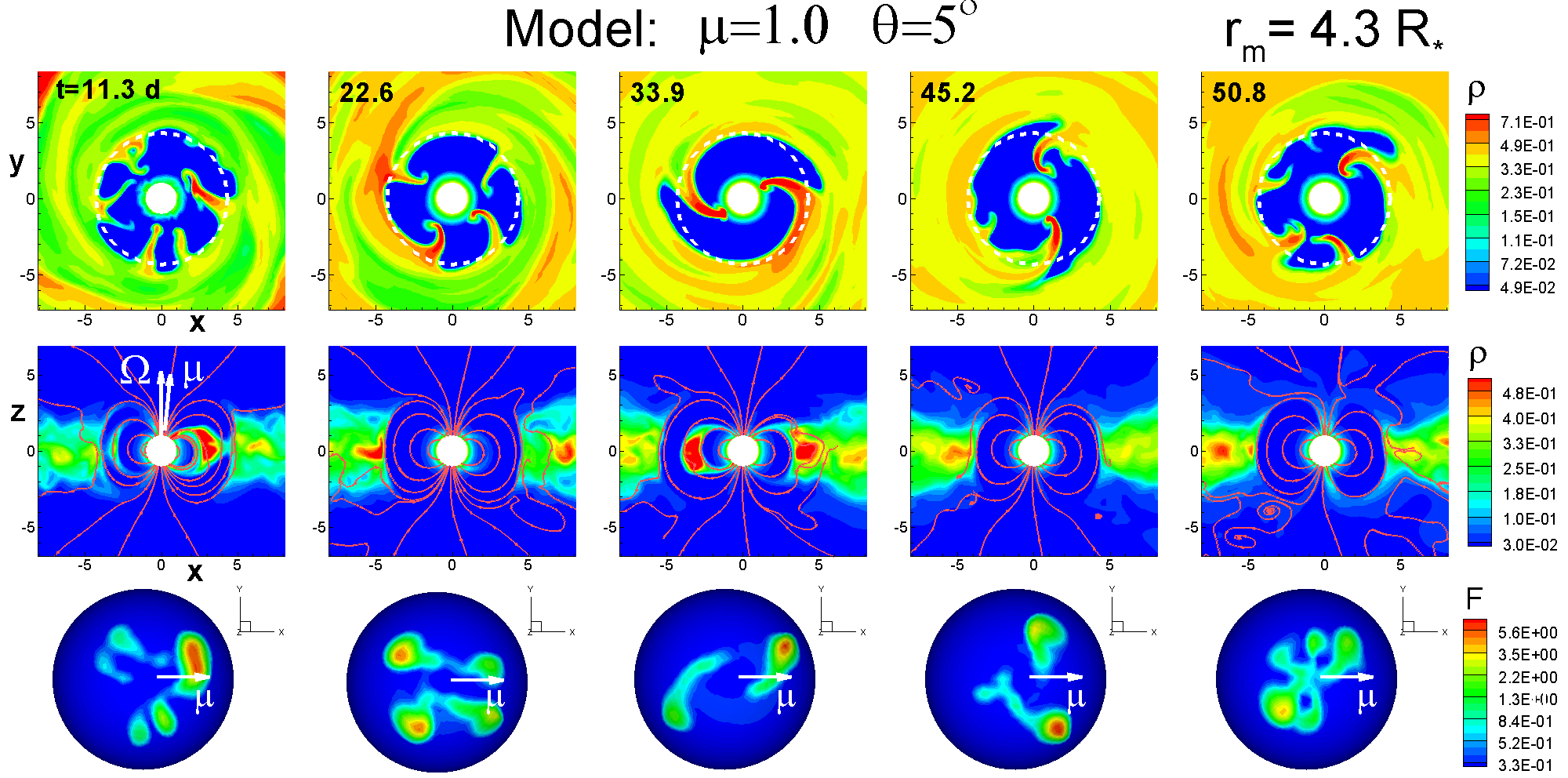}
     \includegraphics[width=0.9\textwidth]{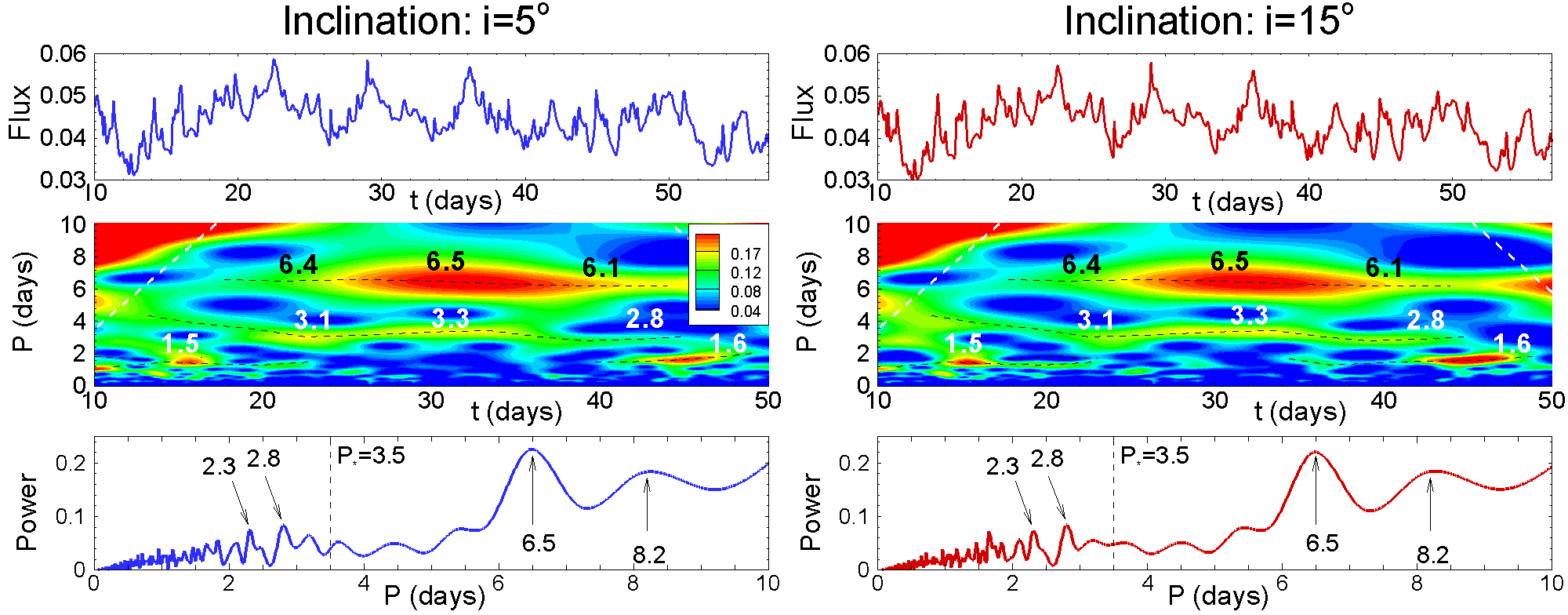}
    \caption{The same as in Fig. \ref{fig:2d-d0.5-t20} but for the model  $\mu 1\theta 5$.
     \label{fig:2d-d1-t5}}
\end{figure*}

%\begin{figure*}
%     \centering
%     \includegraphics[width=0.65\textwidth]{figures/2d-d1-t10.png}
% \includegraphics[width=0.7\textwidth]{figures/flux-w-Fur-d1-t10.png}
%    \caption{The same as in Fig. \ref{fig:2d-d0.5-t20} but for model  $\mu 1\theta 10$.
%     \label{fig:2d-d1-t10}}
%\end{figure*}

\begin{figure*}
     \centering
     \includegraphics[width=0.7\textwidth]{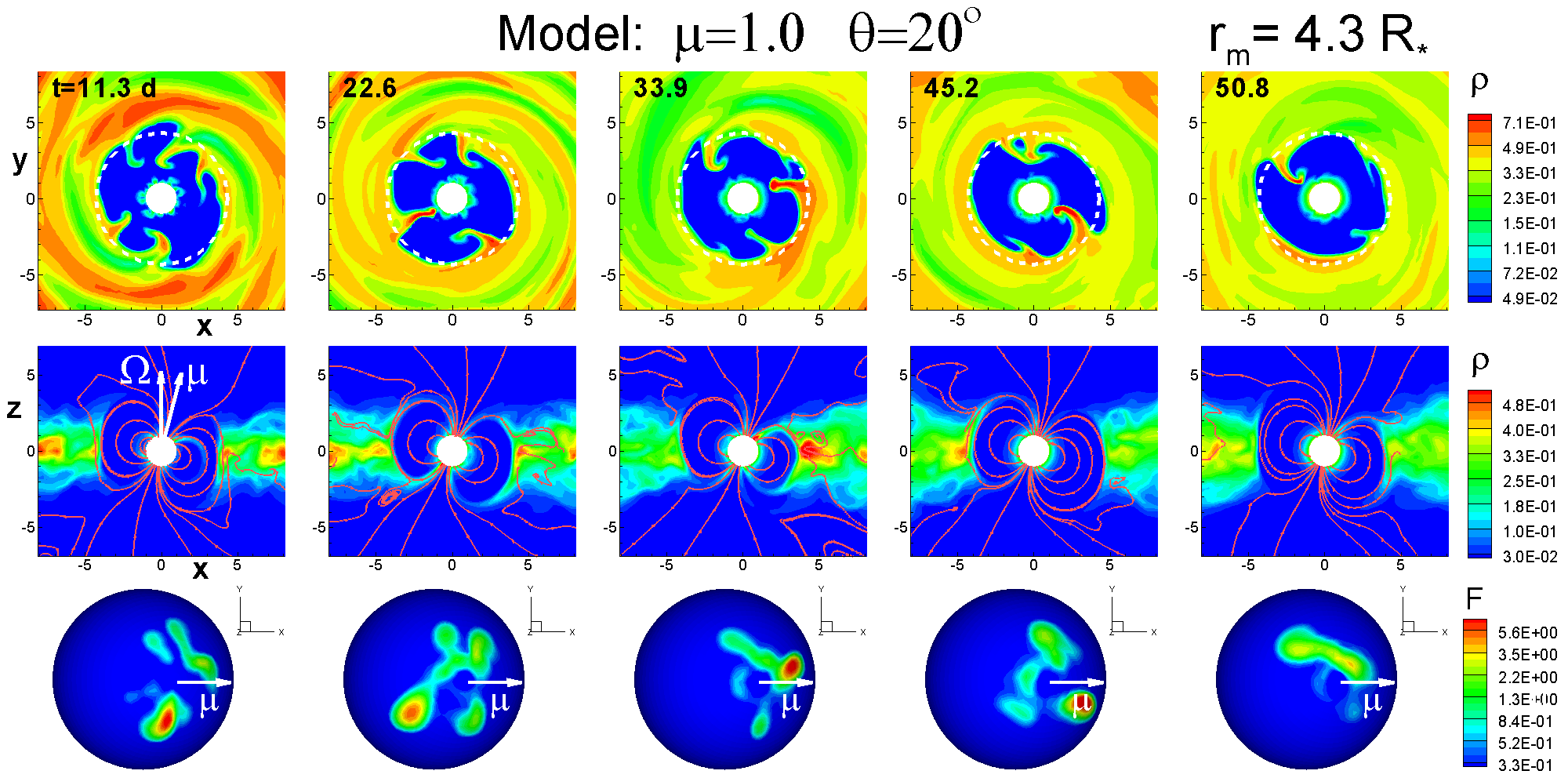}
  \includegraphics[width=0.9\textwidth]{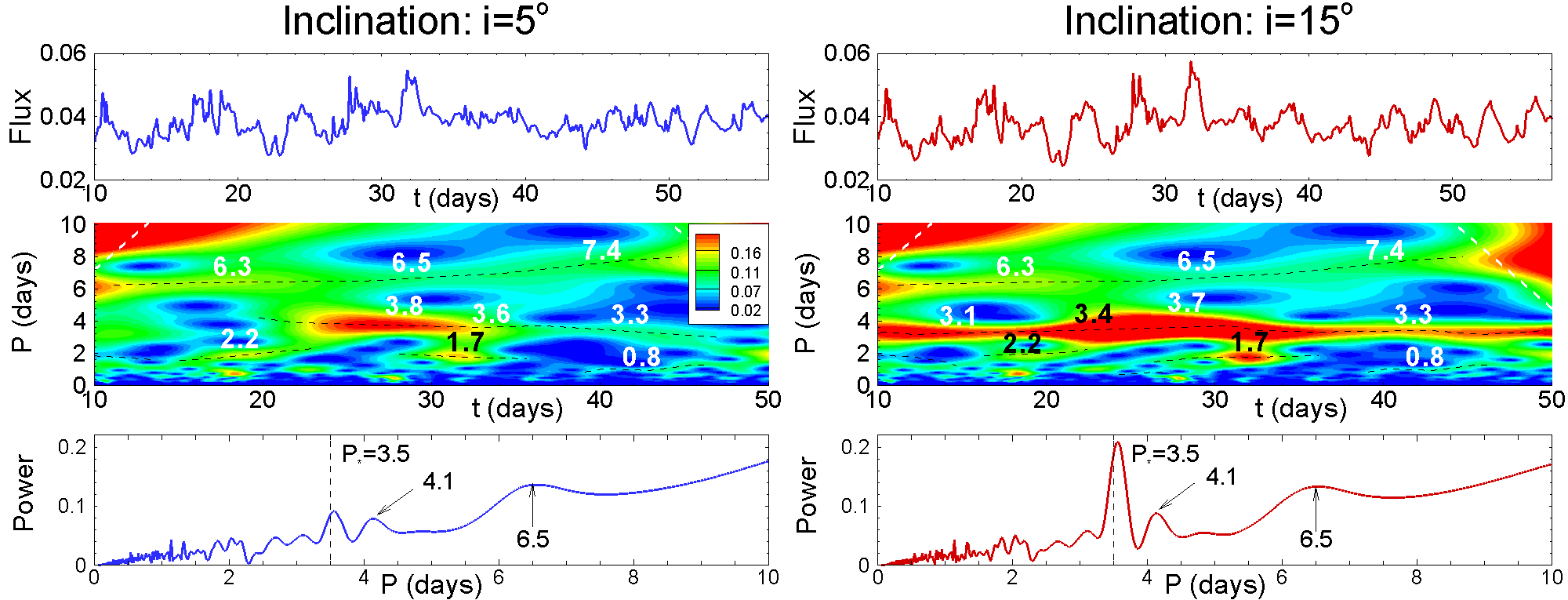}
    \caption{The same as in Fig. \ref{fig:2d-d0.5-t20} but for the model  $\mu 1\theta 20$.
     \label{fig:2d-d1-t20}}
\end{figure*}

\begin{figure}
     \centering
     \includegraphics[width=0.49\textwidth]{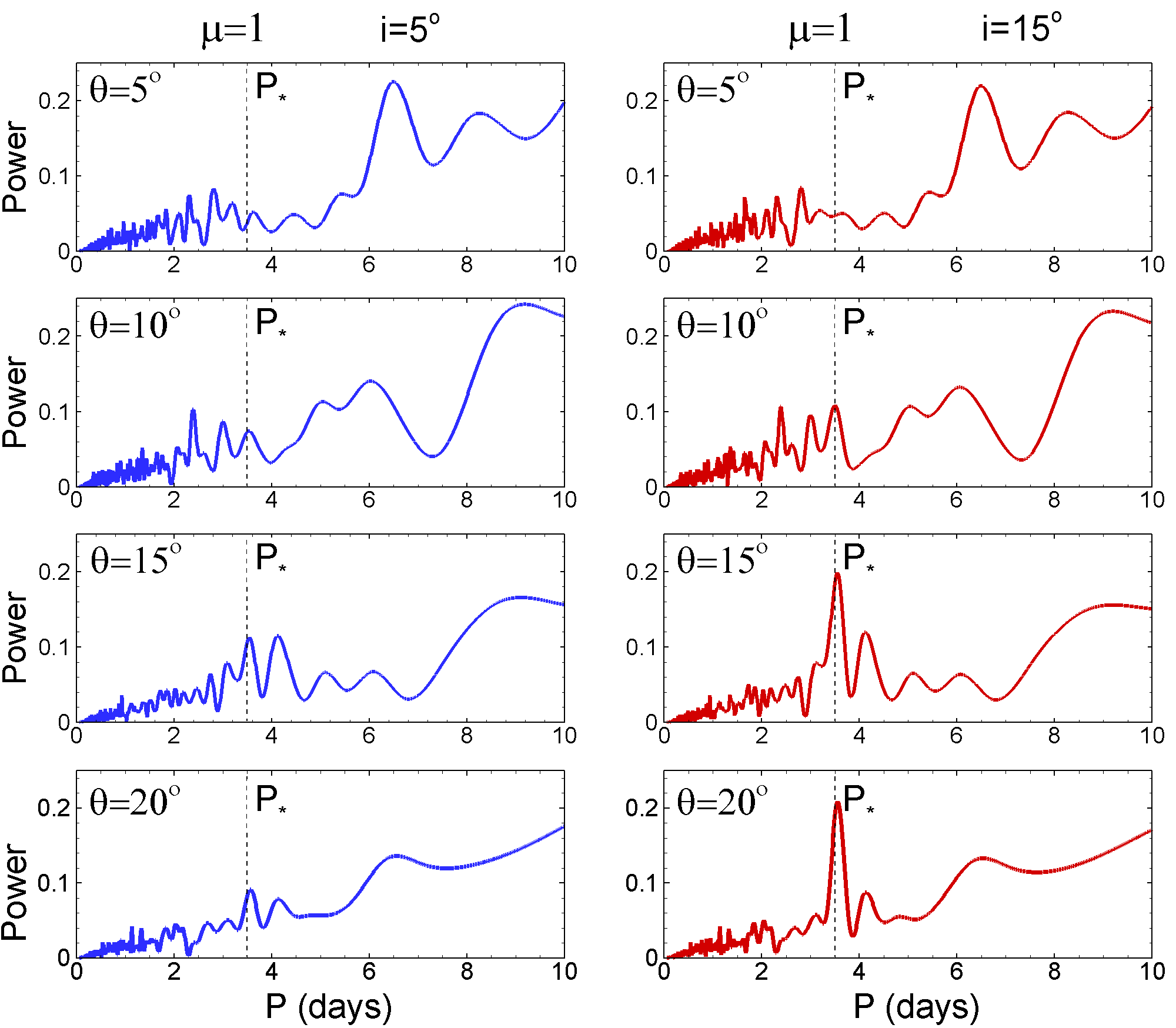}
    \caption{Fourier spectrum from light curves obtained in models with $\mu=1$ but different tilt angles of the dipole $\theta=5^\circ, \theta=10^\circ, \theta=15^\circ$ and $\theta=20^\circ$. The observer's inclination angles are $i=5^\circ$ (left panel) and  $i=15^\circ$ (right panel).
 \label{fig:fur-d1-tdiff}}
\end{figure}

\begin{figure*}
     \centering
     \includegraphics[width=0.45\textwidth]{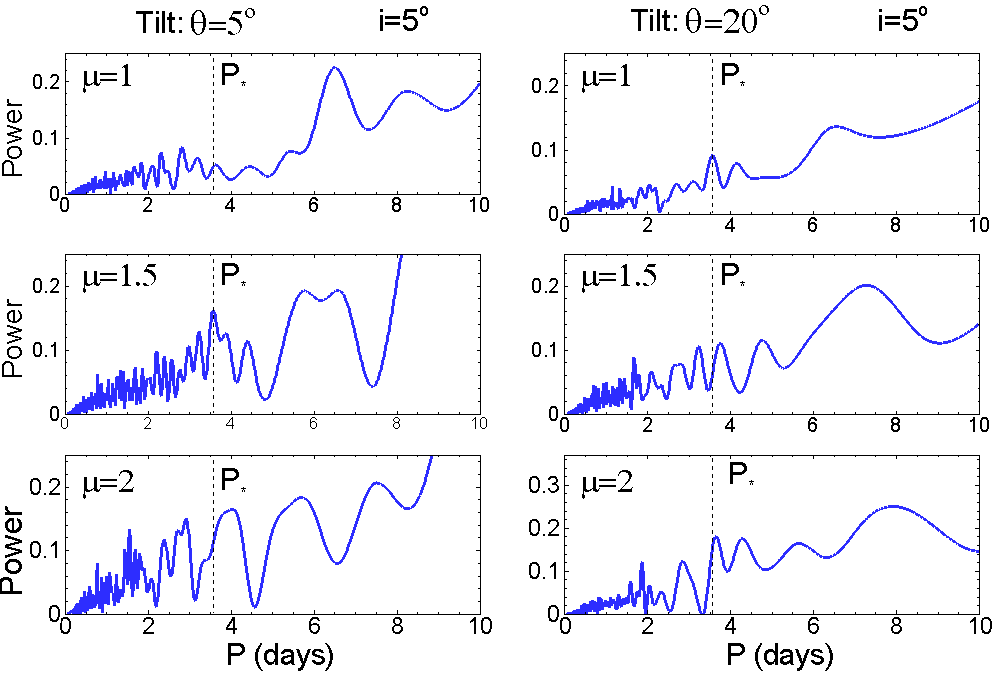}
     \includegraphics[width=0.45\textwidth]{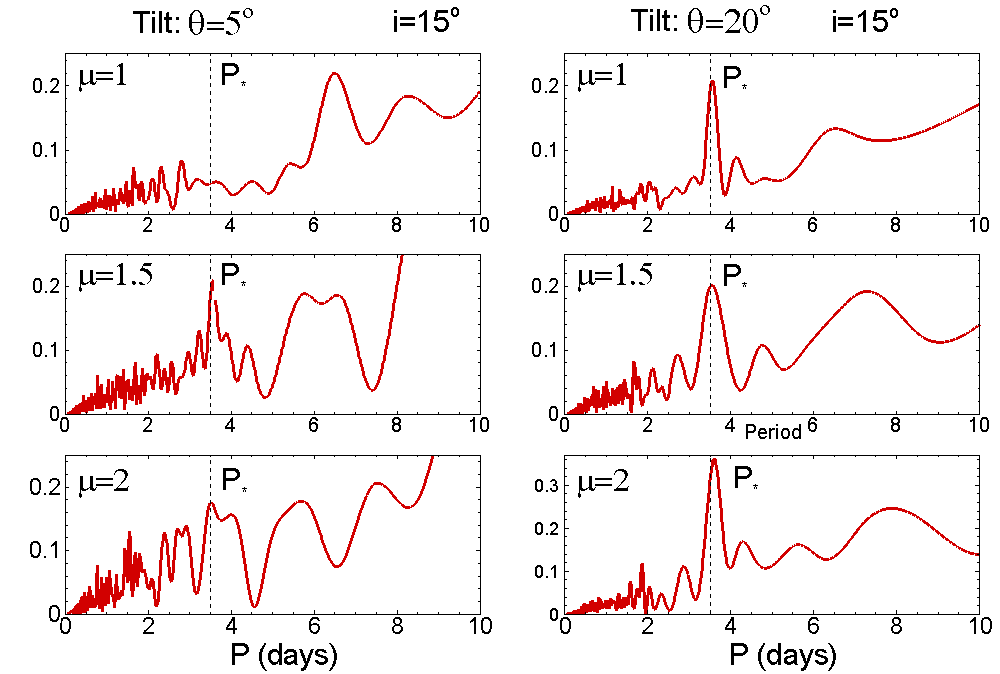}
    \caption{Fourier spectrum from light curves obtained in different models and shown at the inclination angle $i=5^\circ$ 
(left two panels, blue lines) and
$i=15^\circ$ (right two panels, red lines). 
 \label{fig:fur-i5-i15}}
\end{figure*}

\subsection{Mid-sized magnetospheres. The dependence on the dipole's tilt $\theta$}

At the magnetospheric parameter $\mu=1$, we obtain slightly larger magnetospheres with approximate magnetospheric radii of $r_m\approx 4.3$. We use this model to test the dependence of the variability on the tilt of the magnetic dipole. We calculated models with several tilts of the dipole:  $\theta=5^\circ, 10^\circ, 15^\circ$, and $20^\circ$. 

\textbf{\large Model $\mu1\theta5$} (tilt of the dipole: $\theta=5^\circ$).  Fig. \ref{fig:2d-d1-t5} shows the results of simulations in the model where the dipole is only slightly tilted about the rotational axis,  $\theta=5^\circ$. We observed that part of the time matter accretes in an unstable regime (with multiple tongues), and part of the time is in a strongly unstable regime (with two tongues). The light curves show some ordered oscillations with a quasiperiod of 6.5 days and multiple short-period bursts with different quasiperiods of 2.8-3.1, 1.5-1.6, and shorter periods. In addition, the longer quasiperiod of 8.2 days is present in the Fourier spectrum, but it is not seen in the wavelet, so it is insignificant.
A quasiperiod of 2.8-4.2 days is present in wavelet spectra but is not visible in Fourier spectra because the quasiperiod changes with time.   Note that spectra look similar at both inclination angles. 
%This is an example where a quasiperiod close to the period of the star is present in the wavelet but is absent in the Fourier spectrum.
We conclude that at a small tilt of the dipole, it is more tricky to find the period of the star from light curves.

\medskip

\textbf{\large{Model $\mu1\theta20$}} (tilt of the dipole: $\theta=20^\circ$).  Fig. \ref{fig:2d-d1-t20} shows the results of simulations in the model with an even larger tilt $\theta=20^\circ$.
The top three rows show a typical unstable regime of accretion. The light curve looks stochastic at an inclination of $i=5^\circ$ and more periodic at $i=15^\circ$. Different quasiperiods are observed in both wavelets, and a quasiperiod of 3.1-3.7 days associated with the star's rotation is persistent at $i=15^\circ$.
Fourier analysis shows periods of 6.5, 3.5, and 4.1 days if $i=5^\circ$ and a clear peak associated with the star's rotation if $i=15^\circ$.

Fig. \ref{fig:fur-d1-tdiff} summarizes plots for Fourier spectra for models calculated for medium magnetospheres ($\mu=1$).  At $i=5^\circ$ (left panels), the peak associated with the star's rotation is not prominent, and it is typically smaller than other peaks. At $i=15^\circ$ this peak is smaller than other peaks at tilts of the dipole of $\theta=5^\circ$ and $\theta=10^\circ$. However, it is seen and has the highest amplitude when $\theta=15^\circ$ and  $\theta=20^\circ$. We suggest that different variability patterns in the light curves of TW Hya 
can be due to variation of the dipole tilt: at small tilts, the period of the star becomes less significant than other quasiperiods, and vice versa.

\subsection{Larger magnetospheres}

We also calculated models with a parameter $\mu=1.5$ that provides larger  magnetospheres.    Here, we briefly discuss these models.

\smallskip

\textbf{\large{Model $\mu1.5\theta5$}}.  Fig. \ref{fig:2d-d1.5-t5} of the Appendix shows the results of simulations. The top row of panels shows that matter accretes in an unstable regime through multiple tongues. The magnetospheric radius is
 $r_m\approx 4.9 R_*$. The second row shows that tongues are stopped at radii $(3-4) R_*$ by a strong dipole field, and matter falls from these radii to the stellar surface. The 3rd row shows that multiple hot spots form at the star's surface. Wavelets show different periods. Fourier analysis shows the period of the star at an inclination angle $i=15^\circ$. However, a peak with a similar amplitude is observed at periods 5.8-7 days.   

\smallskip

\textbf{\large{Model $\mu1.5\theta20$}}.  Fig. \ref{fig:2d-d1.5-t20} of the Appendix shows  that accretion is still in the unstable regime. However, more matter flows above the main magnetosphere, like in a stable regime (see the second row of panels). 
Hot spots look irregular, like in other models. At $i=15^\circ$, wavelet shows persistent quasiperiod associated with rotation of the star. Fourier analysis shows a peak associated with the period of the star at $i=15^\circ$ and no evident period at $i=5^\circ$.

\smallskip

\textbf{\large{Model $\mu2\theta5$}}. At an even larger magnetosphere (obtained at $\mu=2$), the magnetospheric radius
is $r_m\approx (5.7-5.9) R_*$, respectively.  Fig. \ref{fig:2d-d2-t5} of the Appendix shows similar unstable accretion 
but at a larger magnetosphere. Wavelet shows quasiperiod close to the period of the star and other periods. Fourier shows several peaks at different periods at both inclination angles. 

\smallskip

\textbf{\large{Model $\mu2\theta20$}} At a larger tilt of the magnetosphere, $\theta=20^\circ$ accretion occurs through both stable funnels and unstable tongues (see top two rows of Fig. \ref{fig:2d-d2-t20}). After $t>30$ days, matter mainly accretes above the magnetosphere, and the flux becomes more stable. Fourier shows a clear peak associated with rotation of the star at $i=15^\circ$, and several major peaks are observed at $i=5^\circ$.

\begin{table*}
\begin{tabular}[]{ c  | c | c | c |c}
 \hline   
\hline
Model & Magn. moment  & Tilt of the dipole                     &      $i=5^\circ$        &  $i=15^\circ$                   \\
\hline
 $\mu0.5\theta20$ & $\mu=0.5$ & $\theta=20^\circ$       &       $P=5.7, 1.9, 1.6, 1.4, 1.0$ days   &    $P={\bf 3.5}, 5.7, 1.9, 1.6, 1.4, 1.0$ days            \\
\hline
 $\mu1\theta5$  & $\mu=1$    & $\theta=5^\circ$           &       $P=6.5, 8.2, 2.8, 2.3$ days   &    $P=6.5, 8.2, 2.8, 2.3$ days                                 \\
 $\mu1\theta10$ & $\mu=1$    & $\theta=10^\circ$         &       $P=6.0, 5.0, 3.0, 2.4, {\bf 3.5}$ days   &    $P=6.0, 5.0, {\bf 3.5}, 2.4, 3.0$ days    \\
 $\mu1\theta15$ & $\mu=1$    & $\theta=15^\circ$         &       $P=4.1, {\bf 3.5}, 2.7, 5.0, 6.1$ days   &    $P={\bf 3.5}, 4.2, 5.1, 6.1$  days    \\
 $\mu1\theta20$ & $\mu=1$    & $\theta=20^\circ$         &       $P=6.5, {\bf 3.5}, 4.1$ days         &    $P={\bf 3.5}, 6.5, 4.1 $  days    \\
\hline
$\mu1.5\theta5$ & $\mu=1.5$  & $\theta=5^\circ$          &       $P=5.8, 6.6, {\bf 3.5}, 3.2, 4.4$ days    &    $P={\bf 3.5}, 5.8, 6.6, 3.2, 4.4$  days    \\
 $\mu1.5\theta20$ & $\mu=1.5$  & $\theta=20^\circ$      &       $P=7.2, 4.7, {\bf 3.7}, 3.2, 2.7, 1.7$ days    &    $P={\bf 3.5}, 7.2, 4.7, 2.7, 1.7$ days    \\
\hline
$\mu2\theta5$ & $\mu=2$  & $\theta=5^\circ$          &       $P=7.5, 5.7, 4.0,  2.9, 2.4, 1.5$ days    &    $P=7.5, 5.7, {\bf 3.5}, 2.9, 2.4, 1.5$ days     \\
 $\mu2\theta20$ & $\mu=2$  & $\theta=20^\circ$        &       $P=7.9, {\bf 3.7}, 4.3, 2.8, 1.8 $  days    &    $P={\bf 3.5}, 7.9, 4.3, 2.8, 1.8$ days    \\
\hline
\end{tabular}
\caption{Main periods observed in Fourier spectra of different models at inclination angles of the observer $i=5^\circ$ and $15^\circ$. Numbers in bold show the period of the star and periods close to this number.
  \label{tab:QPOs}}
\end{table*}

\subsection{Variation of the magnetic field and photometric variability}

Fig. \ref{fig:fur-d1-tdiff} compares Fourier spectra for models  with the same magnetospheric parameter $\mu=1$ (corresponding to  $r_m\approx 4.3 R_*$) but at different tilts of the dipole magnetosphere $\theta$. One can see that at a very small  tilt, $\theta=5^\circ$ -  there is no peak associated with the period of the star at both inclination angles of the observer. When  $\theta=10^\circ$ -  the peak associated with the period of the star is present, but it is smaller than other peaks.   At $\theta=15^\circ$ and $20^\circ$, and inclination angle of $i=15^\circ$, it dominates, but comparable with other peaks at $i=5^\circ$. This result is important: if the tilt angle of the dipole moment slightly varies with time, then observers will detect or not detect the period of the star from their light curves. 

Fig.  \ref{fig:fur-i5-i15} shows Fourier spectra for models with different magnetospheric radii (parameters $\mu$) and tilts of the dipole $\theta=5^\circ$, and $\theta=20^\circ$.  The right panels of the figure show that at a larger tilt,  $\theta=20^\circ$, and $i=15^\circ$
 period of the star dominates over other QPO  peaks. At $\theta=5^\circ$ and $i=15^\circ$, and larger magnetospheres ($\mu=1.5$ and $2$) the peak associated with the period of the star is larger than in model with smaller magnetosphere ($\mu=1$), but still comparable with peaks of other QPOs.  At a small inclination angle of $i=5^\circ$ (two left columns), the peak associated with the period of the star has either a very small amplitude or is comparable with other peaks.
This analysis shows that in models with different sizes of the magnetosphere, the period of the star is less visible at a smaller tilt of the dipole magnetosphere.

\begin{table*}
\begin{tabular}[]{ l|c  }
\hline
     B-field  (Gauss)           &         B-field  (Gauss)         \\
    polar            &     equatorial            \\
\hline
%   \tilde M=0.16
$B_*=1,000$      &   $B_{\rm eq}=500$     \\
\hline
$B_*=800$       &  $B_{\rm eq}=400$       \\
\hline
%$\dot M$                     &      $1.06\times 10^{-7}  \frac{\rm M_{\odot}}{yr}$       \\ 
$B_*=600$  &  $B_{\rm eq}=300$       \\
\hline
\end{tabular}
\begin{tabular}[]{c  }
\hline
    $\mu=0.5$                   \\
    $r_m\approx 3.2 R_*$   \\
\hline
%   \tilde M=0.16
 $6.64\times 10^{-9} $       \\
\hline
 $4.26\times 10^{-9} $       \\
\hline
%$\dot M$                     &      $1.06\times 10^{-7}  \frac{\rm M_{\odot}}{yr}$       \\ 
 $2.40\times 10^{-9} $       \\
\hline
\end{tabular}
\begin{tabular}[]{ c }
\hline
$\mu=1.0$               \\
$r_m\approx 4.3 R_*$   \\
%   \tilde M=0.35
\hline
                   $8.15\times 10^{-9} $       \\
\hline
                   $2.33\times 10^{-9}  $       \\
\hline
                   $1.31\times 10^{-9}   $       \\
\hline
\end{tabular}
\begin{tabular}[]{c }
 \hline   
 $\mu=1.5$    \\
 $r_m\approx 4.9 R_*$    \\
%   \tilde M=0.51 and 0.6  Taking average 0.55 \pm 0.05
\hline
      $2.00\times 10^{-9}  $       \\
\hline
      $1.28\times 10^{-9}  $       \\
\hline
      $7.19\times 10^{-10}  $       \\
\hline
\end{tabular}
\begin{tabular}[]{ c }
\hline
      $\mu=2.0$       \\
      $r_m\approx 5.6 R_*$    \\
%   \tilde M=0.66 and 0.84  Taking average 0.75 \pm 0.09
\hline
         $1.95\times 10^{-9} $       \\
\hline
         $1.24\times 10^{-9} $       \\
\hline
        $7.02\times 10^{-10} $       \\
\hline
\end{tabular}
\caption{The accretion rate $\dot M  (\rm M_{\odot}/yr)$ obtained in models with  different values of  the magnetospheric parameter $\mu$ and equatorial magnetic field of the star $B_{\rm eq}$. 
 \label{tab:mag-mdot}}
\end{table*}

Observations of TW Hya show that the wavelet and Fourier spectra vary from one observing season to the next and show different ``periods," which are not real periods of the star.  
We suggest that the dipole component of the field changes from year to year: both the strength of the  field and its tilt may vary with time due to convection processes inside the star. Our models show that the tilt of the dipole is the major factor: at small tilts $\theta \lesssim 10^\circ$, amplitudes of different QPOs are typically larger than the amplitude of the stellar period. At a larger tilt angle of the dipole, 
$\theta \gtrsim 10^\circ$, QPOs associated with the period of the star often dominate.

\begin{figure*}
     \centering
    \includegraphics[width=0.7\textwidth]{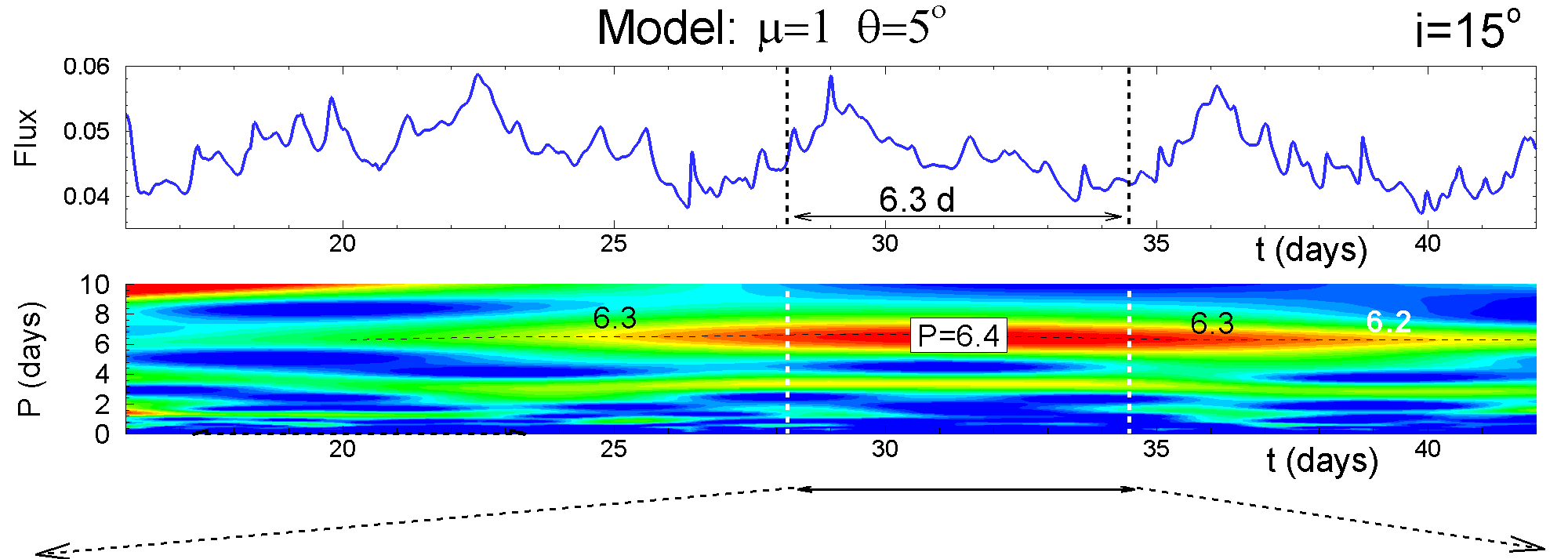}
     \includegraphics[width=0.7\textwidth]{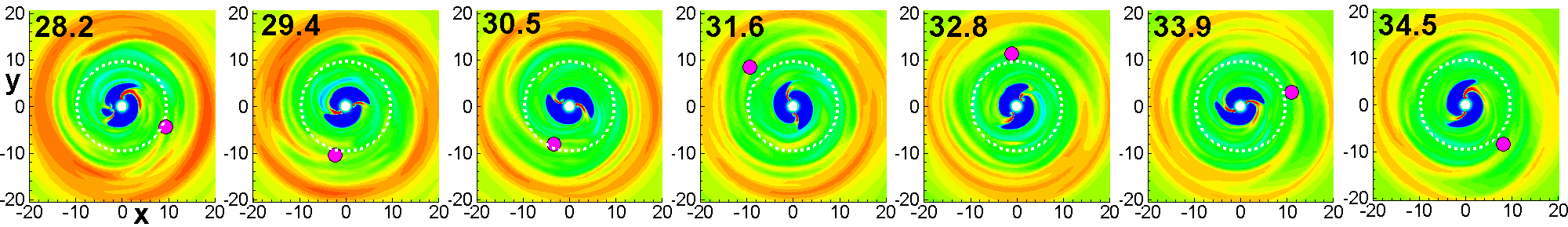}
    \caption{\textit{Top panel:} A part of the light curve shown in Fig. \ref{fig:2d-d1-t5} for the model $\mu1\theta5$, where the long-period QPO is observed.   \textit{Middle panel:} wavelet corresponding to this part of the light curve where we concentrate on the long period, 6.3-6.5 days QPO.
\textit{The bottom row of panels:} density distribution in the equatorial plane at different moments in the interval of 28.2-34.5 days, which approximately correspond to one rotation of the large-scale density wave about the star. The wave is marked with a pink circle.
     \label{fig:waves}}
\end{figure*}

\begin{figure}
     \centering
    \includegraphics[width=0.4\textwidth]{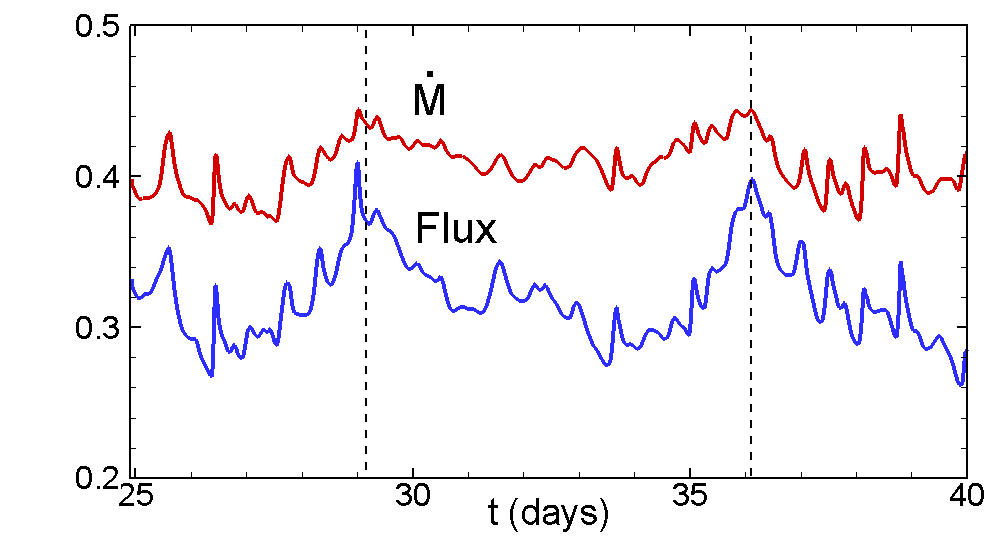}
    \caption{Accretion rate onto the star $\dot M$ (red curve) and the observed flux from spots $\dot E_{\rm obs}$ (blue curve). The latter was multiplied by the factor 7 to show them side by side.  Vertical lines show approximate positions of maxima.
     \label{fig:mdot-edot}}
\end{figure}

\subsection{Best matching models}

The above analysis shows that models with midsize magnetospheres  $4.3 R_*$ and $4.9 R_*$ show wavelets similar to those obtained with \textit{TESS} and ground-based telescopes.  These models were calculated using parameters $\mu=1$ and 1.5. However, at other times, models with smaller or larger magnetospheres may be applied. For example, in years where  the ``period" of $\sim 2$ days dominated, the magnetosphere may be smaller, like in our model $\mu=0.5$ (where $r_m\approx 3.2 R_*$). If the period of the star strongly dominates, and there are signs of high-velocity accretion onto the stellar surface and to higher latitudes (closer to the magnetic pole), then we expect a larger magnetosphere, which may correspond to our model where $\mu=2$ with $r_m\approx 5.6 R_*$ or even larger\footnote{We note that it takes progressively longer time to simulate models with larger magnetospheres.}.  In models with $r_m/R_*=3.2, 4.3, 4.9, 5.6$ and our corotation radius $r_{\rm cor}/R_*=9.6$, we obtain ratios $r_m/r_{\rm cor}\approx 0.33, 0.45, 0.51, 0.58$, respectively. The strength of instability decreases when this ratio increases. 

The magnetospheric radius depends on the strength of the magnetic field and the accretion rate $r_m\sim B_*^{4/7}/{\dot M}^{2/7}$.
Different combinations of $B_*$ and $\dot M$ are possible. Table \ref{tab:mag-mdot} shows combinations of these parameters for different sizes of the magnetosphere $r_m$. In the table, we calculate the accretion rate from the model using the time-averaged value of the dimensionless accretion rate and multiply it by $\dot M_0$  taken from Tab. \ref{tab:refval} of the Appendix.
One can see that if the polar field is $B_*=1,000$ G then the accretion rate should be $\dot M \approx 8.15\times 10^{-9} {\rm M_\odot/yr}$ or 
 $\dot M \approx 2.0\times 10^{-9} {\rm M_\odot/yr}$ in models with $\mu=1$ and $\mu=1.5$, respectively.
In the latter case, the accretion rate approoximately corresponds to that found from optical observations of \citet{HerczegEtAl2023} ($\dot M \approx 2.51\times 10^{-9} {\rm M_\odot/yr}$ ) and $r_m\approx 4.9 R_*$. The former case  ($\mu=1$) may be relevant if the accretion rate is larger   (see Sec. 2.4) and $r_m\approx 4.3$. Tab. 3 shows that if the magnetic field is smaller, then at the same accretion rates, $r_m$ is expected to be smaller, and the unstable regime stronger.  For example, at $B_*=600$ G and $\dot M \approx 2.4\times 10^{-9} {\rm M_\odot/yr}$, the magnetospheric radius $r_m\approx 3.2 R_*$, and stronger unstable regime is expected.

Note that these results depend on the radius of the star. In our models, we took  $R_*=0.93 R_\odot$, and fixed the value $r_{\rm cor}/R_*=9.6$.  If the stellar radius is larger (see Tab. \ref{tab:TWHya-param}) then the results are different.
For example, if we take $R_*=1.16 R_\odot$, and $M_*=0.8 M_\odot$ \citep{BaraffeEtAl2015}, then the corotation radius is smaller $R_{\rm cor}\approx 7.9 R_*$. Test simulations at this radius and $\mu=1$ and $1.5$ show, that the magnetospheric radius is approximately the same as in models with larger $R_{\rm cor}$.
However, the ratios 
 are larger:  $r_m/r_{\rm cor}\approx 0.40, 0.54, 0.62, 0.71$  in models with $\mu=0.5, 1, 1.5, 2$, respectively, and unstable regime is not as strong as in our main models. 

\begin{figure*}
     \centering
     \includegraphics[width=0.7\textwidth]{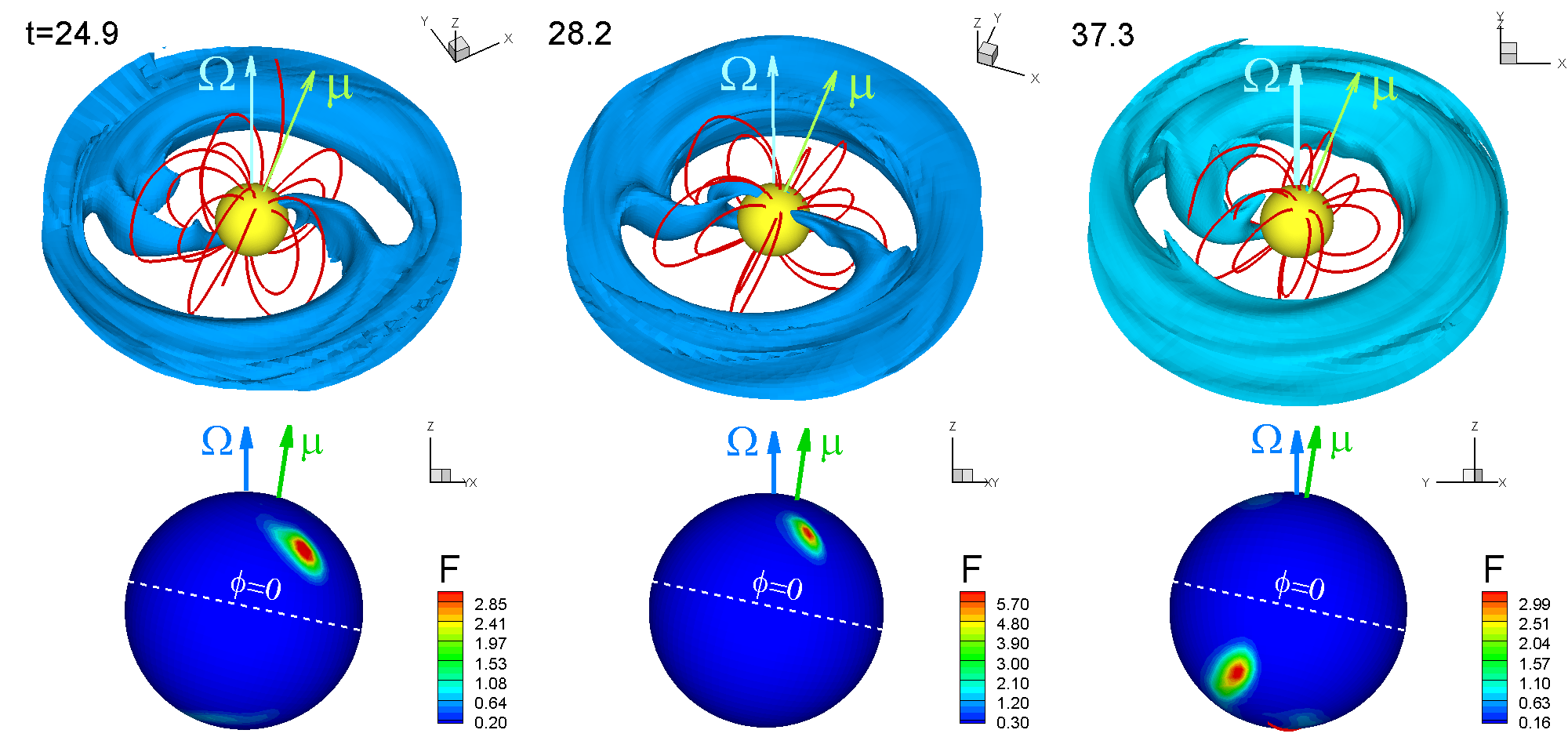}
    \caption{\textit{Top panels:} 3D plots of matter flow in the model $\mu0.5\theta20$ at sample times.  The color background shows one of the density levels. Red lines are sample field lines.  \textit{Bottom panels:}  Distribution of the energy flux at the star's surface observed at an inclination angle $i=90^\circ$. 
     \label{fig:3d-spots-d0.5-t20}}
\end{figure*}

\begin{figure*}
     \centering
     \includegraphics[width=0.7\textwidth]{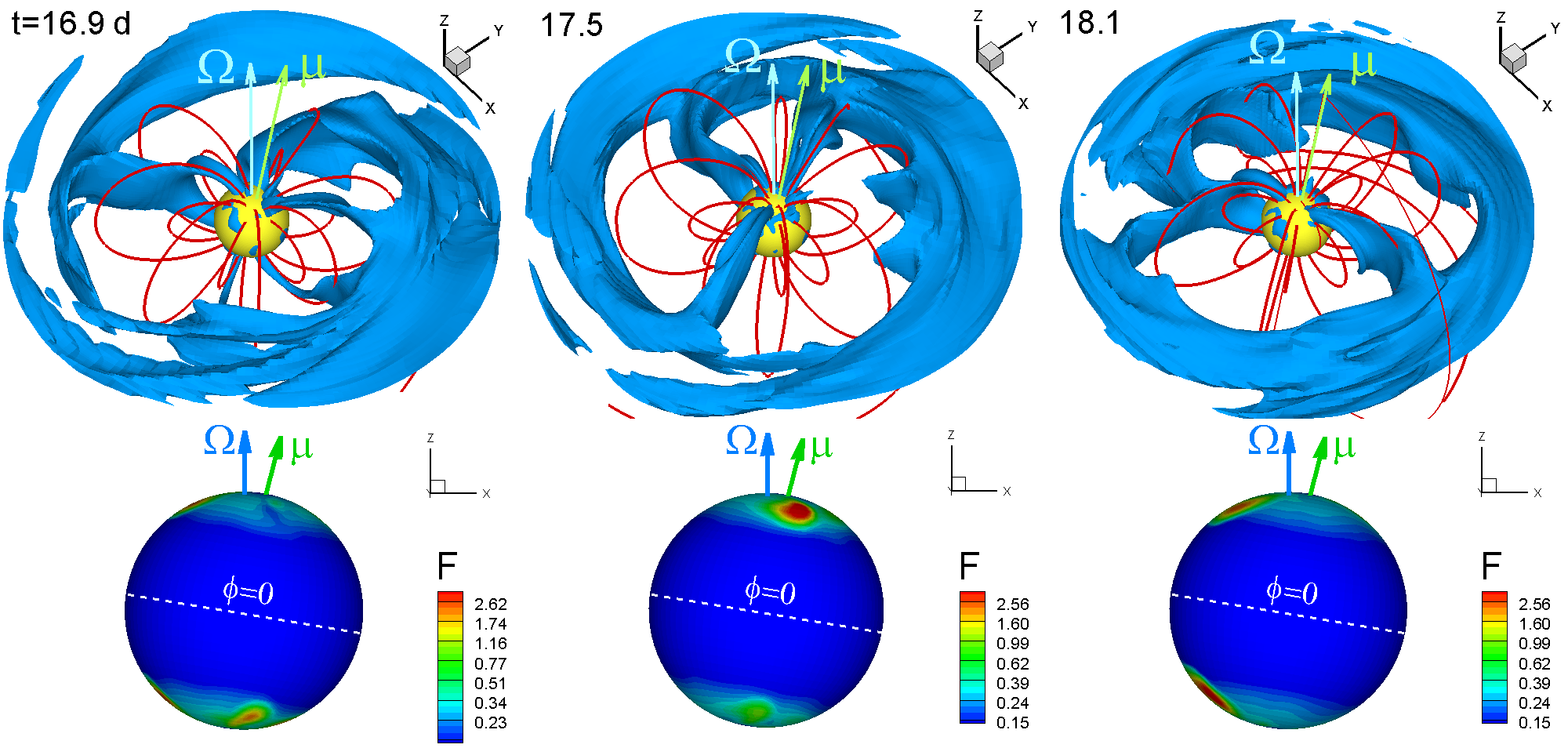}
    \caption{The same, but in the model $\mu1\theta10$. 
     \label{fig:3d-spots-d1-t10}}
\end{figure*}

\subsection{Long-period QPOs: waves in the disc}
\label{sec:waves}

The light curves also show the longer-period QPOs with quasiperiods of 6.4-6.6 days, 7.5 days, and 8.2 days in models $\mu1\theta5$,  $\mu1\theta10$, and  $\mu1\theta15$, respectively.  These QPOs can be explained by waves in the accretion disc. 3D MHD simulations show that a star with a tilted dipole magnetosphere excites  different types of density and bending waves in the accretion disc \citep{RomanovaEtAl2013}. Simulations show that if the magnetospheric radius is comparable with the corotation radius, then the bending wave forms near the closed rotating magnetosphere and rotates with an angular velocity of the star (see Figs. 1-3 from \citealt{RomanovaEtAl2013})\footnote{Formation of a bending waves has been predicted theoretically (e.g., \citealt{TerquemPapaloizou2000}) and proposed to explain variability in AA Tau (e.g., \citealt{BouvierEtAl1999,BouvierEtAl2007b}).}. 
However, if the magnetospheric radius is much smaller than the corotation radius (like in TW Hya), then slowly rotating waves form at the outer Lindblad resonance (see Fig. 12 of the same paper for 3D and 2D plots and Figs. 13-15 for the location of these waves at different corotation radii). 

We demonstrate the presence of such a wave using a model 
$\mu1\theta5$ where the slowly rotating wave is clearly observed.  The light curve in Fig. \ref{fig:2d-d1-t5} shows an episode of long-period QPO  with a quasi-period of 6.4-6.6 days. We analyzed this time interval in greater detail (see Fig. \ref{fig:waves}).  We took one of the quasiperiods and plotted the density distribution in the disc frequently (see bottom panels of Fig. \ref{fig:waves}). We detected the presence of a density wave, which had approximately one complete rotation during this interval of time. The light curve originates from accretion onto the star. That is why we checked the variability in the accretion rate. Fig. \ref{fig:mdot-edot} shows that the accretion rate also has oscillations with approximately the same quasi-period. It means that the remote density wave modulates the accretion rate, which leads to variation in the light curve. This analysis shows that the long-period QPOs observed in light curves of TW Hya can be caused by large-scale density wave that forms beyond the corotation radius. It is possible that many long-period QPO features are connected with large-scale density or bending waves in the disc.

\citet{DonatiEtAl2024} reports the detection of
a radial velocity signal of semi-amplitude 11.1 m s$^{-1}$
at a period
of 8.3 days in the spectrum
of TW Hya. They suggested that the origin may be attributed to either a non-axisymmetric density structure in
the inner accretion disc or a $0.55 M_J$ candidate close-in planet at an orbital distance of $0.075\pm 0.001$ au.  
We suggest that this period can be associated with the wave in the disc.
However, the period associated with density wave may vary in time (e.g., due to inhomogeneities in the disc). 
If this period has a high precision, then the hypothesis of the  planet is much more probable.

\subsection{Properties of hot spots}

Unstable tongues propagate some distance in the equatorial plane (pushing magnetic field lines aside), then encounter the stronger field of the inner magnetosphere 
(at some radius $r_{\rm m,in}$) and are lifted above and below the magnetosphere, forming short-lived funnel-looking  streams. They deposit matter at some distance from magnetic poles. In models with smaller magnetospheres,  tongs are lifted at $r_{\rm m,in}\approx (1-2) R_*$, and spots are located at $\sim (35^\circ-45^\circ)$ from magnetic poles  (see Fig. \ref{fig:3d-spots-d0.5-t20} for a model with $\mu=0.5$). 
At larger magnetospheres, they are lifted at larger distances from the star,  $r_{\rm m,in}\approx (2-4) R_*$, 
and spots are located at  $\phi\lesssim 30^\circ$\footnote{\citet{ZhuEtAl2024} obtained the similar result in independent 3D MHD simulations.} from magnetic poles (see Fig. \ref{fig:3d-spots-d1-t10} for model with $\mu=1$).

Matter  accreting from temporary funnel streams is accelerated by gravity and falls to a star with a free-fall velocity. If the matter falls from the distance   $r_{\rm m,in}$ then
its velocity at the surface of the star is 
\begin{eqnarray}
\nonumber V_{\rm ff}=\bigg[\frac{2GM_*}{R_*}\bigg(1-\frac{R_*}{r_{\rm m,in}}\bigg)\bigg]^{1/2} \\
\approx 569 \frac{km}{s} \bigg[\bigg(\frac{M_*}{0.79 M_\odot}\bigg) \bigg(\frac{R_*}{0.93 R_\odot}\bigg)
\bigg(1-\frac{R_*}{r_{\rm m,in}}\bigg)\bigg]^{1/2} ~.
\end{eqnarray}
If matter falls from the distance of $r_{\rm m,in}=2R_*$ then  $V_{\rm ff}\approx 402$ km/s. It is 464 km/s in case of $r_{\rm m,in}=3R_*$, and 493 km/s in case of  $r_{\rm m,in}=4R_*$. 

The spots have an inhomogeneous structure: the energy flux per unit area is larger in the central regions of the spots and smaller on the outskirts.
The inhomogeneity of the energy distribution in hot spots has been predicted in 3D MHD simulations of stars accreting in stable regime \citep{RomanovaEtAl2004} and confirmed in observations by \citet{EspaillatEtAl2021} and \citet{SinghEtAl2024}. 
In the unstable regime, spots are also inhomogeneous (see bottom panels in Figs. \ref{fig:3d-spots-d0.5-t20} and \ref{fig:3d-spots-d1-t10}). The area covered by the highest energy flux is much smaller than that covered by the lower energy fluxes. 
This could be reflected in fluxes of light curves observed in different wavebands. Fig. \ref{fig:wavelength-dep} shows that the stellar magnitude is the largest (and the flux is the smallest) in the u-band. The flux systematically increases with the wavelength. It is the largest in $i$ and $z$ wavebands.

The stochasticity in light curves arises from the fact that tongues and funnel streams are short-lived features: they deposit matter onto the stellar surface rapidly and form temporary hot spots on the surface. Typically, more powerful tongues live longer and provide longer-living spots on the surface of the star, while 
less powerful tongues form spots that live for a shorter period of time.  
This may explain QPOs observed in wavelets: QPOs with larger power last longer, often a few days, while QPOs with smaller power last shorter time. 

%The shortest-period QPOs observed in our simulations correspond to 0.3-0.5 days. 

\section{Discussion}
\label{sec:Discussion}

 \subsection{Comparisons with \textit{MOST-ASAS} observations }

\citet{RucinskiEtAl2008} and Siwak et al.  (2011, 2014, 2018)
analyzed results of TW Hya observations obtained with \textit{MOST} and \textit{ASAS} telescopes. 
Most of their results are similar to results obtained in our analysis of light curves obtained with \textit{TESS}, ground-based telescopes,
and in numerical simulations. 
%Here, we compare some of their results with results of 
%observations and simulations shown in our paper.
 The common feature of all wavelets is the presence of QPOs with different 
quasi-periods. The duration of QPOs decreases when the quasi-period decreases. 
 \citet{RucinskiEtAl2008} and \citet{SiwakEtAl2014,SiwakEtAl2018} noted that in a few instances period of QPOs decreases with time. Our observational and simulation data show that the period of QPOs may decrease, or increase, or be approximately the same. 
%  We note that in \textit{MOST} observations of  2011, the period of the main QPO slightly decreases then increases.
Additional work is required to track QPO periods with similar amplitudes, which can be done in the future.   \citet{SiwakEtAl2011} noted that a shorter-period oscillations are often observed during higher-amplitude bursts
(see e.g. Fig. 1 from their paper). We observed similar features in Sec. 9 and 36 of TESS light curves but less so in Sector 63. 
In numerical simulations, this phenomenon is observed in models with larger-sized magnetospheres, e.g., in models $\mu2\theta5$ and 
$\mu2\theta20$.   \citet{SiwakEtAl2018} noted that the time duration of QPOs may be equal to a few dynamical rotations. Using one of our models ($\mu0.5\theta5$) where 1-2 unstable tongues were observed, we demonstrate that QPO periods can be associated with 3-4 Keplerian rotations of the inner disc, which is in accord with their hypothesis.  In models with a larger number of tongues, the picture of 
QPOs is more complex, though the strongest tongue may develop a QPO with the period of the inner disc.

\subsection{Comparisons with other works}
\label{sec:comparisons}

\textbf{\citet{DonatiEtAl2011,DonatiEtAl2024}} measured the magnetic field in TW Hya in different observational sets, and obtain different strengths and tilts of the dipole component. In more recent paper, (\citealt{DonatiEtAl2024}, hereafter D24) they  have shown that 
the magnetic field of TW Hya represents a slightly tilted dipole field     
with a strength of $B_d=990-1190$ G. The left panels of Fig. 5 of their paper show that the radial component of the field (its stronger part) is located  at  $\phi\approx 30^\circ-40^\circ$ off the pole. 
The authors estimated the magnetospheric radius  $r_m\approx 4.5^{+2.0}_{-1.1} R_*$, using the formulae of \citet{BessolazEtAl2008} ($B=1.1$ kG and ${\rm log} (\dot M)=-8.65$). They concluded that matter flows from the inner disc to the star in a stable funnel stream and accretes close to the magnetic pole.
Our simulations (see Tab. 3) show that at comparable values of $B_d=1$kG and similar accretion rate, $\dot M=2\times 10^{-9}\dot M/{\rm yr}$ the magnetospheric radius is $r_m\approx 4.9 R_*$  which is in a good agreement with that derived by D24. Our model shows that at these radii, accretion is unstable. However, 
in models with large magnetospheres (at $\mu=1.5$ and $2.0$), the unstable tongues are lifted above the inner magnetosphere (like in a stable regime) and form spots at high latitudes within $\phi\sim 30^\circ-40^\circ$ off the magnetic pole .  
However, instead of one funnel stream, there are several streams and spots that reside at relatively high latitudes. More detailed observations are needed to distinguish between one or several accretion spots
located at similar latitudes.

For estimates, D24 accepted parameters: $M_*=(0.8\pm0.01) M_\odot$, $R_*=(1.16\pm0.13) R_\odot$   \citep{BaraffeEtAl2015}
and corotation radius $r_{\rm cor}=(7.9\pm 0.3) R_*$.
We performed special simulation runs using their corotation radius, $r_{\rm cor}\approx 7.9R_*$ and our model  parameters  $\mu=1$ and $1.5$. We obtain from simulations $r_m\approx 4.3 R_*$  at $\mu=1$ and
$r_m\approx 4.8 R_*$  at $\mu=1.5$ (this radius is expected to be even larger if $\mu=2$, see trend in Tab. 3). These radii are close to  those obtained in our main models  (where $r_{\rm cor}=9.6 R_*$ ). The ratios are  $r_m/r_{\rm cor}\approx 0.54$ and 0.61, respectively. They are slightly larger than in our main models and correspond to milder unstable regimes.  In a milder regime, the unstable tongues are lifted above the inner magnetosphere at larger distances from the star, 
 and spots form closer to the magnetic pole at  $\phi\lesssim 30^\circ$  from the  pole. These results are in line with observations of D24.

In earlier observations of TW Hya, \citet{DonatiEtAl2011} derived predominantly octupolar magnetic field with a much smaller dipole component of a few hundred Gauss. At a weaker dipole component, the magnetospheric radius is smaller, and a star is expected to be in a strongly unstable regime. The light curve is expected to be more stochastic, and spots form further away from the magnetic pole.  We do not possess the detailed light curve during their observational period. Future observations of the magnetic field structure with simultaneous photometric observations would be interesting for further understanding the dependence of the light curve on the structure of the dipole magnetic field. 

\smallskip

\textbf{\citet{SiciliaEtAl2023}} analyzed several spectral lines that are expected to form near the footprint of the funnel flow (such as HeI 5016 A and FeII 5018) and found a periodic signal in radial velocities with a period of the star.  They concluded that matter producing these spectral lines flows in a stable funnel stream, while stochastic photometric variability 
is produced by some spots that form in random places and have a random strength. We suggest that both phenomena can be explained by a model of accretion in a mildly unstable regime, in which spots form at a higher latitude and provide chaotic components in the light curve. At the same time, due to the tilt of the dipole and the inclination angle of the observer, the whole set of spots (located around the magnetic pole) rotates about the rotational axis of the star with the angular velocity of the star and produces periodic components in spectra.
 We note that in Fig. 5 of these authors (the folded phase-velocity plot 
for HeI 5016 A and FeII 5018 lines) there is a significant scatter of velocities, which may be connected to multiple funnel streams that fall near the magnetic pole but not in the same location. 

\smallskip

\textbf{\cite{GRAVITY2020}} used interferometry to measure the size of the magnetosphere and radius of the TW Hya star (see details in their paper).   They measured the radius of the $Br\gamma$ emitting region as 
$R_{Br\gamma}=(3.49\pm0.20) R_*$ and suggested that this radius corresponds to the magnetospheric radius $r_m$. They measured the radius of the star as 
$R_*=(1.29\pm0.19)R_\odot$. At this radius  (and other parameters:  $M_*=0.8 M_\odot$, $P_*=3.566$ d), we obtain a corotation radius $r_{\rm cor}=7.06 R_*$ that is smaller than that accepted in our paper ($r_{\rm cor}=9.6 R_*$). 
We calculated an additional model taking $r_{\rm cor}/R_*=7.06$ and magnetospheric parameter $\mu=1$. We observed an unstable regime of accretion.  From simulations, we measured the magnetospheric radius $r_m\approx 3.9 R_*$ and the ratio $r_m/r_{\rm cor}\approx 0.63$. This corresponds to a mildly unstable regime.

\begin{figure*}
     \centering
   \includegraphics[width=0.4\textwidth]{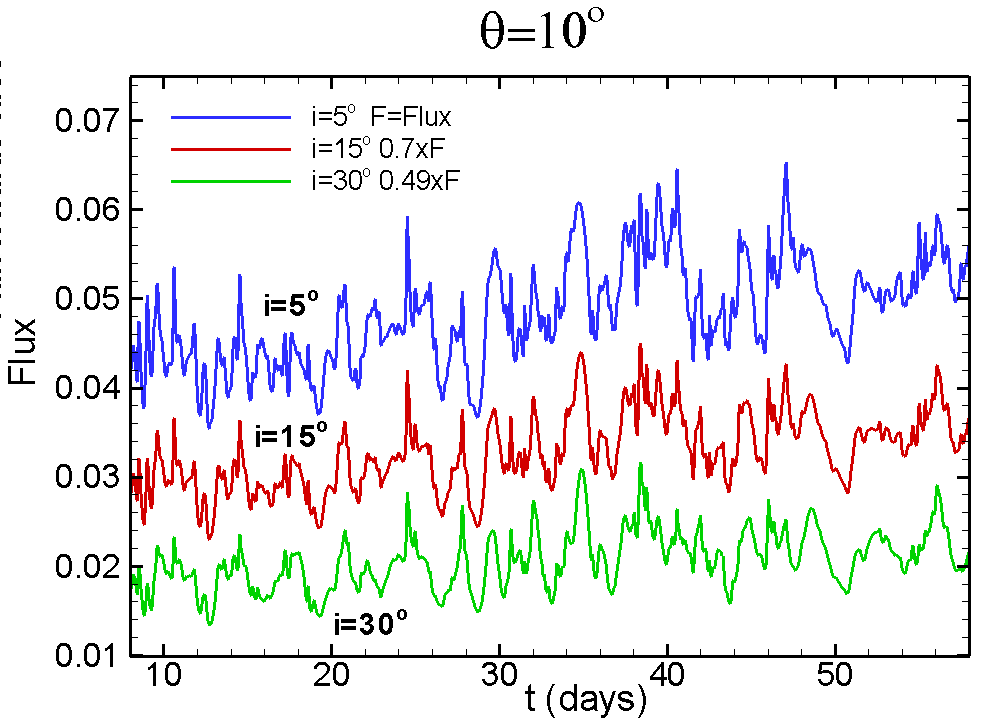}  
     \includegraphics[width=0.4\textwidth]{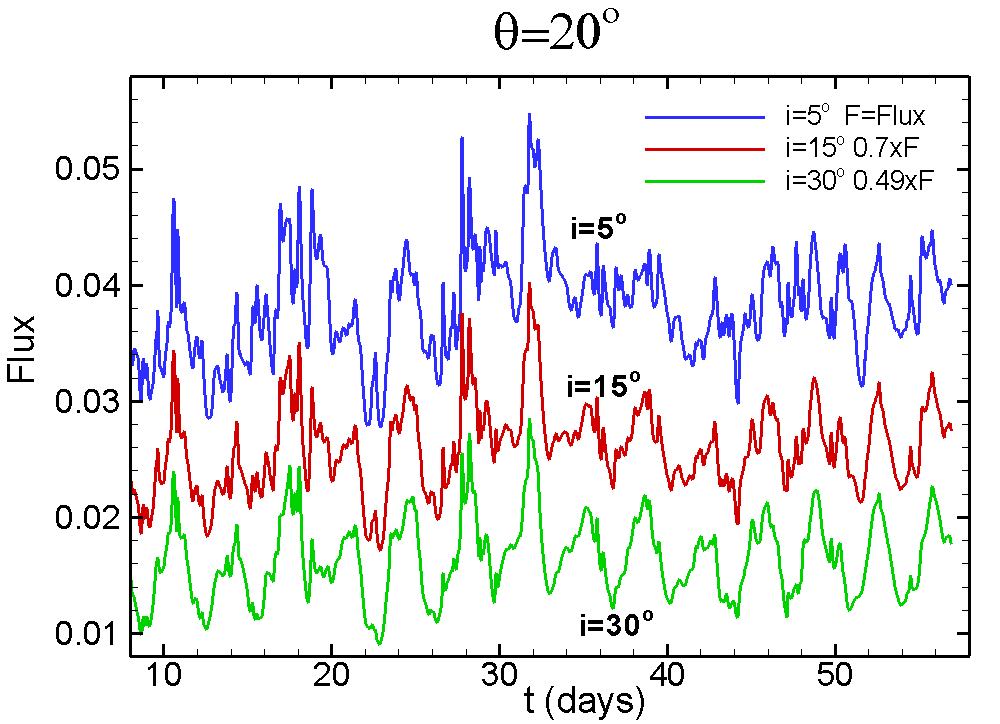}  
    \caption{Light curves in the model with $\mu=1$ and tilts of the dipole $\theta=10^\circ$ (left panel) and $\theta=20^\circ$ (right panel) 
shown for different inclination angles of the observer: $i=5^\circ, 15^\circ$, and $30^\circ$. Fluxes are scaled for better visibility.
     \label{fig:incl-3-d1-t20}}
\end{figure*}

\subsection{More complex fields}

The magnetic field of TW Hya may have strong octupolar and higher-order components of the field 
 \citep{DonatiEtAl2011,DonatiEtAl2024}. 
Earlier 3D MHD
simulations of models where the octupole component has been included show that initially, matter flows from the disc to the star along the dipole field, but closer to the star the octupole component redirects matter to a new position on the surface of the star determined by the octupole field (e.g., \citealt{LongEtAl2008,LongEtAl2011,LongEtAl2012,RomanovaEtAl2011}, see also  \citealt{GregoryEtAl2010}). In this  case, 
the interchange instability at the disc-magnetosphere boundary will also produce unstable tongues and multiple flares in the variability curve. However, the location of hot spots will be determined by the octupolar field. 
In another research, the authors show evidence of accretion onto a low-latitude, almost equatorial spot using X-ray observations and argue that matter may accrete to one of the poles of the complex field \citep{ArgiroffiEtAl2017}. It is possible if the dipole component is very weak  (e.g.,\citealt{GregoryEtAl2010}). Or if some unstable tongues penetrate deeper into a small dipole magnetosphere.
A more detailed answer requires modeling accretion to a star with a dipole plus a more complex field, which can be done in the future.

\subsection{Other possible origins of stochasticity}

Stochastic-looking light curves can also originate in stable two-funnel magnetospheric accretion if matter flowing into the funnel has inhomogeneities of different sizes caused by turbulence.  \citet{RobinsonEtAl2021} used a one-dimensional accretion model from a turbulent disk, showing that it may explain stochastic accretion (see also \citealt{RobinsonEtAl2017}). This model requires more detailed 3D simulations with well-resolved turbulence in the disc. 

3D MHD simulations of MRI-driven turbulent matter onto a star with a very weak (dynamically unimportant) magnetic field (where the disc matter accretes onto the surface of the star in the boundary layer regime) show that the disc has turbulent cells of different sizes (see Figs. 3 and 4 from \citealt{RomanovaEtAl2012}; hereafter R12).  The accretion rate curve shows high-amplitude peaks associated with the accretion of the largest turbulent cells and more frequent small-amplitude peaks associated with smaller-sized cells  (see Fig. 6 from R12).  
In similar models but with larger, dynamically important  magnetospheres,  matter accretes predominantly through the large-scale azimuthally-elongated turbulent cells if the tilt of the dipole is large $\theta=30^\circ$ (see Figs. 7-9 from R12). 
 In models with a small tilt of the dipole $\theta=2^\circ$, turbulent matter penetrates the magnetosphere through interchange instability (see right panel of Fig. 7  from R12).

The irregular photometric variability of CTTS can also be caused by
variable circumstellar extinction. 
Such obscuration by dust is more typical for stars with a high inclination angle, where the line of sight intersects with the dusty disc winds or warp in the disc  (e.g., \citealt{BouvierEtAl2007a}). 
However, TW Hya is observed almost pole-on (e.g., \citealt{QiEtAl2004}),  and therefore, variability associated with dust extinction is expected to be insignificant.

%\smallskip

\subsection{Possible role of winds. Two-channel accretion}

Many spectral lines of TW Hya
%(e.g., Balmer lines)
show blueshifted absorption (e.g., \citealt{AlencarBatalha2002,LamzinEtAl2004,HerczegEtAl2023,WendebornEtAl2024c}), which points to the presence of winds. Winds can originate at the surface of the star  (e.g., \citealt{MattPudritz2005}) and may form polar winds (e.g., \citealt{Cranmer2009,DupreeEtAl2012}) or at the disc-magnetosphere boundary
(e.g., \citealt{ShuEtAl1994,GoodsonEtAl1997,RomanovaEtAl2009,ZanniFerreira2013,LiiEtAl2014}). 
The variability in blueshifted absorption typically does not correlate with variability in spectral lines or with photometric variability (e.g., \citealt{AlencarBatalha2002}). \citet{DonatiEtAl2024} notes that no periodicity was observed in the blue wing of the He 1083 nm line (see their Fig 8), which means that it either comes from an erratic stellar wind or the inner disc. 
The origin of winds is yet to be understood. 

Recent 3D MHD simulations of accretion onto a magnetized star confirmed accretion through instabilities (e.g., \citealt{TakasaoEtAl2022, ParfreyTchekhovskoy2023,ZhuEtAl2024}).  
\citet{ZhuEtAl2024} used the code with a high grid resolution (Cartesian with mesh refinement) and  have shown that unstable filaments have a finer substructure which have not been seen in current work.  This may explain the short-period QPOs observed by the \textit{TESS} telescope.  \citet{TakasaoEtAl2022} report on observations of unstable accretion at a wider range of $r_m/r_{\rm cor}$ compared with condition $r_m/r_{\rm cor}\lesssim 0.71$ found by \citet{BlinovaEtAl2016}. They observed inflation and reconnection of the field lines connecting a star and the disc  (e.g., \citealt{LovelaceEtAl1995}). 
\citet{TakasaoEtAl2022} observed the formation of conical-shaped winds in the case of faster-rotating stars, which may lead to large-scale outflows. However, ejections are less ordered than in axisymmetric simulations   (e.g., \citealt{ZanniFerreira2013,LiiEtAl2014}). In cases of slowly-rotating stars, 
 \citet{TakasaoEtAl2022} repot on the formation of slow turbulent winds which fail to live the system and accrete back to the star. These ``failed" winds may be another source of variability in CTTSs.

Simulations also point to two-channel accretion, where some matter accretes in several unstable tongues in the equatorial plane while some matter
accretes from the top layers of the disc in funnel streams  \citep{ZhuEtAl2024}.
Earlier,  \citet{BachettiEtAl2010} studied two-channel accretion in 3D simulations and applied it to accreting millisecond pulsars  (which represent a scaled version of CTTSs).
They  noticed that in 
the unstable regime moving spots at the stellar surface may produce two frequencies:  the higher frequency is associated with unstable tongues, driven by the inner disc, and the lower frequency is caused by the moving spots, resulting from funnel streams originating at larger distances from the star. 
They used this model to explain two QPO frequencies observed in accreting millisecond pulsars.
In our current models, we also see both accretion through instabilities and funnel accretion, which starts at the surface layers of the disc. The latter becomes more significant in models with milder unstable regime and  larger-sized magnetospheres. 

\subsection{Projection to other CTTSs}

TW Hya is seen almost pole-on. In models, we can look at the star from different observer angles. We observed that at larger inclination angles, the light curve becomes more periodic or quasi-periodic. At the same time, the unstable accretion continues, providing QPO flares on different time scales. 
For example, the light curve in the model  $\mu=1$ with the tilted angle of the dipole $\theta=20^\circ$ looks stochastic at  
 $i=5^\circ$, but at  $i=15^\circ$, and $i=30^\circ$, it becomes more quasiperiodic  (see right panel of Fig. \ref{fig:incl-3-d1-t20}). 
On the other hand, at a small tilt of the dipole  $\theta=10^\circ$, the light curve looks stochastic at all above inclination angles (see left panel of the same plot). It becomes more periodic at larger inclination angles.  Overall, we expect that the period of the star becomes more visible when the inclination angle increases because the whole set of spots is located around magnetic poles and rotates around the rotational axis together with the magnetic axis  (see also 
\citet{StaufferEtAl2014}).   At the same time, the stochastic component stays.

If a star is seen at high inclination, one
can expect to observe stochastic occultations of the photosphere
by dust lifted above the disk plane near the base of the accretion
tongues whenever one passes in front of our line of sight, leading
to light curves dominated by aperiodic extinction events (e.g., \citealt{McGinnisEtAl2015,PetrovEtAl2019})\footnote{Earlier, this idea was proposed by \citealt{BouvierEtAl1999} for stellar light obscuration by inner bending wave in the disc.}.

\section{Conclusions}
\label{sec:Conclusions}

The conclusions of our work are the following:

\textbf{1.} Wavelet analysis of light curves obtained with the \textit{TESS} telescope in Sectors 9, 36, and 63  show multiple QPOs ranging from 8-9 days to less than 2 hours. QPOs with periods close to the period of the star (3.56 days) are often present where the period varies in the range of 3 to 5 days. QPOs with longer periods of 7.5-8.8 days are always present.  Multiple short-period QPOs are present and last for a shorter time. 

\textbf{2.} A subset of ODYSSEUS light curves covering 70 days in 2022 (part of Epoch 2) was selected due to more frequent observations and analyzed in detail.  The light curve looks stochastic. Wavelet analysis shows multiple QPOs similar to those observed with \textit{TESS}.   QPOs with the stellar period are seen in both wavelet and Lomb-Scargle periodograms.  No stellar period is seen in Epoch 1.

%Lomb-Scargle periodogram shows multiple quasiperiods for Epoch 1 observations, where period of the star has a small %amplitude compared with many other QPOs. In Epoch 2, the period of the star dominates, while QPOs with different periods %are also present. 

\textbf{3.} We developed 3D MHD models of a star with TW Hya parameters and several magnetosphere sizes from $3.2R_*$ to $5.6R_*$. In all models, matter accretes in the unstable regime, where unstable tongues form moving hot spots, which produce stochastic-looking light curves and QPOs with different quasiperiods. In some models, the light curves and wavelets are strikingly similar to those obtained from observations by \textit{TESS} and ground-based telescopes.

\textbf{4.} In models with smaller tilts of the dipole magnetosphere $\theta=5^\circ$ and $10^\circ$, variability is more stochastic, and the amplitude of 
QPO associated with the stellar period is either smaller or comparable to other QPOs.     At larger tilts $\theta=15^\circ-20^\circ$, the QPO associated with a stellar period typically dominates in Fourier and wavelet spectra.  

\textbf{5.}  The magnetic field of TW Hya varies from year to year (e.g., \citealt{DonatiEtAl2011,DonatiEtAl2024}). We suggest that the difference in light curves obtained in different years may be connected with variations in the strength and tilt of the dipole field.

\textbf{6.} In a model with a small magnetosphere ($r_m\approx 3.2 R_*$), matter accretes in one or two funnel streams, which rotate with  period of the inner disc. This quasi-period may be mistakenly accepted to be period of the star. 

\textbf{7.} We show that persistent long-period QPOs can be caused by modulation of matter flow by density waves that form beyond the corotation radius.

\textbf{8.} Unstable tongues are stopped by the inner parts of the magnetosphere, form funnel-like streams, and hit the star at 
an angle of $\phi\sim 20^\circ-30^\circ$ (from magnetic pole)  in models with larger magnetospheres, and 
$\phi\sim 35^\circ-45^\circ$,  in models with smaller magnetospheres. 

\textbf{9.} Light curves may have both: properties of the ordered magnetospheric accretion (due to matter flow around the closed magnetosphere) and stochasticity due to accretion of multiple tongues. In cases of a larger inclination angle of the observer, the light curves may become more ordered but with a stochastic component. 

\section*{Acknowledgments}
The authors thank Dr. Michal Siwak for a careful reading of our paper and the insiteful, helpful report. Resources supporting this work were provided by the NASA High-End
Computing (HEC) Program through the NASA Advanced Supercomputing
(NAS) Division at Ames Research Center and the NASA Center for
Computational Sciences (NCCS) at Goddard Space Flight Center. MMR and RVEL were supported
in part by the NSF grant AST-2009820.  They thank Dr. Alexander Koldoba for earlier developed codes and Bez Thomas for technical support.  CCE and JW were supported by NSF AST-2108446, HST AR-16129, and NASA ADAP
80NSSC20K0451.

\section{Data Availability}

The data underlying this article will be shared on reasonable request to the corresponding author (MMR).

\appendix

\section{Details of the numerical model}
\label{sec:model}

Reference parameters of the mass $M_0$, scale $R_0$, velocity $v_0$, and periods of rotation $P_0$ were described in Sec. \ref{sec:setup}.
Other reference parameters depend on the magnetic field of the star $B_*$ (at the equator) and dimensionless parameter $\mu$ , which
 determines the final size of the magnetosphere.  
 We determine the reference magnetic
field $B_0$ and magnetic moment $\mu_0=B_0 R_0^3$ such that
$\mu_0=\mu_*/\mu$, where
 the  magnetic moment of the star
 $\mu_*=B_* R_*^3$.  
Then $B_0=\mu_0/R_0^3$.  
We determine the reference density from pressure balance at $R_0$: 
$\rho_0=B_0^2/v_0^2$. 
The  reference value for the accretion rate
$
\dot M_0=\rho_0 v_0 R_0^2 ~,
$
for energy flux (used to convert dimensionless energy fluxes shown in plots for light curves to dimensional ones) 
$
\dot E_0=\dot M_0 v_0^2 ~ ,
$
and for energy flux per unit area (used to show energy flux distribution in hot spots on the surface of the star)
$
F_0={\dot E_0}/{R_0^2}~.
$
Tab. \ref{tab:refval} of the Appendix shows reference values for different $\mu$ and $B_*$.

\begin{table*}
\begin{tabular}[]{l|l|l|l|l|l|l}
\hline
\hline
& Magn. field at the pole & {$B_{\star}$ (G)}                                                    &      1000                           &      800                            &   600      \\
\hline 
& Equatorial field  & {$B_{\star}$ (G)}                                                              &      500                           &      400                            &   300      \\
\hline 
\hline
& Reference density                   & $\rho_0$ (g cm$^{-3}$)                                 &   $3.24\times 10^{-12}$  &   $2.08\times 10^{-12}$   &  $1.17\times 10^{-12}$    \\
$\mu=0.5$  & Reference accretion rate       & $\dot M_0$ ($M_\odot yr^{-1}$)       &  $4.15\times 10^{-8}$     &  $2.66\times 10^{-8}$     &    $1.50\times 10^{-8}$    \\
& Reference energy flux         & $\dot E_0$ (erg s$^{-1}$)                                   &   $1.50\times 10^{33}$   &   $9.58\times 10^{32}$    &    $5.39\times 10^{32}$   \\
& Ref. flux per unit area  & $ F_0$ (erg s$^{-1}$cm$^{-2}$)                                &  $4.37\times 10^{10}$      &    $2.80\times 10^{10}$   &    $1..57\times 10^{10}$     \\
\end{tabular}
\begin{tabular}[]{l|l|l|l|l|l|l}
\hline
&Reference density                   & $\rho_0$ (g cm$^{-3}$)                                  &   $8.11\times 10^{-13}$  &   $5.19\times 10^{-13}$  &  $2.92\times 10^{-13}$    \\
$\mu=1.0$&Reference accretion rate       & $\dot M_0$ ($M_\odot yr^{-1}$)          &  $1.03\times 10^{-8}$     &  $6.65\times 10^{-9}$     &    $3.74\times 10^{-9}$    \\
&Reference energy flux         & $\dot E_0$ (erg s$^{-1}$)                                    &   $3.74\times 10^{32}$   &   $2.39\times 10^{32}$    &    $1.35\times 10^{32}$   \\
&Ref. flux per unit area  & $F_0$ (erg s$^{-1}$cm$^{-2}$)                                  &  $1.09\times 10^{10}$    &     $7.00\times 10^{9}$   &    $3.94\times 10^{9}$     \\
\end{tabular}
\begin{tabular}[]{l|l|l|l|l|l|l}
\hline
&Reference density               & $\rho_0$ (g cm$^{-3}$)                                     & $3.60\times 10^{-13}$    &  $2.31\times 10^{-13}$  &  $1.30\times 10^{-13}$    \\
 $\mu=1.5$     &Reference accretion rate       & $\dot M_0$ ($M_\odot yr^{-1}$)   &  $4.62\times 10^{-9}$     &  $2.96\times 10^{-9}$    &    $1.66\times 10^{-9}$    \\
&Reference energy flux         & $\dot E_0$ (erg s$^{-1}$)                                   &  $1.66\times 10^{32}$     &  $1.06\times 10^{32}$    &    $5.99\times 10^{31}$   \\
&Ref. flux per unit area  & $F_0$ (erg s$^{-1}$cm$^{-2}$)                                 &  $4.86\times 10^{9}$     &  $3.11\times 10^{9}$     &    $1.75\times 10^{9}$   \\
\end{tabular}
\begin{tabular}[]{l|l||l|l|l|l|l}
\hline
&Reference density               & $\rho_0$ (g cm$^{-3}$)                                   & $2.03\times 10^{-13}$    &  $1.30\times 10^{-13}$  &  $7.30\times 10^{-14}$    \\
  $\mu=2.0$ &Reference accretion rate       & $\dot M_0$ ($M_\odot yr^{-1}$)    &  $2.60\times 10^{-9}$     &  $1.66\times 10^{-9}$   &   $9.36\times 10^{-10}$    \\
&Reference energy flux         & $\dot E_0$ (erg s$^{-1}$)                                 &  $9.35\times 10^{31}$     &  $5.99\times 10^{32}$  &    $3.99\times 10^{31}$   \\
&Ref. flux per unit area  & $F_0$ (erg s$^{-1}$cm$^{-2}$)                               &  $2.70\times 10^{9}$       &  $1.75\times 10^{9}$     &    $1.75\times 10^{9}$   \\
\hline
\end{tabular}
\caption{Reference values for models with different parameters $\mu$ and values of the equatorial magnetic field of the star $B_*$. Note that the magnetic field provided by observations 
(e.g., \citealt{DonatiEtAl2024}) corresponds to the field at the magnetic pole and is twice as large compared with the equatorial field given in the table. 
\label{tab:refval}}
\end{table*}

 3D MHD equations  were solved numerically using a Godunov-type
numerical scheme,  where the seven-wave Roe-type approximate Riemann solver is used. The ``cubed-sphere'' coordinate system
rotates with the star \citep{KoldobaEtAl2002}. 
The energy equation is written in the form of entropy balance, and
the equation of state is that of an ideal gas. Viscosity is modeled using the $\alpha$-model
\citep{ShakuraSunyaev1973}, and is incorporated only into the
disc so that it controls the accretion rate through the disc. We use a small
$\alpha$-parameter $\alpha=0.02$ in all simulation runs.

 At both the inner and outer
boundaries, most variables $A_j$ are taken to have free
boundary conditions at both the inner and outer boundaries ${\partial A_j}/{\partial r}=0$. At the stellar
surface, accreting gas can cross the surface of the star
without creating a disturbance in the flow. These conditions
neglect the complex physics of interaction between the accreting
gas and the star. The cross-section of the inflowing matter at the distance of
stellar radius is interpreted as a spot, and the energy flux distribution in the flow provides us with
the energy distribution in hot spots.  
The magnetic field is
frozen onto the surface of the star. That is, the normal component
of the field, $B_n$, is fixed, while
    the other components of the
magnetic field vary. At the outer boundary, matter flows freely out of the region. We forbid the back flow of matter from the outer boundary
 into the simulation region.

\begin{figure*}
     \centering
     \includegraphics[width=0.7\textwidth]{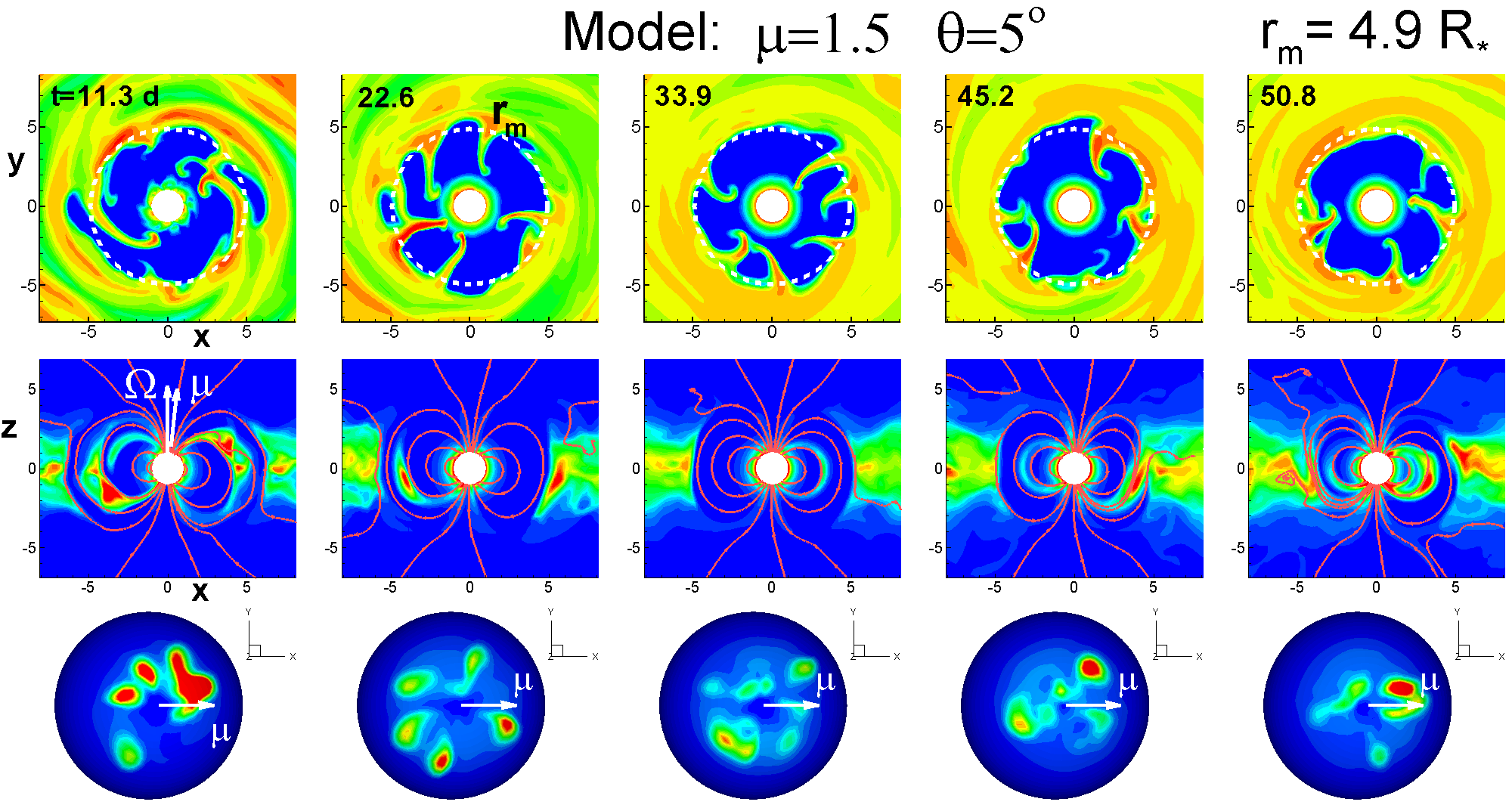}
  \includegraphics[width=0.9\textwidth]{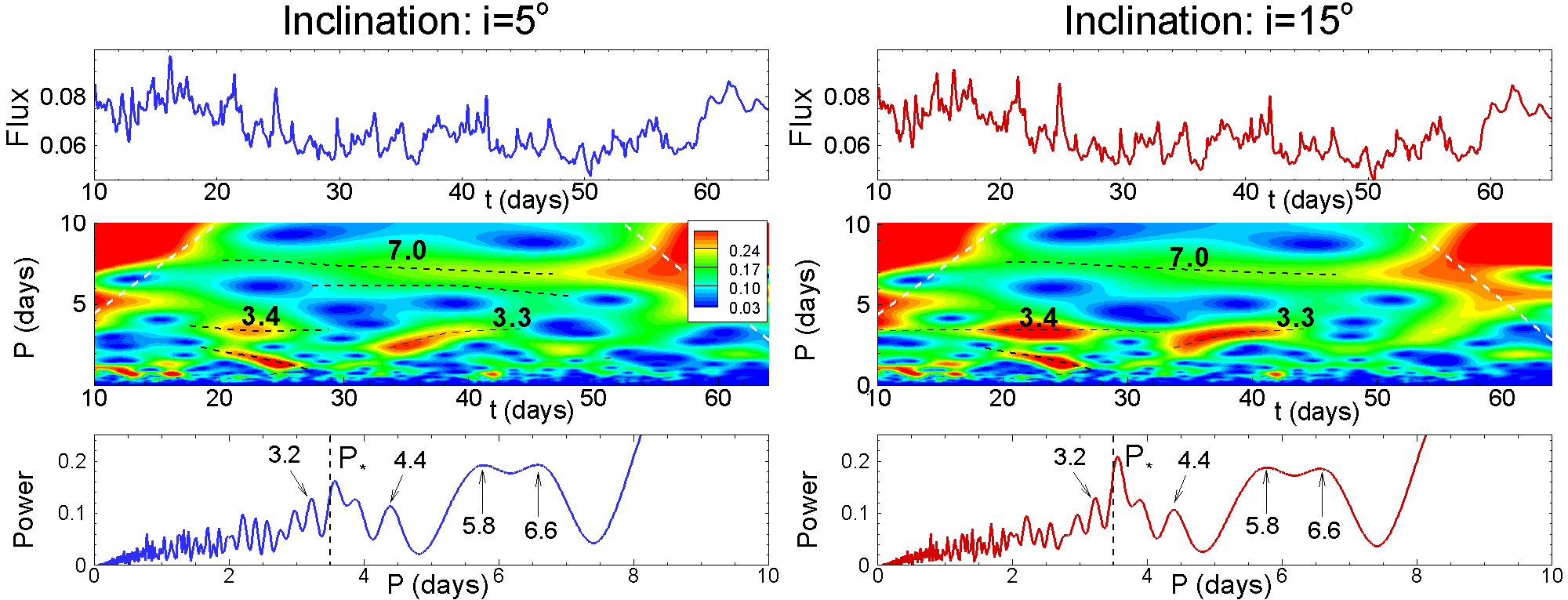}
    \caption{The same as in Fig. \ref{fig:2d-d0.5-t20} but for the model  $\mu 1.5\theta 5$.
     \label{fig:2d-d1.5-t5}}
\end{figure*}

\begin{figure*}
     \centering
     \includegraphics[width=0.7\textwidth]{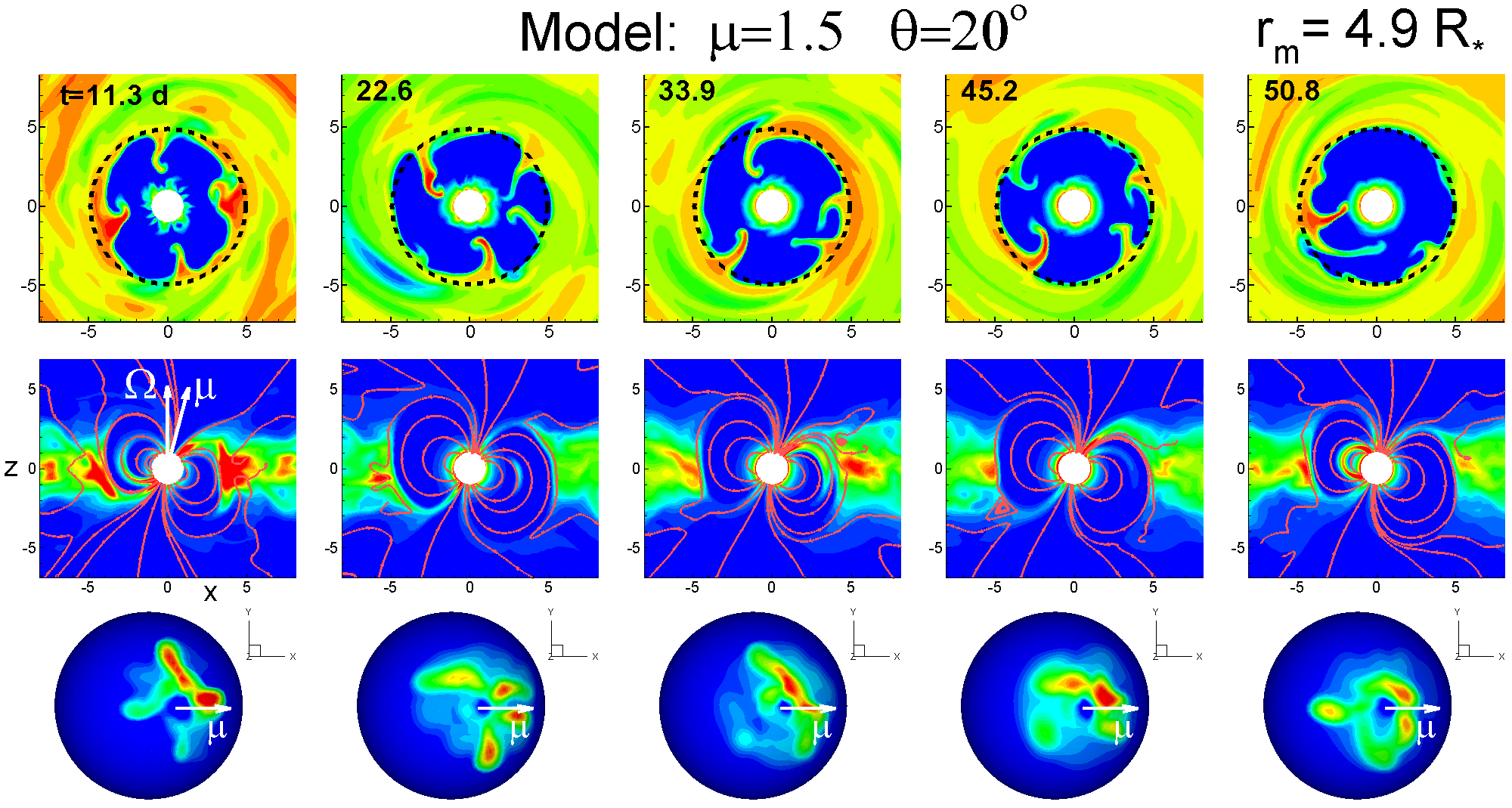}
 \includegraphics[width=0.9\textwidth]{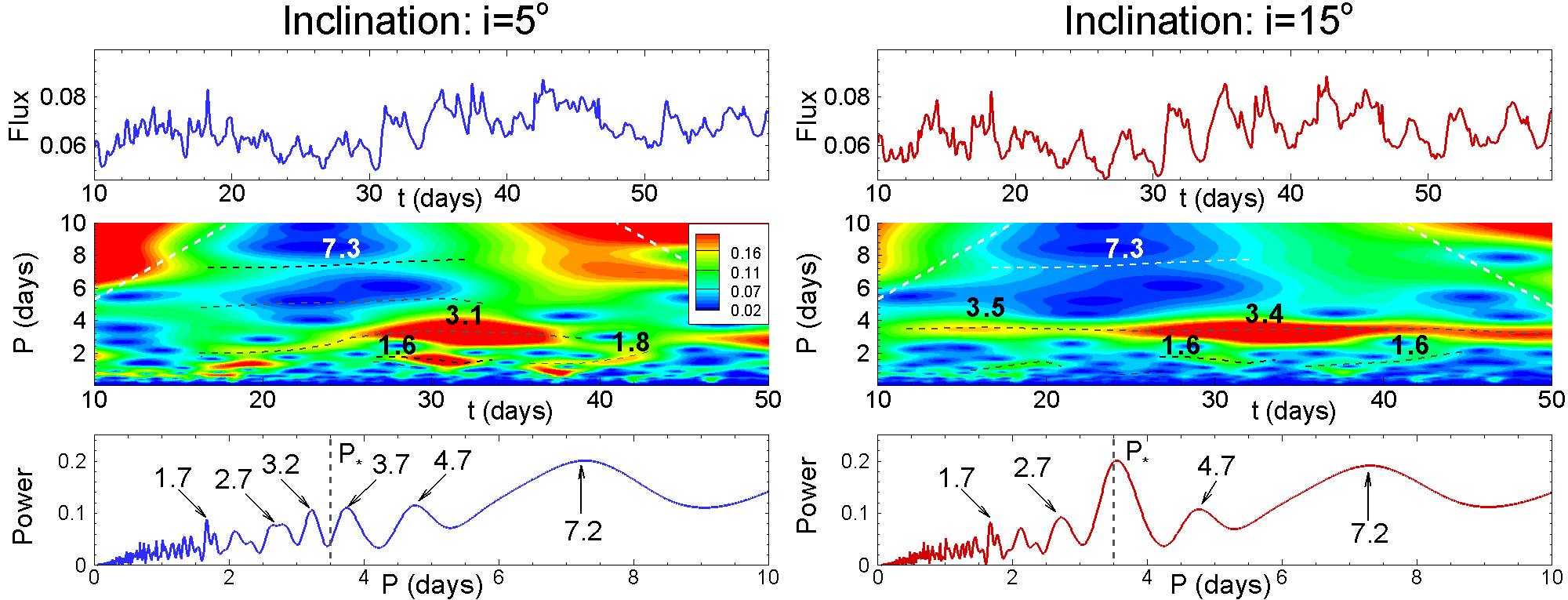}
    \caption{The same as in Fig. \ref{fig:2d-d0.5-t20} but for the model  $\mu 1.5\theta 20$.
     \label{fig:2d-d1.5-t20}}
\end{figure*}

\begin{figure*}
     \centering
     \includegraphics[width=0.65\textwidth]{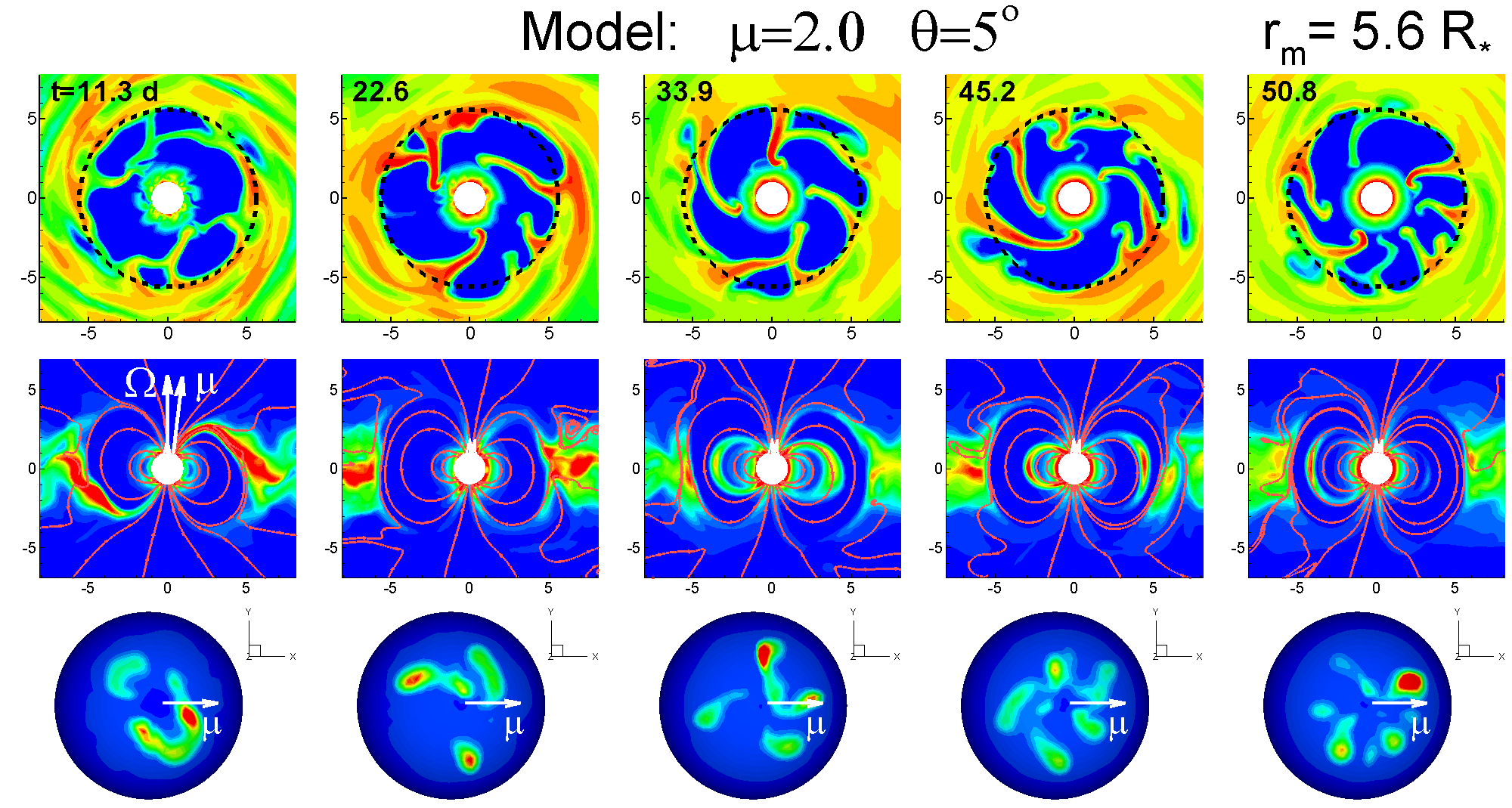}
  \includegraphics[width=0.7\textwidth]{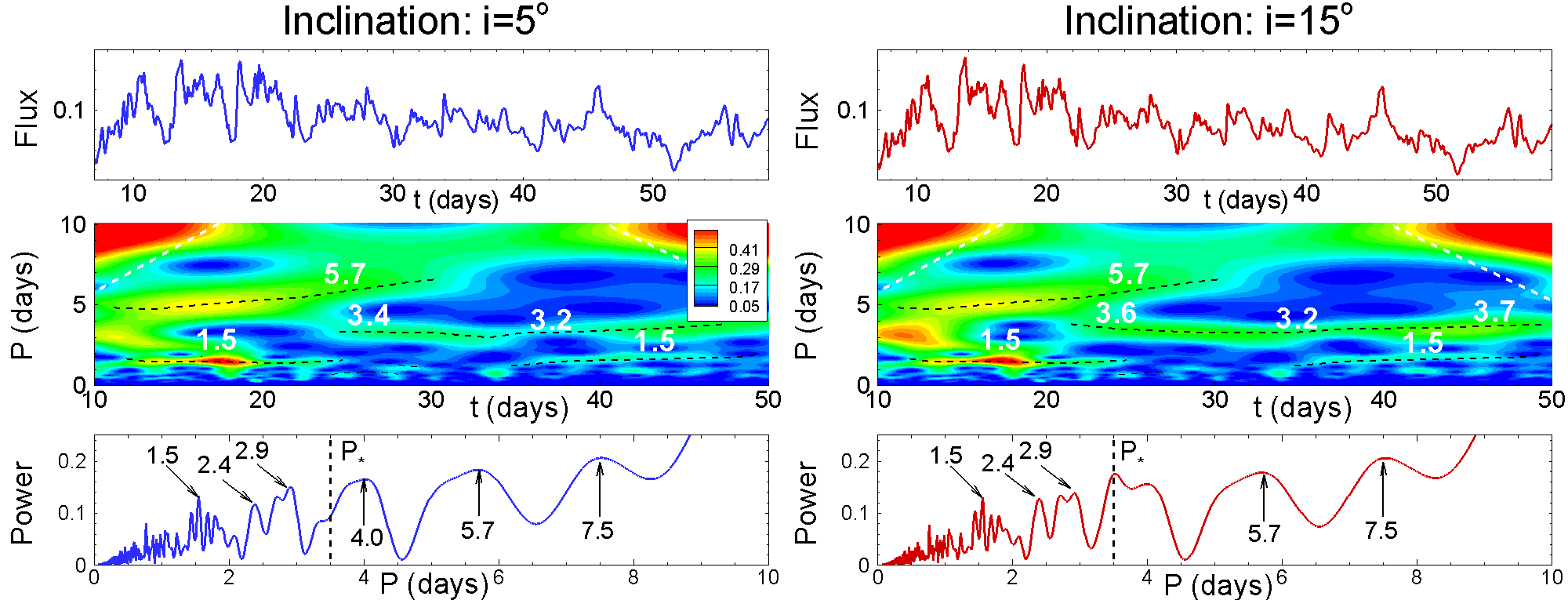}
    \caption{The same as in Fig. \ref{fig:2d-d0.5-t20} but for the model  $\mu 2\theta 5$.
     \label{fig:2d-d2-t5}}
\end{figure*}

\begin{figure*}
     \centering
     \includegraphics[width=0.65\textwidth]{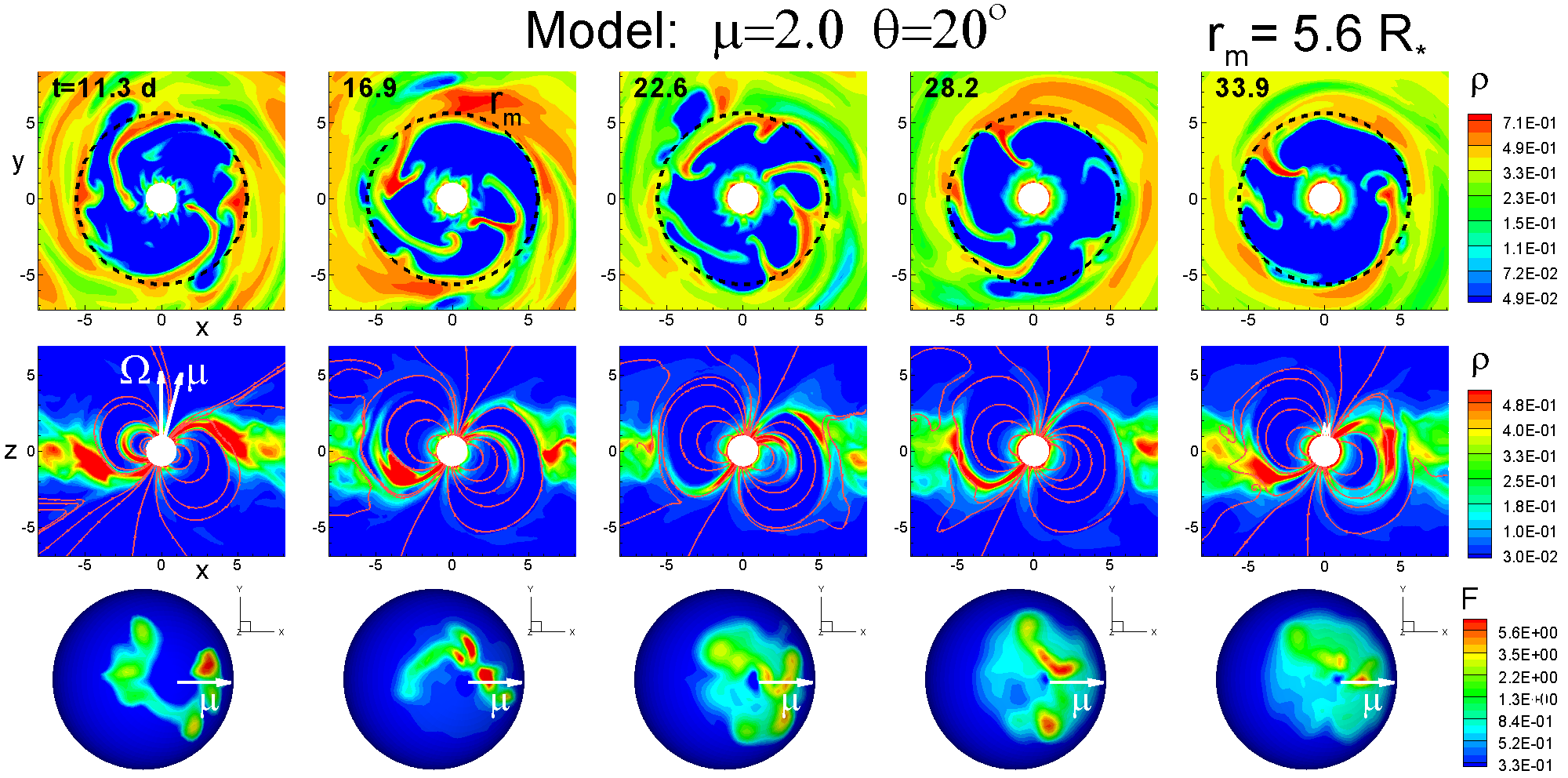}
 \includegraphics[width=0.7\textwidth]{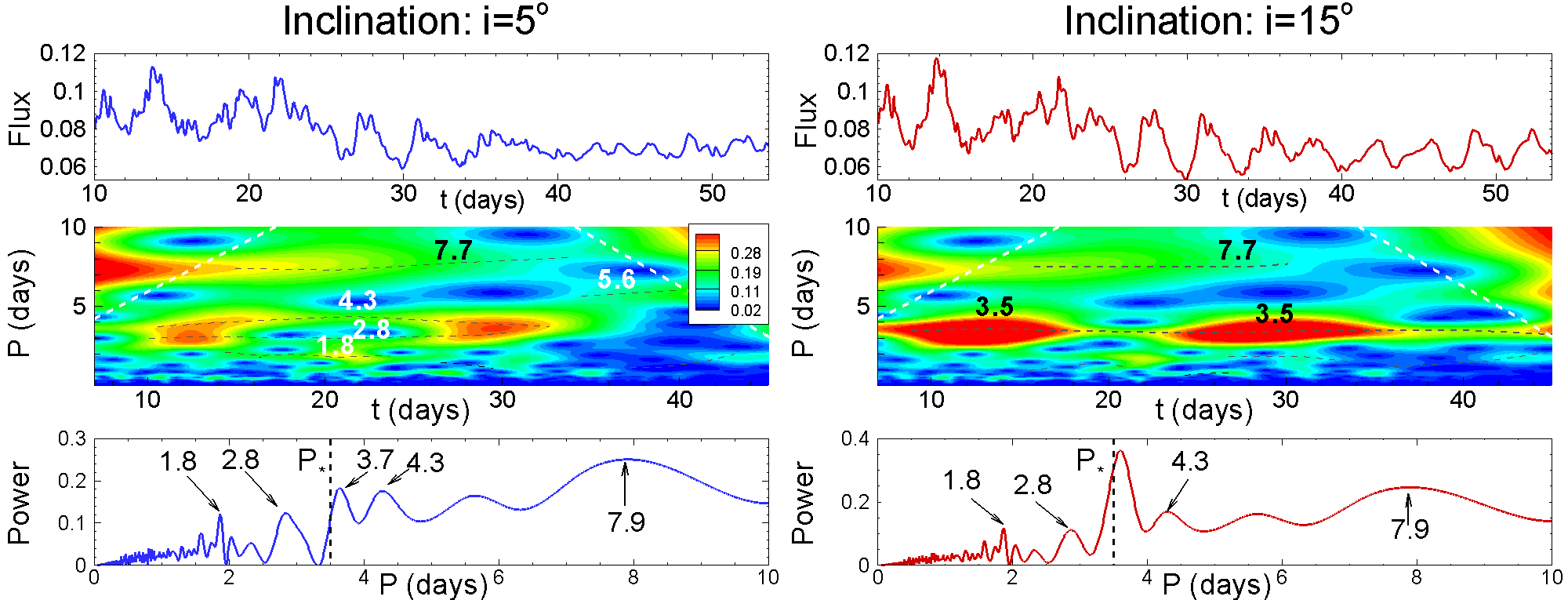}
    \caption{The same as in Fig. \ref{fig:2d-d0.5-t20} but for the model  $\mu 2\theta 20$.
     \label{fig:2d-d2-t20}}
\end{figure*}

\bibliographystyle{mn2e}

\end{document}